\documentclass[aps,prd,twocolumn,10pt,superscriptaddress,nofootinbib]{revtex4-2}
\usepackage{amsmath,amssymb}
\usepackage{ragged2e}
\usepackage{caption}
\usepackage{subcaption}
\usepackage{booktabs}
\usepackage{array}
\usepackage{rotating}
\usepackage{tabularx}
\usepackage{multirow}
\usepackage{bigdelim}
\usepackage{soul}
\usepackage{orcidlink}
\usepackage{lineno}
\usepackage{hyperref}
\usepackage[table]{xcolor}
\usepackage{enumitem}
\usepackage{mathrsfs}
\begin{document}

\title{
 Comisso-Asenjo Mechanism in Rotating 
 \texorpdfstring{$\mathcal{N}=2,U(1)^2$}{N=2,U(1)2} Gauged Supergravity Black Holes:
 Extended Comparison With Kerr Black Hole
 }

\author{Abhinav Jaguri \orcidlink{0009-0008-4197-6006}}
\email{Abhinavjaguri1234@gmail.com}
\affiliation{Department of Physics, School of Sciences, H.N.B. Garhwal University, Birla Campus, Srinagar Garhwal, Uttarakhand – 246174, India}

\author{Hemwati Nandan \orcidlink{0000-0002-1183-4727}}
\email{hemwati.nandan.physics@gmail.com}
\affiliation{Department of Physics, School of Sciences, H.N.B. Garhwal University, Birla Campus, Srinagar Garhwal, Uttarakhand – 246174, India}

\author{Pankaj Sheoran \orcidlink{0000-0001-8283-8744}}
\email{pankaj.sheoran@vit.ac.in}
\affiliation{Department of Physics, School of Advanced Sciences (SAS), Vellore Institute of Technology (VIT), Ranipet–Katpadi Road, Vellore - 632 014, Tamil Nadu, India}

\author{Sanjar Shaymatov \orcidlink{0000-0002-5229-7657}}
\email{sanjarshaymatov@gmail.com}
\affiliation{Institute of Fundamental and Applied Research, National Research University TIIAME, Kori Niyoziy 39, Tashkent 100000, Uzbekistan}
 \affiliation{University of Tashkent for Applied Sciences, Str. Gavhar 1, Tashkent 100149, Uzbekistan}
 \affiliation{Tashkent State Technical University, 100095 Tashkent, Uzbekistan}

\begin{abstract}
In this paper, we investigate energy extraction via the Comisso-Asenjo (CA) magnetic reconnection process near a coupled $\mathcal{N} = 2,\,U(1)^2$ gauged supergravity Black Hole (BH). Our study focuses on the combined impact of the independent parameter set $p_i \in (N_g, g, v,e)$ with the spin parameter $a$ on the extracted energy ($\epsilon_{\pm}$), efficiency ($\eta$) and the extracted power ($\mathcal{P}_{CA}$), aiming to find the optimal combinations where it is possible to extract energy from a coupled $\mathcal{N} = 2,\,U(1)^2$ gauged supergravity BH with higher efficiency in certain cases at lower spin $(a\sim 0.39)$ than the Kerr extremal spin case ($a\sim1$). Using the spacetime parameters, we explore various cases that lead to distinct spacetimes and provide an extended comparison with the Kerr Black Hole (KBH). In addition, we examine the influence of the orientation angle ($\xi$) and the magnetization parameter ($\sigma_0$) on both the efficiency and the extracted power. Investigating low $[\,\forall p_i < 0.2 \land N_g < 0.08\,]$, mid $[\,\exists p_i  \geq 0.5 \land N_g \in (0.08,0.15)\,]$, high $[\,\exists p_i > 0.7 \land N_g \in (0.16, 0.23)\,]$, and mixed $[\,\forall p_i \in (0,1) \land N_g \in (0,0.23)\,]$ parameter combinations, we explore only the extremal cases for all spacetime parameters and demonstrate that the extremal Kerr efficiency limit ($\eta >1.495$) can be exceeded. Statistical Kendall's Tau approach allowed us to identify the key independent parameters that act as boosters or dampers in the energy extraction process and enabled us to visualize the intricate relationship between $(N_g, g, v,e)$ the and the physical outputs ($a_{ext},r_E,r_{ergo},\epsilon_{\pm}, \eta, \mathcal{P}_{CA}, R_{\eta}, R_{\mathcal{P}}$) individually. Furthermore, we show that the observable Lundquist number $S_{\rm obs}$ in rotating BH spacetimes acquires an observer-dependent angular dependence through the lapse function $(\alpha)$. This leads to deviations from the standard Sweet-Parker scaling when expressed in terms of observable quantities.
\end{abstract}


    \maketitle

\flushbottom

\section{Introduction}
\label{sec: Introduction}

Among the most critical predictions of Einstein's General Theory of Relativity, one of the essential predictions was the existence of the most energetic and densest astrophysical objects in the universe,  the Astrophysical Black Holes (BHs). These objects were invisible to researchers until recent advancements in the Light Interferometry Gravitational Observatory (LIGO) and Event Horizon Telescopes (EHTs) \cite{Abbott2009LIGO, Accadia2012Virgo, Akiyama2019EHT, Abbott2016GW150914, Abbott2016TestsGR, Abbott2017GW170817}, creating a whole new avenue in observational astronomy and enhancing the study of BHs. Being energetic and massive, most high-energy astrophysical events can be associated with these objects. Notably, they are thought to power relativistic jets and outflows, such as gamma-ray bursts (GRBs). Possibly because the relativistic jets can originate either from the Potential energy of the matter falling inside the accreting BH, or from the Electromagnetic (EM) field energy originating near the vicinity of a BH {\cite{Combes2021AGN, Misner1973Gravitation, Bardeen1972Rotating, Blandford1977BZ, Blandford1982BP, Tchekhovskoy2011EfficientJets}, this remains a prominent subject of ongoing scientific investigation to understand how energy extraction mechanisms can be associated with the phenomena.

The first of the well-known mechanisms of energy extraction was proposed by Penrose, who envisioned a framework to address the Energy extraction from a rotating Black Hole \cite{Penrose1969, PenroseFloyd1971Extraction}. When a particle splits into two parts after entering the ergosphere of a rotating BH, with one generated carrying negative energy as seen by the observer at infinity, which the other escapes to infinity carrying more energy than the original energy, resulting in a less total energy and mass of a BH. Although theoretically appealing, this process is believed to be hard to achieve in astrophysical scenarios \cite{Wald1974EnergyLimits, Bardeen1972, Piran1975Penrose, Schnittman2014Revised, Berti2009SpinLimit} due to the extreme velocity requirement $(>0.5c)$  for particles involved, which leads to insufficient energy extraction. However, keeping the Penrose framework, several theoretical alternative energy extraction mechanisms are proposed, such as Superradiance Scattering of a massive bosonic field \cite{Teukolsky1974Perturbations, Starobinsky1973Amplification, Press1972Bomb, Brito2015Superradiance}, collisional Penrose process \cite{Daughton2009, Piran1975Collisional, Banados2009HighEnergy, Schnittman2014Revised}, the Blankford-Znajek (BZ) process\cite{Blandford1977Znajek, Komissarov2004Electrodynamics, Tchekhovskoy2011BZ, McKinney2004, Komissarov2001ForceFree}, and magnetic hydrodynamics Penrose process \cite{Wagh1986Revival, Wagh1985Magnetic, Parthasarathy1986MPP, Dadhich2001Energy, Tursunov:2019oiq, Shaymatov24PhRvD.110d4042S, Shaymatov24EPJC...84.1015S, Shaymatov:2022eyz, Khamidov25JCAP...03..053X}.

Above all, the processes that gained wide acceptance were the BZ mechanism, which explains the highly energetic jets observed in active galactic nuclei [AGNs] and gamma ray Bursts [GRBs] \cite{McKinney2004Luminosity, Hawley2006Jets, Komissarov2004Electrodynamics, Tchekhovskoy2008GRBMHD, Lee2000BZCentralEngine, Tchekhovskoy2011EfficientJets, Komissarov2009Activation}. The principle behind the BZ mechanism is that BH, immersed in a magnetic field generated by the accretion disk, induces electric field lines that accelerate charge particles in the surrounding plasma, which explains these phenomena. However, the explanation for frequent jet bursts of flares near the BH vicinity that have been observed for different wavelengths in different energy ranges was not plausible. 

Later, Comisso and Asenjo provided a mechanism where fast magnetic reconnection within the ergosphere of a rapidly rotating BH results in an energy extraction rate and reconnection efficiency \cite{Comisso2021}. The study showed that rapid plasma movement enabled by magnetic reconnection can more effectively extract energy, particularly in cases of high spin and strong magnetic fields, potentially exceeding the energy efficiency of the BZ process. 
Various scenarios were investigated via the CA process, such as rotating non-Kerr BH \cite{Liu2022NonKerr, Liu2022NonKerrReconnection, Li2023HairyBH, Wei2022NonKerrReconnection, Zhou2023ParametricDeviation, Wang2023Johannsen, Zhang2023DeformedBH, Eshtursunov:2026kgr}, rotating magnetized BHs \cite{Zhang2024KonoplyaRezzollaZhidenko, Yang2022MagnetisedKerr, Chen2023WaldField, Liu2023MagnetizedNonlinear, Zhou2024MagnetizedAdS}, rotating BHs with a $U(1)$ charge \cite{Shaymatov:2024PRD, Wu2022KerrNewmanReconnection, Zhang2023ChargedAdS, Li2024NonlinearCharge, Huang2023ModifiedKN} or a string charge \cite{Khodadi2022, Wei2023HairyReconnection, Zhou2023EGBReconnection, Wang2023DilatonReconnection, Jiang2024StringBH}. A similar analysis has also been recently extended to higher-dimensional BHs \cite{Eshtursunov:2026ows}. While our analysis applies uniformly across these spacetimes, higher efficiencies are generally associated with larger values of the spin parameter $a$, which seems less physical in the astrophysical scenarios. In this work, we therefore explore the possibility of achieving higher efficiencies and higher extracted power even for lower spin values $(a\sim0.4)$.  Further, we will try to investigate the non-trivial dependence of observable Lundquist number $S_{obs}$ on the spacetime geometry, and try to predict the deviation from the standard Sweet-Parker prediction (\cite{Shen2024}) from the perspective of the observer. Also, we derive the explicit relation between the observable reconnection rate $R_{obs}$ and the observable Lundquist number $S_{obs}$ in the Dyonic Kerr-Newman-NUT-AdS BH and later compare it with the Kerr BH within the constrained limits.

What motivates us to study the Dyonic $\mathcal{N}=2, U(1)^2$ gauged Supergravity BH (Rotating Kerr-Newman-NUT-AdS Supergravity BH) metric is that this spacetime incorporates mass($M$), rotation($a$), electric($e$) and magnetic charges($v$), the NUT parameter($N_g$), and the gauge coupling constant($v$) within a single consistent solution, which opens a broad system for a systematic investigation of each physical ingredient that influences the reconnection efficiency and associated power extraction. The complexity and the intricate relationship between ($N_g, g, v, e, a_{ext}, r_E, r_{ergo}, \epsilon_{\pm}, \eta, \mathcal{P}_{CA}, R_{\eta}, R_{\mathcal{P}}$) needs to be analyzed to get the global influence each independent parameter has on the energy extraction process. For this, several parameters are set to zero as limiting cases. In particular, by constraining suitable parameters, the geometry reduces successively to some parameters like Kerr-Newman-AdS BH, Kerr-Newman BH, Kerr-AdS BH, and ultimately to the Kerr spacetime, as will be summarized later in the comparative Table~\ref{Table: Various spacetimes}. This base structure will allow us to perform magnetic reconnection for all classes of Spacetime within the same formalism by simply taking the appropriate limits, and ultimately reach Kerr geometry naturally as a reference. For the main part of our calculations, we vary all the parameters ($N_g,g,v,e$) and, for each parameter set, we perform all calculations for extremal cases ($r_+\equiv r_E, a\equiv a_{ext}$).  Now, after extracting output($\epsilon_{\pm}, \eta, \mathcal{P}_{CA}$), we systematically analyze the intricate connection between the independent inputs and extracted energy outputs ($\epsilon_{\pm}, \eta, \mathcal{P}_{CA}$). We then follow a computational approach to extract a rank-based correlation map~\cite{pearson1896mathematical, spearman1904proof, kendall1938new, kendall1990rank}, allowing us to identify multidirectional relationships reflecting the discrete and interconnected values in the table, providing physical insight and instantly highlighting the driving forces (boosters) or dampers within the system.   

This paper will focus on magnetic reconnection in a Dyonic Kerr-Newman-NUT-AdS BH, using the Comisso-Ansenjo framework for energy extraction. The presence of different parameters affects the efficiency, extracted power, and dynamics of the extraction process. Our purpose is to explore the benefits we can attain through the manipulation of available parameters. The paper is organized into 6 sections and 2 appendices, which are as follows:
Section (\ref{sec: Introduction}) is the Introduction and motivation for this paper. In section (\ref{sec: Spacetime}), we briefly introduce Kerr-Newman-NUT-AdS Spacetime and examine the influence of BH parameters ($Ng,g,v,e,a$) on the event horizon and ergoregion structure of the spacetime. Additionally, we discussed whether the causality condition breaks as parameters vary widely, as shown in the table~\ref{Table: Various Parameters and different conditions}. In section (\ref{sec: Geodesic}), we calculate the geodesics for the photon orbit and the ISCO. Section (\ref{sec: Reconnection}) introduces and analyzes energy extraction rate via the Comisso-Ansenjo mechanism extensively by varying parameters and calculating the energy extraction rate, efficiency rate, and power extraction rate. Later, we compare our results with various spacetimes in Section (\ref{sec: Efficiency and Power}) and present a tabular comparison with Kerr magnetic reconnection efficiency. (\ref{tab:extraction_bounds_combined}). In the section (\ref{sec: Observable reconnection rate}), we analyze the relativistic effects on the reconnection rate from the observer's perspective and derive the explicit relation between the observable Reconnection rate $R_{obs}$ and the observable Lundquist number $S_{obs}$, and compare its limiting behavior with Kerr BH in different regimes. Finally, in section (\ref{sec: summary}), we summarize the results with the conclusion. We assume normalized units $G=c=\hslash=1$ throughout this work.

\section{\texorpdfstring{Dyonic Kerr-Newman-NUT-AdS Black Hole Metric Structure}
                 {N=2 Gauged Supergravity Black Hole (SBH) Metric Structure}}
                 \label{sec: Spacetime}


Chow and Compère first formulated a class of axisymmetric and Stationary dyonic anti-de Sitter rotating spacetime metric as $\mathcal{N}= 2 $, $U(1)^2$ gauged supergravity BH or Dyonic Kerr-Newman-NUT-AdS BH in \cite{Chow2014}. The metric line element in Boyer--Lindquist coordinates is given by \cite{Rudra2019}
\begin{equation}
\begin{aligned}
   ds^2 = &-\frac{R_g}{\mathcal{B} - a A}\left(dt - \frac{A}{\Xi} d\phi \right)^2 
+ \frac{\mathcal{B} - a A}{R_g} dr^2 \\
&\quad \,+ \frac{\mathcal{B} - a A}{\Theta_g} d\theta^2
+ \frac{\Theta_g a^2 \sin^2\theta}{\mathcal{B} - aA} 
\left(dt - \frac{\mathcal{B}}{a\Xi} d\phi \right)^2\, , 
\label{metric-main} 
\end{aligned}
\end{equation}
where the metric functions are:
\begin{equation}\label{parameters}
\begin{aligned}
R_g \quad& =\, r^2 - 2 M r + a^2 + e^2 - N_g^2 \\
&\quad +g^2[r^4 + (a^2 + 6N_g^2 - 2v^2)r^2 + 3N_g^2(a^2 - N_g^2)
],
\\[4pt]
\Theta_g \quad&= \,1 - a^2 g^2 \cos^2\theta - 4 a^2  N_g  \cos\theta,\\[4pt]
A\quad &= \,a \sin^2\theta + 4N_g\, \sin^2\frac{\theta}{2},\\[4pt]
\mathcal{B}\quad &= \,r^2 + (N_g + a)^2 - v^2,\\[4pt]
\Xi \quad &= \,1 - a^2 g^2 - 4 a N_g g^2 .
\end{aligned}
\end{equation}

It should be noted that our notation follows Refs.~\cite{Rudra2019, Chow2014}, where the solution is characterized by six independent parameters: the mass $(M)$, rotation $(a)$, electric charge $(e)$, magnetic charge $(v)$, gauge coupling constant $(g)$, and NUT charge or Gravitomagnetic Mass $(N_g)$.
The energy extraction process occurs in a region lying between the Horizon and the Static Limit Surface (SLS); this region is called the Ergoregion. So it's crucial to evaluate both Horizon and SLS. The horizon structure of the spacetime is the null hypersurface of constant radius, determined by the condition ($g^{\mu\nu}\partial_{\mu}r \partial_{\nu}r = 0$). By taking the tensor sum, we obtain:
\begin{equation}
    g^{\mu\nu}\partial_{\mu}r \partial_{\nu}r \equiv R_g(r)=0,
\end{equation}
where $R_{g}(r)$ is the radial function. Rearranging $R_{g}(r)$ in polynomial form yields
\begin{equation}
\begin{aligned}
    g^2 r^4& + \Big(1 +  g^2(a^2 + 6N_g^2 - 2\nu^2)\Big) r^2 \\
& - 2Mr + \Big(a^2 + e^2 - N_g^2 + 3g^2 N_g^2(a^2 - N_g^2)\Big)=0\, .
\end{aligned}
\end{equation}

Dividing by $g^2$ (for $g\neq0$), the equation reduces to a depressed quartic
\begin{equation}\label{horizon}
r^4 + \zeta r^2 + \beta r + \Upsilon =0,
\end{equation}
where
\begin{align}
\zeta &= \frac{1 + g^2(a^2 + 6N_g^2 - 2\nu^2)}{g^2}, \\
\beta  &= -\frac{2M}{g^2}, \\
\Upsilon &= \frac{a^2 + e^2 - N_g^2 + 3g^2N_g^2(a^2 - N_g^2)}{g^2}.
\end{align}

Using Ferrari’s method, the four roots are obtained as
\begin{equation}
r =
\frac{1}{2}
\left(
\pm R
\pm \sqrt{-R^2 - \zeta \pm \frac{\beta}{R}}
\right),
\label{quarticroots}
\end{equation}


\begin{figure}[t]
     \includegraphics[scale=0.43,trim={0cm 0cm 0cm 0.68cm}, clip=true]{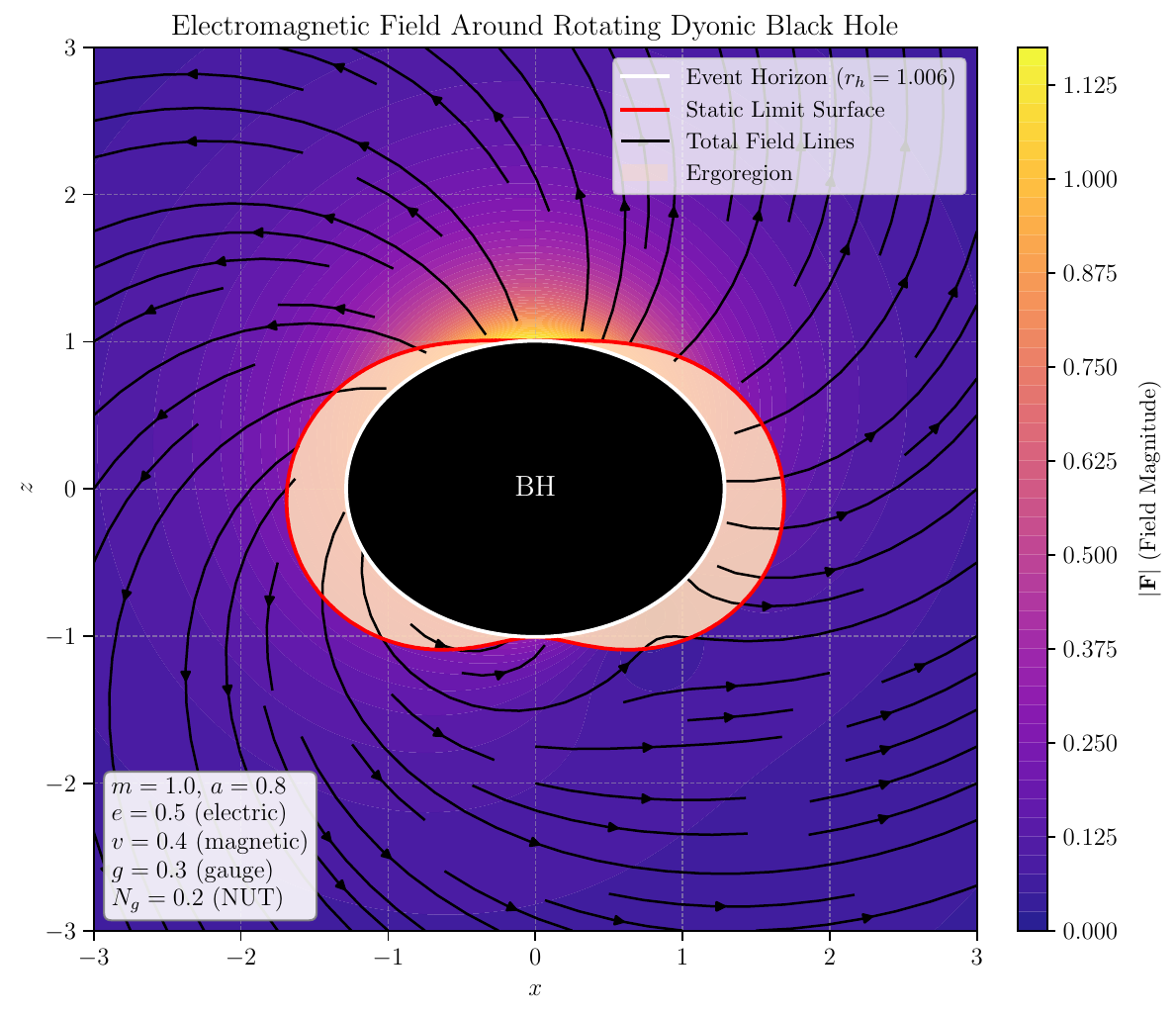}
     \caption{\justifying Deformation near the south pole of the Dyonic Kerr-Newman-NUT-AdS BH, exhibiting explicit dependence on the polar angle. The color map shows the total field density. Parameters chosen here are only for illustrative purposes to represent the geometry of the BH, and the geometry is projected onto the $(x,z)$ plane.}
    \label{fig: dyonic Black Hole deformed}
\end{figure}

where
\begin{equation}
R = \sqrt{\frac{\zeta^2}{4} - \Upsilon + y_0},
\end{equation}
and $y_0$ is a real solution of the resolvent cubic. The detailed derivation of Eq.~(\ref{quarticroots}) and the explicit expression of $y_0$ are presented in Appendix~\ref{Analytic_roots}.

Our primary choice of parameter values corresponds to the extremal configuration, where the outer horizon ($r_+$) and the Cauchy horizon ($r_-$) coincide.
The extremal configuration corresponds to the degeneracy condition
\begin{equation}
R_g(r_E)=0, 
\qquad 
\frac{dR_g}{dr}\Big|_{r=r_E}=0,
\label{eq: Horizon condition}
\end{equation}
which ensures ($r_+ = r_- = r_E$). We can observe that for $R_g\equiv R_g(N_g,g,v,e)$, we obtain a unique value of $\{r_E,a_{ext}\}$ by solving Eq.~(\ref{eq: Horizon condition}), unlike KBH. By solving these two conditions simultaneously, the extremal radius is obtained as
\begin{equation}
r_E^2 =
\frac{
- \mathfrak{B}
+
\sqrt{\mathfrak{B}^2 + 12 g^2 C}
}{6 g^2}
,
\label{rEexplicit}
\end{equation}
where
\begin{align}\label{SimplifyBnC}
\mathfrak{B} &= 1 + g^2(a^2 + 6N_g^2 - 2\nu^2)=g^2\zeta, \\
C &= a^2 + e^2 - N_g^2 + 3g^2N_g^2(a^2 - N_g^2)=g^2\Upsilon.
\end{align}

\begin{figure*}[htbp]
  \includegraphics[scale=0.45]{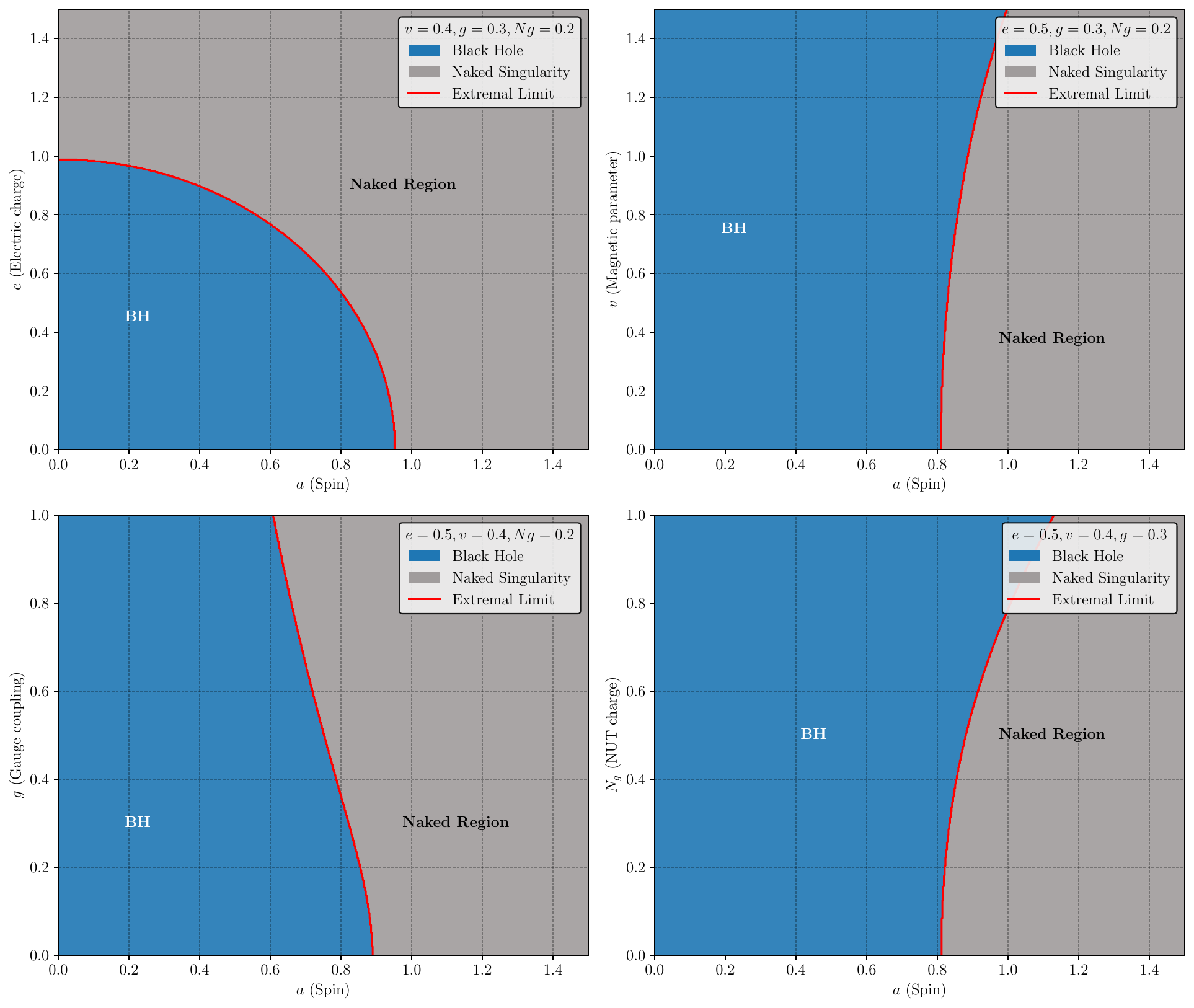} 
  \caption{      \justifying Extremal curve in the $(Ng,g,e,v)$ vs $a$ parameter space obtained from the condition (\ref{eq: Horizon condition}).}
 \label{Plot: extremal curve vs a}
\end{figure*}

The positive branch of Eq.~(\ref{rEexplicit}) gives the radius of the physical horizon. The detailed derivation of this expression and the corresponding extremal mass is presented in Appendix~\ref{Extremality Condition}. Fig.~\ref{Plot: extremal curve vs a} shows the red line as an extremal limit ($r_E, a_{ext}$) with different parameter variations, whereas the blue shaded region shows the region inside the horizon limit, while the gray region shows the naked singularity region.

Within the Ergoregion, the timelike killing vector ($\xi_0^{(t)}= \partial_t$) becomes spacelike, implying that no observer can remain stationary. Consequently, the outer boundary of this region is known as the Static Limit Surface (SLS), where the killing vector $\xi_{\mu}$ becomes null and satisfies $\xi_\mu \xi^\nu = g_{tt}=0$, resulting in the following 
\begin{equation}
R_{g}(r) - \Theta_{g}(\theta)\, a^{2} Sin^{2}\theta = 0\, ,
\end{equation}
or in the form of a radial function
\begin{equation}\label{SLS_compact}
g^2 r^4 + \mathfrak{B} r^2 -2Mr + C_\theta =0,
\end{equation}
where
\begin{equation}
C_\theta =
C - a^2 \Theta_g(\theta)\sin^2\theta\, .
\end{equation}
Thus, the static limit surface Eq.~(\ref{SLS_compact}) has the same structure as the horizon Eq.~(\ref{horizon}) [also see Eq.~(\ref{SimplifyBnC})], with the replacement $C \rightarrow C_{\theta}$.
The SLS radius is obtained as
\begin{equation}
r_{SLS}^2 =
\frac{-\mathfrak{B} + \sqrt{\mathfrak{B}^2 + 12 g^2 C_\theta}}{6g^2}.
\end{equation}
Substituting the explicit forms of $\mathfrak{B}$ and $C_\theta$, the static limit surface is obtained as
\begin{equation}
\begin{aligned}
&r_{SLS}^2 =
\frac{1}{6g^2}\Bigg[
-\Big[1 + g^2(a^2 + 6N_g^2 - 2\nu^2)\Big] \\
& \quad + \Bigg(\Big[1 + g^2(a^2 + 6N_g^2 - 2\nu^2)\Big]^2\nonumber+12g^2\\
&\Big[
a^2 + e^2 - N_g^2
+ 3g^2 N_g^2(a^2 - N_g^2)
- a^2 \Theta_g(\theta)\sin^2\theta
\Big]\Bigg)^\frac{1}{2}\Bigg]\, ,
\end{aligned}
\end{equation}
\begin{table*}[t]
\scriptsize
\centering 
\caption{      \justifying  Parameter combinations and corresponding spacetimes. 
 ($\bullet$ = non-zero, $\circ$ = zero). BH = black hole.}
\label{Table: Various spacetimes}
\renewcommand{\arraystretch}{1.55}
\begin{tabular}{|c|c|c|c|c|c|c|}
\hline
$M$ & $a$ & $e$ & $v$ & $g$ & $N_g$ & Spacetype \\ 
\hline\hline
$\circ$ & $\circ$ & $\circ$ & $\circ$ & $\circ$ & $\circ$ & Minkowski \cite{Minkowski1908} \\ \hline
$\bullet$ & $\circ$ & $\circ$ & $\circ$ & $\circ$ & $\circ$ & Schwarzschild BH \cite{Schwarzschild1916} \\ \hline
$\circ$ & $\bullet$ & $\circ$ & $\circ$ & $\circ$ & $\circ$ & Unphysical rotating Case \\ \hline
$\circ$ & $\circ$ & $\bullet$ & $\circ$ & $\circ$ & $\circ$ & Electric case \\ \hline
$\circ$ & $\circ$ & $\circ$ & $\bullet$ & $\circ$ & $\circ$ & Magnetic case \\ \hline
$\circ$ & $\circ$ & $\circ$ & $\circ$ & $\bullet$ & $\circ$ & Pure AdS case \cite{deSitter1917} \\ \hline
$\circ$ & $\circ$ & $\circ$ & $\circ$ & $\circ$ & $\bullet$ & Taub-NUT case \cite{Taub1951,Newman1963} \\ \hline
$\bullet$ & $\bullet$ & $\circ$ & $\circ$ & $\circ$ & $\circ$ & Kerr BH \cite{Kerr1963} \\ \hline
$\bullet$ & $\circ$ & $\bullet$ & $\circ$ & $\circ$ & $\circ$ & Reissner-Nordström BH (elec) \cite{Reissner1916,Nordstrom1918} \\ \hline
$\bullet$ & $\circ$ & $\circ$ & $\bullet$ & $\circ$ & $\circ$ & Reissner-Nordström BH (mag) \cite{Reissner1916,Nordstrom1918} \\ \hline
$\bullet$ & $\circ$ & $\circ$ & $\circ$ & $\bullet$ & $\circ$ & Schwarzschild-AdS BH \cite{Kottler1918} \\ \hline
$\bullet$ & $\circ$ & $\circ$ & $\circ$ & $\circ$ & $\bullet$ & Schwarzschild-NUT \cite{Newman1963} \\ \hline
$\circ$ & $\circ$ & $\bullet$ & $\bullet$ & $\circ$ & $\circ$ & Dyonic case \\ \hline
$\circ$ & $\circ$ & $\circ$ & $\circ$ & $\bullet$ & $\bullet$ & AdS-Taub-NUT case\cite{Hawking1999} \\ \hline
$\bullet$ & $\bullet$ & $\bullet$ & $\circ$ & $\circ$ & $\circ$ & Kerr-Newman BH (elec) \cite{Newman1965} \\ \hline
$\bullet$ & $\bullet$ & $\circ$ & $\bullet$ & $\circ$ & $\circ$ & Kerr-Newman BH (mag) \cite{Newman1965} \\ \hline
$\bullet$ & $\bullet$ & $\circ$ & $\circ$ & $\bullet$ & $\circ$ & Kerr-AdS BH \cite{Carter1968} \\ \hline
$\bullet$ & $\bullet$ & $\circ$ & $\circ$ & $\circ$ & $\bullet$ & Kerr-NUT BH \cite{Demianski1966} \\ \hline
$\bullet$ & $\circ$ & $\bullet$ & $\bullet$ & $\circ$ & $\circ$ & Dyonic Reissner-Nordström BH \cite{Gibbons1982,MTW1973} \\ \hline
$\bullet$ & $\circ$ & $\bullet$ & $\circ$ & $\bullet$ & $\circ$ & Reissner-Nordström-AdS BH (elec) \cite{Romans1992} \\ \hline
$\bullet$ & $\circ$ & $\bullet$ & $\circ$ & $\circ$ & $\bullet$ & Reissner-Nordström-NUT BH (elec.) \cite{Brill1966} \\ \hline
$\bullet$ & $\circ$ & $\circ$ & $\bullet$ & $\bullet$ & $\circ$ & Reissner-Nordström-AdS BH (mag) \cite{Romans1992} \\ \hline
$\bullet$ & $\circ$ & $\circ$ & $\bullet$ & $\circ$ & $\bullet$ & Reissner-Nordström-NUT BH (Mag.) \cite{Brill1966} \\ \hline
$\bullet$ & $\circ$ & $\circ$ & $\circ$ & $\bullet$ & $\bullet$ & Schwarzschild-NUT-AdS BH \cite{Awad2000} \\ \hline
$\bullet$ & $\bullet$ & $\bullet$ & $\bullet$ & $\circ$ & $\circ$ & Dyonic Kerr-Newman BH \cite{Carter1968} \\ \hline
$\bullet$ & $\bullet$ & $\bullet$ & $\circ$ & $\bullet$ & $\circ$ & Kerr-Newman-AdS BH (elec) \cite{Caldarelli2000} \\ \hline
$\bullet$ & $\bullet$ & $\bullet$ & $\circ$ & $\circ$ & $\bullet$ & Kerr-Newman-NUT BH \cite{Demianski1966} \\ \hline
$\bullet$ & $\bullet$ & $\circ$ & $\bullet$ & $\bullet$ & $\circ$ & Kerr-Newman-AdS BH (mag) \cite{Caldarelli2000} \\ \hline
$\bullet$ & $\bullet$ & $\circ$ & $\bullet$ & $\circ$ & $\bullet$ & Kerr-Newman-NUT BH \cite{Demianski1966} \\ \hline
$\bullet$ & $\bullet$ & $\circ$ & $\circ$ & $\bullet$ & $\bullet$ & Kerr-NUT-AdS BH \cite{Chen2006} \\ \hline
$\bullet$ & $\circ$ & $\bullet$ & $\bullet$ & $\bullet$ & $\circ$ & Dyonic Reissner-Nordström-AdS BH \cite{Romans1992} \\ \hline
$\bullet$ & $\circ$ & $\bullet$ & $\bullet$ & $\circ$ & $\bullet$ & Dyonic Reissner-Nordström-NUT BH \cite{Brill1966} \\ \hline
$\bullet$ & $\circ$ & $\bullet$ & $\circ$ & $\bullet$ & $\bullet$ & Reissner-Nordström-NUT-AdS BH (elec.) \cite{Awad2000} \\ \hline
$\bullet$ & $\circ$ & $\circ$ & $\bullet$ & $\bullet$ & $\bullet$ & Reissner-Nordström-NUT-AdS BH (Mag.) \cite{Awad2000} \\ \hline
$\bullet$ & $\bullet$ & $\bullet$ & $\bullet$ & $\bullet$ & $\circ$ & Dyonic Kerr-Newman-AdS BH \cite{Caldarelli2000} \\ \hline
$\bullet$ & $\bullet$ & $\bullet$ & $\bullet$ & $\circ$ & $\bullet$ & Dyonic Kerr-Newman-NUT BH \cite{Demianski1966} \\ \hline
$\bullet$ & $\bullet$ & $\bullet$ & $\circ$ & $\bullet$ & $\bullet$ & Kerr-Newman-NUT-AdS BH (elec.) \cite{Chen2006} \\ \hline
$\bullet$ & $\bullet$ & $\circ$ & $\bullet$ & $\bullet$ & $\bullet$ & Kerr-Newman-NUT-AdS BH (Mag.) \cite{Chen2006} \\ \hline
$\bullet$ & $\circ$ & $\bullet$ & $\bullet$ & $\bullet$ & $\bullet$ & Dyonic Reissner-Nordström-NUT-AdS BH \cite{Awad2000} \\ \hline
$\bullet$ & $\bullet$ & $\bullet$ & $\bullet$ & $\bullet$ & $\bullet$ & {Rotating $U(1)^2$ Gauged SBH (Dyonic Kerr-Newman-NUT-AdS BH)} \cite{Rudra2019,Chow2014} \\ \hline
\end{tabular}
\end{table*}
and therefore
\begin{equation}
r_{SLS}(\theta) = \sqrt{r_{SLS}^2}\, .
\end{equation}
It is worth noting that, unlike the event horizon, the static limit surface explicitly depends on the polar angle and therefore exhibits an angular shape, as seen in Fig.~\ref{fig: dyonic Black Hole deformed}. At the poles $(\theta = 0)$, we have $\sin\theta = 0$, which yields
\begin{equation}
r_{SLS} = r_{+}\, .
\end{equation}
Hence, the static limit surface coincides with the outer horizon at the poles. At the equatorial plane $(\theta = \pi/2)$, the term  $a^2 \Theta_g(\theta)\sin^2\theta$ attains its maximum contribution, and consequently the ergoregion becomes maximal as visible in Fig.~\ref{fig: dyonic Black Hole deformed}.

The complexity of the metric leads to conditions in which spacetime becomes nonphysical, i.e., cases where $r_{SLS}$ becomes smaller than $r_+$ or a singularity forms outside the horizon. So, for the $r_{SLS}$ to lie strictly outside the outer horizon and the singularity remain inside the horizon at any point in spacetime, two conditions must be imposed \cite{Rudra2019, Chow2014} respectively as: 
\begin{equation}\label{singularity}
\Theta_{g}(\theta) > 0, \qquad r_{+} >r_{sing}\, ,
\end{equation}
where the singularity condition is determined by a geometric factor for which the metric collapses to zero. In the case of our spacetime, that geometric scale factor is $\mathcal{B}-a A$, which, when equated to zero, gives:
\begin{equation}
r^2 - v^2 + (N_g + a\cos\theta)^2 = 0\, ,
\end{equation}
\begin{equation}
r_{sing} = \pm \sqrt{v^2 - (N_g + a \cos\theta)^2}\, .
\label{eqn: Singularity condition for Ng}
\end{equation}

\begin{table*}[htbp]
\centering
\caption{      \justifying Extremal configurations for $(r_E, a_{ext},r_{erg},\Omega_+,\Omega_-)$, and causality conditions for different sets of spacetime parameters $(e, v, g, N_g)$ satisfying (\ref{singularity}), whereas $M=1$. \textit{(All calculated values are rounded to three decimal places.)} 
$(\bullet)\equiv Causality\, Satisfied $, $(\circ)\equiv Causality\,violated$. }
\label{Table: Various Parameters and different conditions}
\renewcommand{\arraystretch}{1.64} 
\setlength{\tabcolsep}{7pt} 

\begin{tabular}{|cl|c|c|c|c|c|c|c|}
\hline
\multicolumn{2}{|c|}{\textbf{Parameters} $\mathbf{(e, v, g, N_g)}$} & $\mathbf{r_E}$ & $\mathbf{a_{\rm ext}}$ & $\mathbf{r_{\rm erg}}$ & $\mathbf{\Omega_+}$ & $\mathbf{\Omega_-}$ & $\mathbf{\Theta_g>0}$ & $\mathbf{r_+>r_{\rm sing}}$ \\
\hline\hline

\ldelim\{{4}{1.2cm}[$\boldsymbol{N_g \uparrow}$] 
& $(0.0, 0.0, 0.0, 0.0)$ & $1.000$ & $1.000$ & $2.000$ & $0.500$ & $-66666.500$ & $\bullet$ & $\bullet$ \\
& $(0.0, 0.0, 0.0, 0.1)$ & $1.000$ & $1.005$ & $2.005$ & $0.452$ & $5.000$ & $\bullet$ & $\bullet$ \\
& $(0.0, 0.0, 0.0, 0.2)$ & $1.000$ & $1.020$ & $2.020$ & $0.410$ & $2.500$ & $\bullet$ & $\bullet$ \\
& $(0.0, 0.0, 0.0, 0.3)$ & $1.000$ & $1.044$ & $2.044$ & $0.372$ & $1.667$ & $\circ$   & $\bullet$ \\
\hline

\ldelim\{{4}{1.2cm}[$\boldsymbol{g \uparrow}$] 
& $(0.0, 0.0, 0.1, 0.0)$ & $0.972$ & $0.990$ & $1.911$ & $0.509$ & $27.209$ & $\bullet$ & $\bullet$ \\
& $(0.0, 0.0, 0.2, 0.0)$ & $0.907$ & $0.966$ & $1.729$ & $0.530$ & $8.325$ & $\bullet$ & $\bullet$ \\
& $(0.0, 0.0, 0.4, 0.0)$ & $0.760$ & $0.902$ & $1.390$ & $0.564$ & $3.329$ & $\bullet$ & $\bullet$ \\
& $(0.0, 0.0, 0.6, 0.0)$ & $0.644$ & $0.840$ & $1.154$ & $0.559$ & $2.152$ & $\bullet$ & $\bullet$ \\
\hline

\ldelim\{{4}{1.2cm}[\shortstack{$\boldsymbol{g \downarrow}$ \\ $\boldsymbol{N_g \uparrow}$}] 
& $(0.0, 0.0, 0.6, 0.1)$ & $0.638$ & $0.837$ & $1.146$ & $0.409$ & $1.165$ & $\bullet$ & $\bullet$ \\
& $(0.0, 0.0, 0.4, 0.2)$ & $0.743$ & $0.904$ & $1.373$ & $0.385$ & $1.162$ & $\bullet$ & $\bullet$ \\
& $(0.0, 0.0, 0.2, 0.3)$ & $0.889$ & $0.997$ & $1.732$ & $0.368$ & $1.278$ & $\circ$   & $\bullet$ \\
& $(0.0, 0.0, 0.1, 0.4)$ & $0.962$ & $1.061$ & $1.963$ & $0.337$ & $1.160$ & $\circ$   & $\bullet$ \\
\hline

\ldelim\{{4}{1.2cm}[$\boldsymbol{v \uparrow}$] 
& $(0.0, 0.1, 0.0, 0.0)$ & $1.000$ & $1.000$ & $2.000$ & $0.503$ & $100.151$ & $\bullet$ & $\bullet$ \\
& $(0.0, 0.3, 0.0, 0.0)$ & $1.000$ & $1.000$ & $2.000$ & $0.524$ & $11.113$ & $\bullet$ & $\bullet$ \\
& $(0.0, 0.5, 0.0, 0.0)$ & $1.000$ & $1.000$ & $2.000$ & $0.571$ & $4.000$ & $\bullet$ & $\bullet$ \\
& $(0.0, 0.7, 0.0, 0.0)$ & $1.000$ & $1.000$ & $2.000$ & $0.662$ & $2.041$ & $\bullet$ & $\bullet$ \\
\hline

\ldelim\{{4}{1.2cm}[\shortstack{$\boldsymbol{v \uparrow}$ \\ $\boldsymbol{N_g \uparrow}$}] 
& $(0.0, 0.1, 0.0, 0.0)$ & $1.000$ & $1.000$ & $2.000$ & $0.503$ & $100.151$ & $\bullet$ & $\bullet$ \\
& $(0.0, 0.3, 0.0, 0.1)$ & $1.000$ & $1.005$ & $2.005$ & $0.472$ & $3.454$ & $\bullet$ & $\bullet$ \\
& $(0.0, 0.5, 0.0, 0.2)$ & $1.000$ & $1.020$ & $2.020$ & $0.456$ & $1.550$ & $\bullet$ & $\bullet$ \\
& $(0.0, 0.7, 0.0, 0.3)$ & $1.000$ & $1.044$ & $2.044$ & $0.451$ & $0.935$ & $\circ$   & $\bullet$ \\
\hline

\ldelim\{{4}{1.2cm}[\shortstack{$\boldsymbol{e \downarrow}$ \\ $\boldsymbol{v \uparrow}$}] 
& $(0.7, 0.1, 0.0, 0.0)$ & $1.000$ & $0.714$ & $1.714$ & $0.476$ & $-1.488$ & $\bullet$ & $\bullet$ \\
& $(0.5, 0.3, 0.0, 0.0)$ & $1.000$ & $0.866$ & $1.866$ & $0.522$ & $-5.412$ & $\bullet$ & $\bullet$ \\
& $(0.3, 0.5, 0.0, 0.0)$ & $1.000$ & $0.954$ & $1.954$ & $0.575$ & $5.963$ & $\bullet$ & $\bullet$ \\
& $(0.1, 0.7, 0.0, 0.0)$ & $1.000$ & $0.995$ & $1.995$ & $0.663$ & $2.073$ & $\bullet$ & $\bullet$ \\
\hline

\end{tabular}
\end{table*}
 
Solving Eq.~(\ref{eq: Horizon condition}) and (\ref{eqn: Singularity condition for Ng}), we get a safe bound for $N_g$, i.e., ($N_g\rightarrow0.235$), where causality condition in Eq.~(\ref{singularity}) is satisfied. Thereby, the lower, mid, mix, and upper bounds for $N_g$ are set separately. The first condition, given in Eq.~(\ref{singularity}), ensures the absence of signature flip in the spacetime geometry, thereby preserving its Lorentzian nature. The second condition guarantees that the curvature singularity remains hidden inside the event horizon, thus avoiding the formation of a naked singularity. The spacetime metric is a generalization of different spacetimes due to various combinations of parameters; thereby, it can be reduced to many spacetimes when certain parameters are limited to zero, as can be seen in the Table~\ref{Table: Various spacetimes}. As the parameters can be varied, the degenerate horizon ($r_+=r_- =r_E$), ergoregion ($r_{SLS}$), and singularity ($r_{sing}$) vary accordingly, enriching the system dynamics.

\section{Geodesic Motion and Orbital Mechanics}
\label{sec: Geodesic}

 The existence of Killing vectors $\xi_{(t)}^{\mu}$ and $\xi_{(\phi)}^{\mu}$ in a stationary and axisymmetric spacetime implies the presence of conserved quantities along geodesics. These conserved quantities correspond to the conserved energy $E$ and angular momentum $L$, respectively, and  are defined as
\begin{equation}
E = - g_{\mu\nu}\,\xi_{(t)}^{\mu} u^{\nu},\qquad
L = g_{\mu\nu}\,\xi_{(\phi)}^{\mu} u^{\nu},
\end{equation}
where $u^{\mu} = \dot{x}^{\mu}\equiv (\dot{t}, \dot{r}, \dot{\theta}, \dot{\phi})$ is the four-velocity of the particle and $\dot{x}^{\mu}\equiv d x^{\mu}/d\tau$, while  $\tau$ is the proper time and dot represents the differentiation with the proper time. For the Dyonic Kerr-Newman-NUT-AdS spacetime considered here, using the metric (\ref{metric-main}) and solving these two first-order equations of the conserved quantities lead to the first-order differential equations for $\dot{t}$ and $\dot{\phi}$:
\begin{equation}
\begin{aligned}
    \dot{t} =
    \frac{\mathcal{B}(E \mathcal{B} - aL\Xi)}{R_g}
    + \frac{A(L\Xi - E A)}{\Theta_g \sin^2\theta}, \\
    \dot{\phi} =
    \frac{a\Xi(E \mathcal{B} - a L\Xi)}{R_g}
    + \frac{\Xi(L\Xi - EA)}{\Theta_g \sin^2\theta}.
\end{aligned}
\end{equation}

To obtain the radial motion, we use the normalization condition for timelike geodesics
\begin{equation}
g_{\mu\nu} u^{\mu} u^{\nu} = - m_0^2\, .
\end{equation}
Substituting the expressions for $\dot{t}$ and $\dot{\phi}$ into the normalization condition and simplifying, we obtain the first- order equations for $\dot{r},\,\dot{\theta}$, which are separable in the coordinates $r$ and $\theta$, the equation involving $\dot{r}$ can be written as,
\begin{equation}\label{effPot}
(\mathcal{B} - aA)^2 \dot{r}^2 =  (E \mathcal{B} - \Xi a L)^2 - R_g \left( K + m_0^2 \mathcal{B} \right)\, ,
\end{equation}
We can then write Eq.~(\ref{effPot}) as $\dot{r}^2 + V_{eff}(r) =0$, where $V_{eff}(r)$ is given by,
\begin{equation}
V_{{eff}} = - \frac{ (E \mathcal{B} - \Xi a L)^2 - R_g \left( K + m_0^2 \mathcal{B} \right)}{(\mathcal{B} - aA)^2}\, .
\end{equation}

For massive particles ($m_0 \neq 0$), the circular orbit conditions are 
\begin{equation}
V_{eff}(r_0) = 0, \qquad \frac{dV_{eff}}{dr}\Bigg|_{r=r_0} = 0\, . 
\label{eqn: rISCO and rPH}
\end{equation}
The red line in Fig.~\ref{Region Plot: spin vs r} shows the dependence of photon radius with spin for different configurations. Similarly, the innermost stable circular orbit (ISCO) is determined by the stability condition. Accordingly, the ISCO radius, $r_{\text{ISCO}}$, is obtained by solving
\begin{equation}
V_{eff}(r_{\text{ISCO}})=0, \quad
V_{eff}'(r_{\text{ISCO}})=0, \quad
V_{eff}''(r_{\text{ISCO}})=0.
\end{equation}
After analyzing the geodesics for particles in the spacetime, we study the localized plasma environment surrounding the BH. We consider a scenario for the circular motion of the equatorial (i.e., $\theta=\pi/2$) for plasma particles. The plasma orbit has a least bound, which is the circular photon orbit as given in Eq.~(\ref{eqn: rISCO and rPH}) for $(m_0=0)$. The orbital velocity is determined by the Keplerian angular velocity, which is defined as  $\Omega_{K} \equiv (\dot{\phi}/\,{\dot{t}})$. Using the general form of the rotating metric  (\ref{metric-main}), the angular velocity of a test particle confined to the equatorial plane satisfies the standard condition, which can be written as \cite{Simpson2023}
\begin{equation}
\frac{d g_{\phi\phi}}{d r}\,\Omega_{K}^{\,2}
\;+\; 2\,\frac{d g_{t\phi}}{d r}\,\Omega_{K}
\;+\; \frac{d g_{tt}}{d r}
=0\, .
\label{KeplerEq}
\end{equation}

This relation follows directly from the normalization of the four-velocity ( $u_{\mu}u^{\mu}=-1$) and the requirement that the radial force on the particle vanishes for a circular orbit. Hence, Eq.~\eqref{KeplerEq} provides the appropriate expression for determining the Keplerian angular velocity in this generalized rotating geometry.  Solving the quadratic equation yields the two branches of the angular velocity, corresponding to co-rotating and counter-rotating circular geodesics, as given in Table~\ref{Table: Various Parameters and different conditions}. Suppose a scenario where $N_g\,=\,v$, then $\Omega_K(\pm)$ becomes
\begin{equation}
\Omega_{K}(\pm)= \frac{\left(a g^2 (a+4 N_g)-1\right)^2}{-a^3 g^2 \pm \frac{r}{\sqrt{\mathcal{X}}} - 6 a^2 g^2 N_g - 8 a g^2 N_g^2 + a + 2 N_g}\, ,
\end{equation}
where the term inside the square root is defined by

\begin{equation}
\mathcal{X} = \frac{N_g^2 \left(1-3 a^2 g^2\right) - e^2 + 3 g^2 N_g^4 + g^2 r^4 + r}{r^2 \left(a g^2 (a+4 N_g)-1\right)^2}\, .
\end{equation}
The fact that $\Omega_K{(\pm)}$ has a radial dependence implies that the reconnection does occur due to the difference in the rotation of the accreting plasma, which creates a shear in the background magnetic field lines, and plays a crucial role in driving the energy extraction process. Fig.~\ref{Region Plot: spin vs r} shows the region plot for the plasma acquiring negative energy in the counter-rotating orbit for decelerating plasma, as we will discuss in the next section.  
\begin{figure*}[htbp]
\centering
{\includegraphics[width=0.32\textwidth]{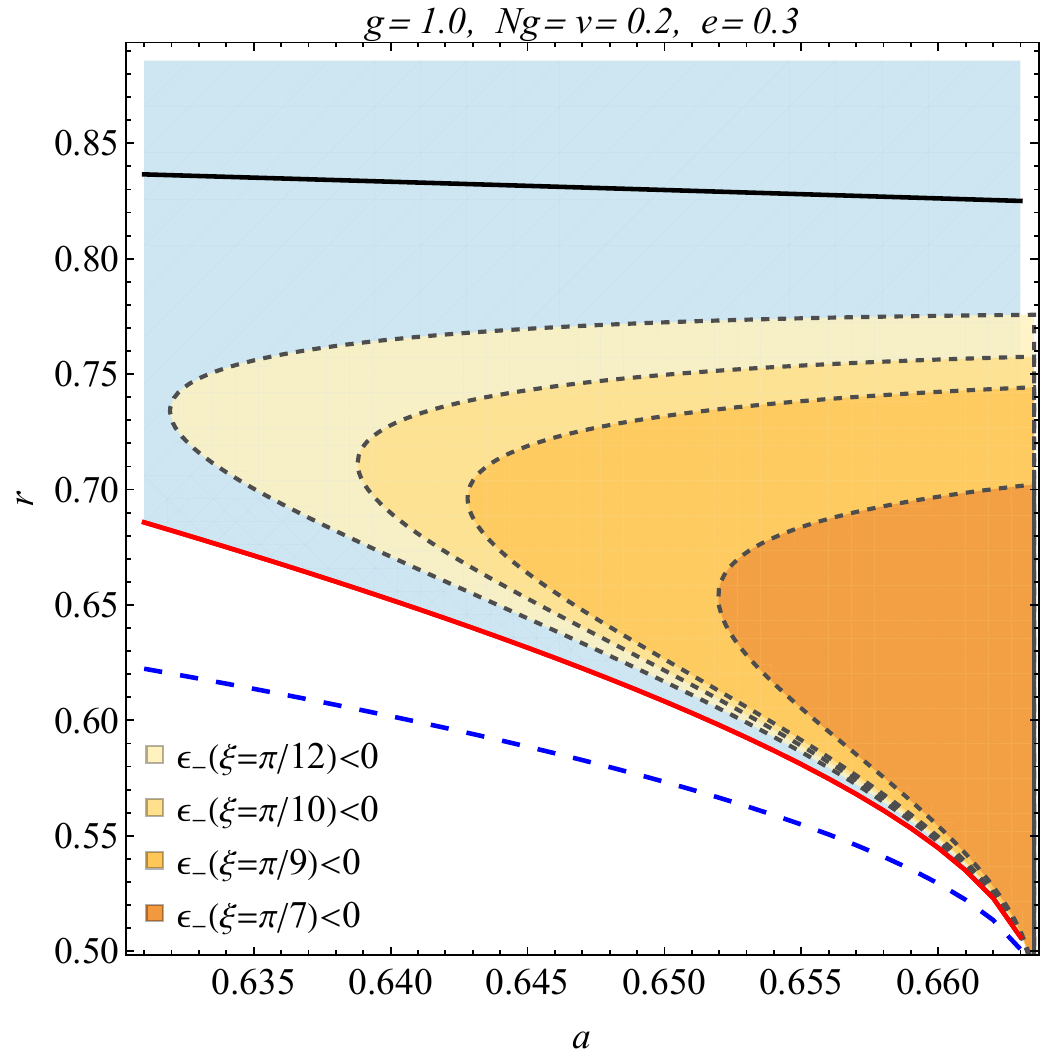}}
{\includegraphics[width=0.32\textwidth]{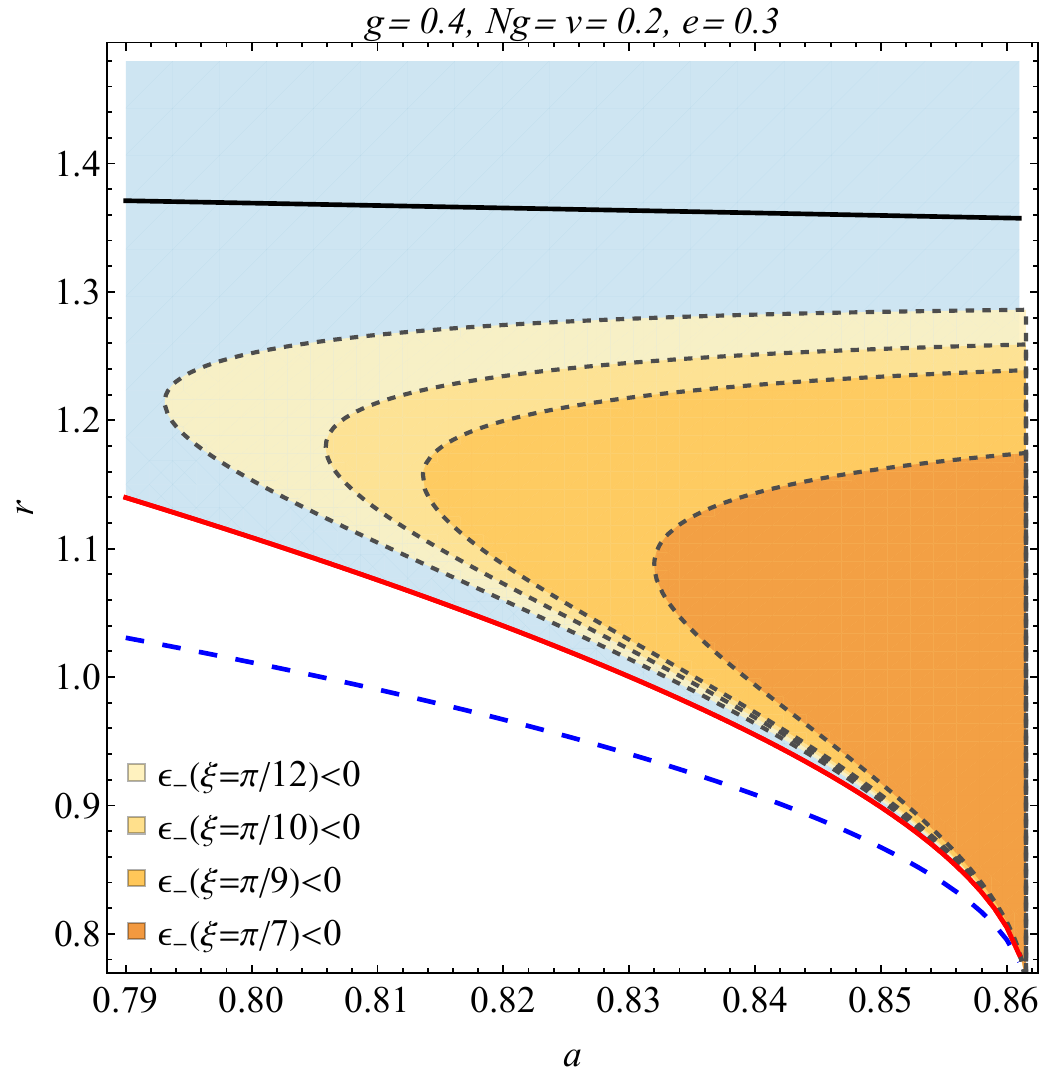}}
{\includegraphics[width=0.32\textwidth]{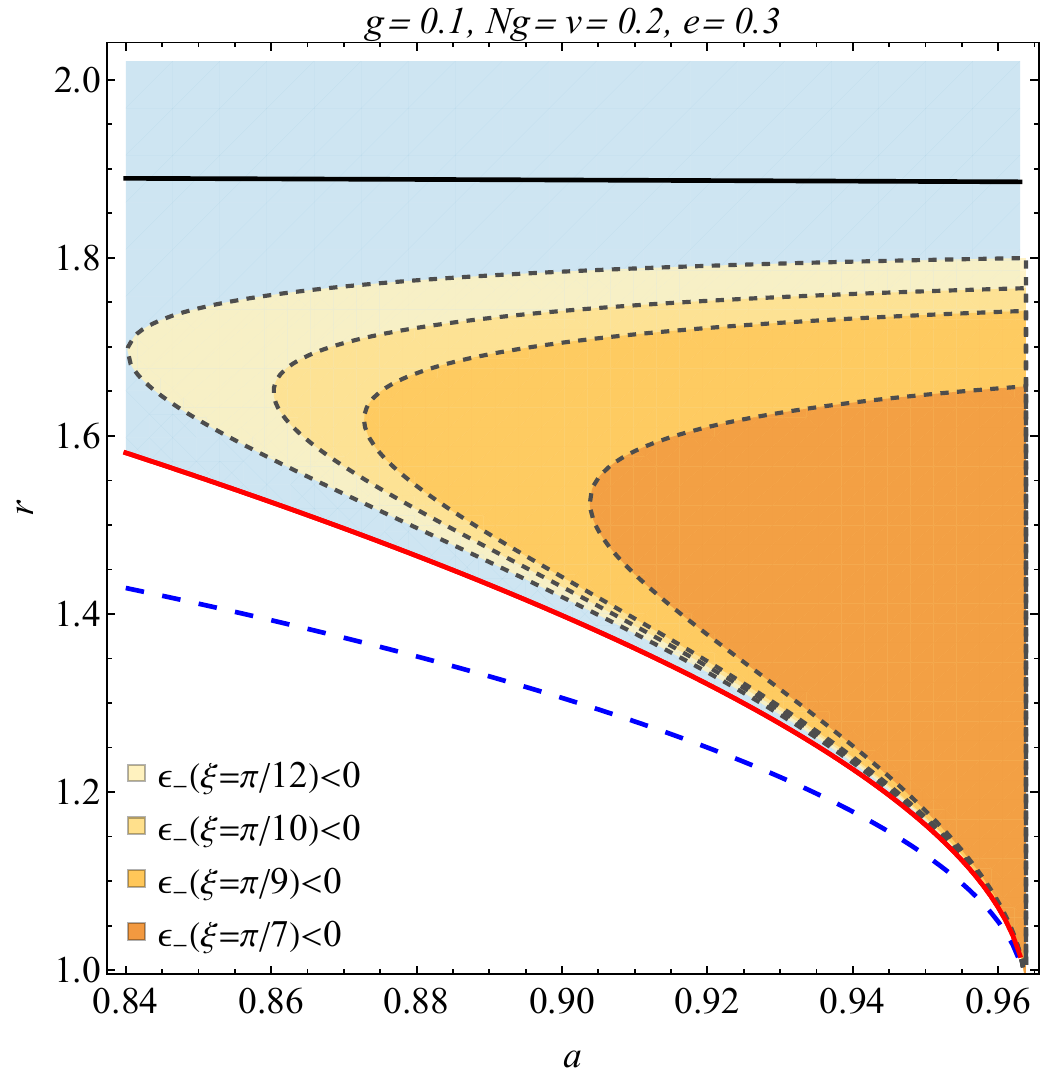}}
\vspace{0.3 cm}

{\includegraphics[width=0.32\textwidth]{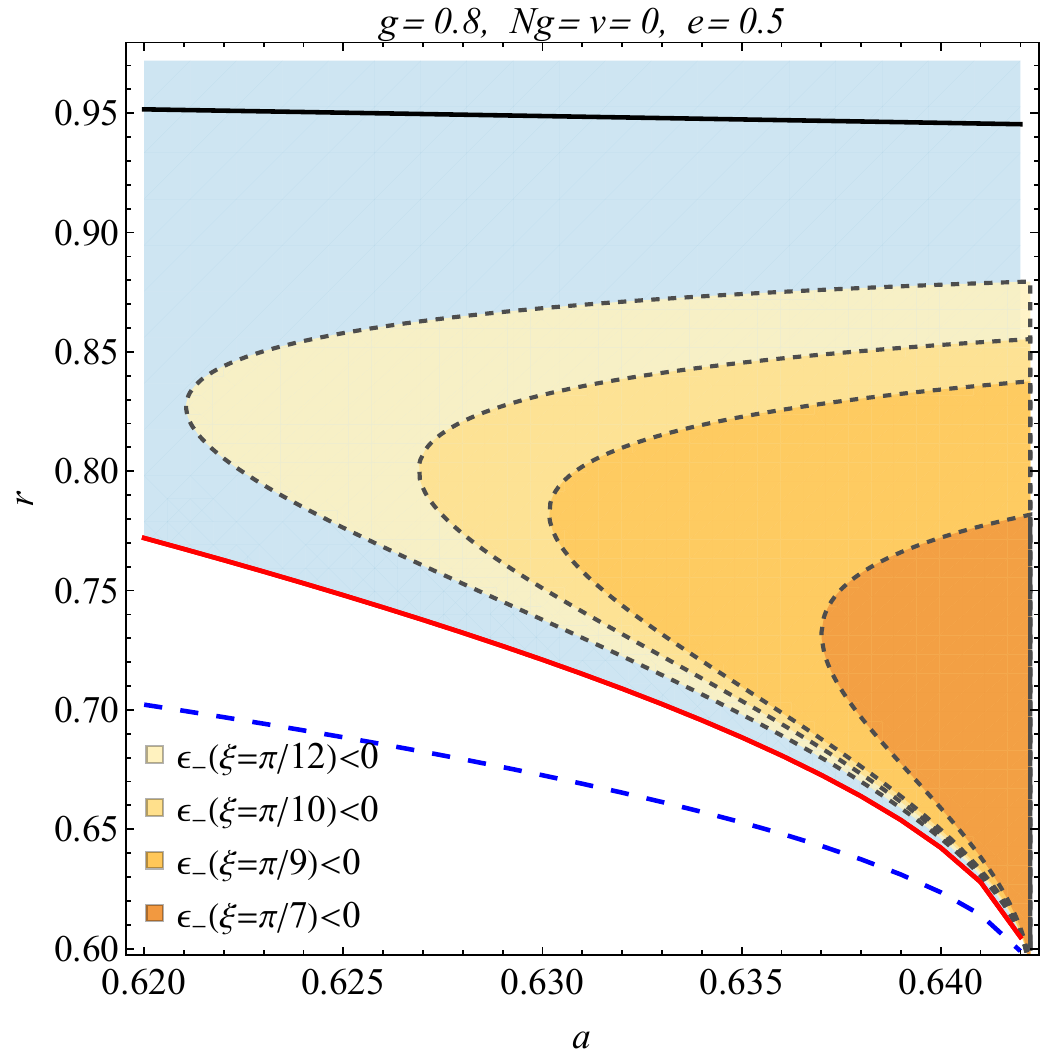}}
{\includegraphics[width=0.32\textwidth]{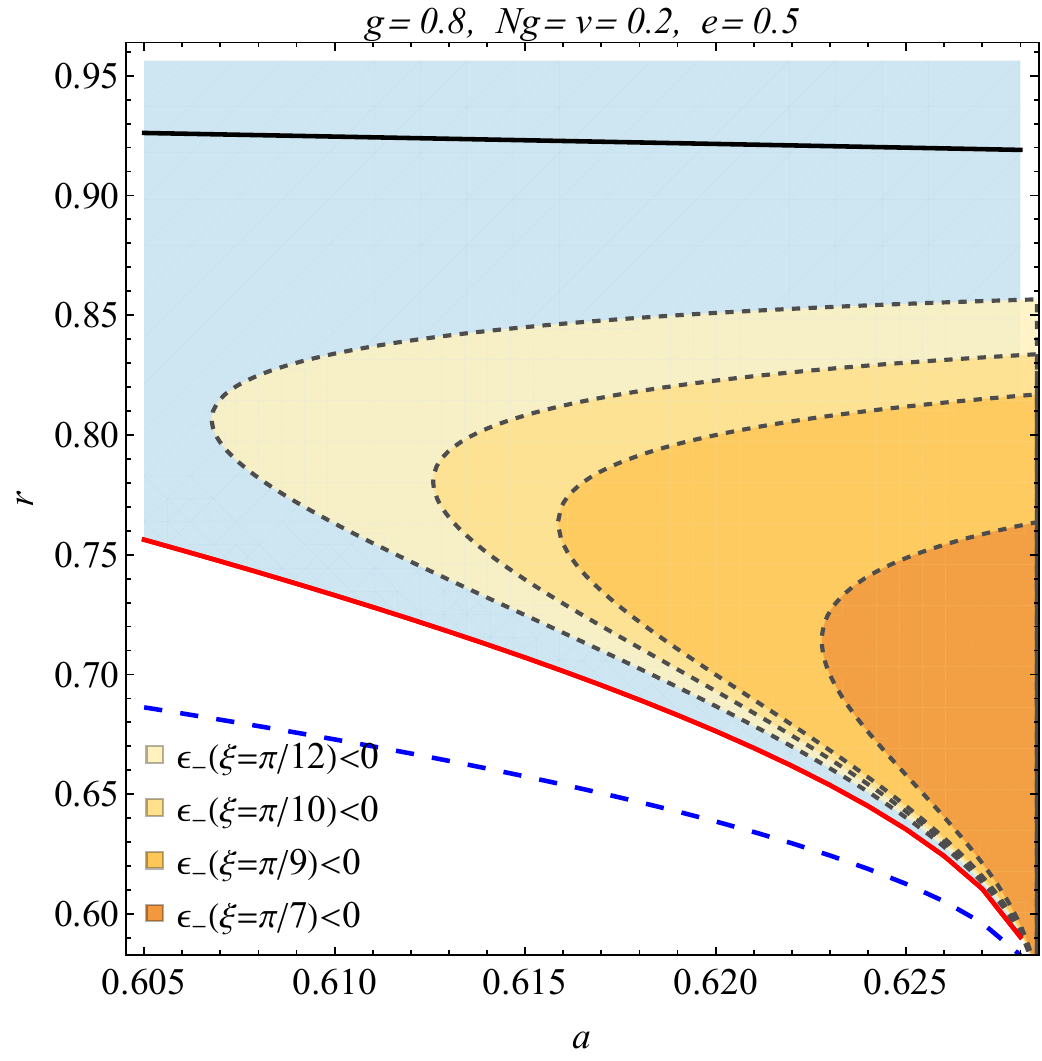}}
{\includegraphics[width=0.32\textwidth]{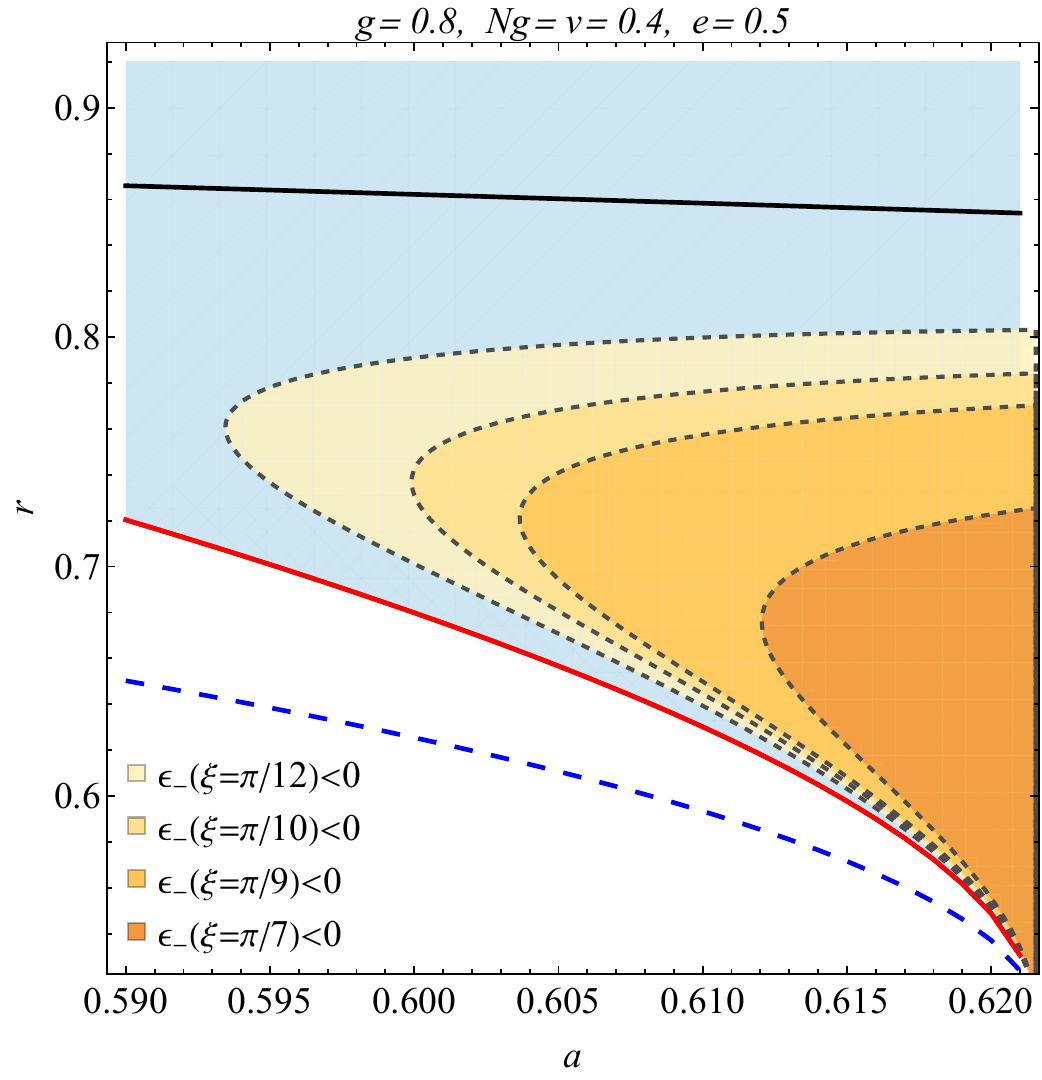}}

\vspace{0.3 cm}

{\includegraphics[width=0.32\textwidth]{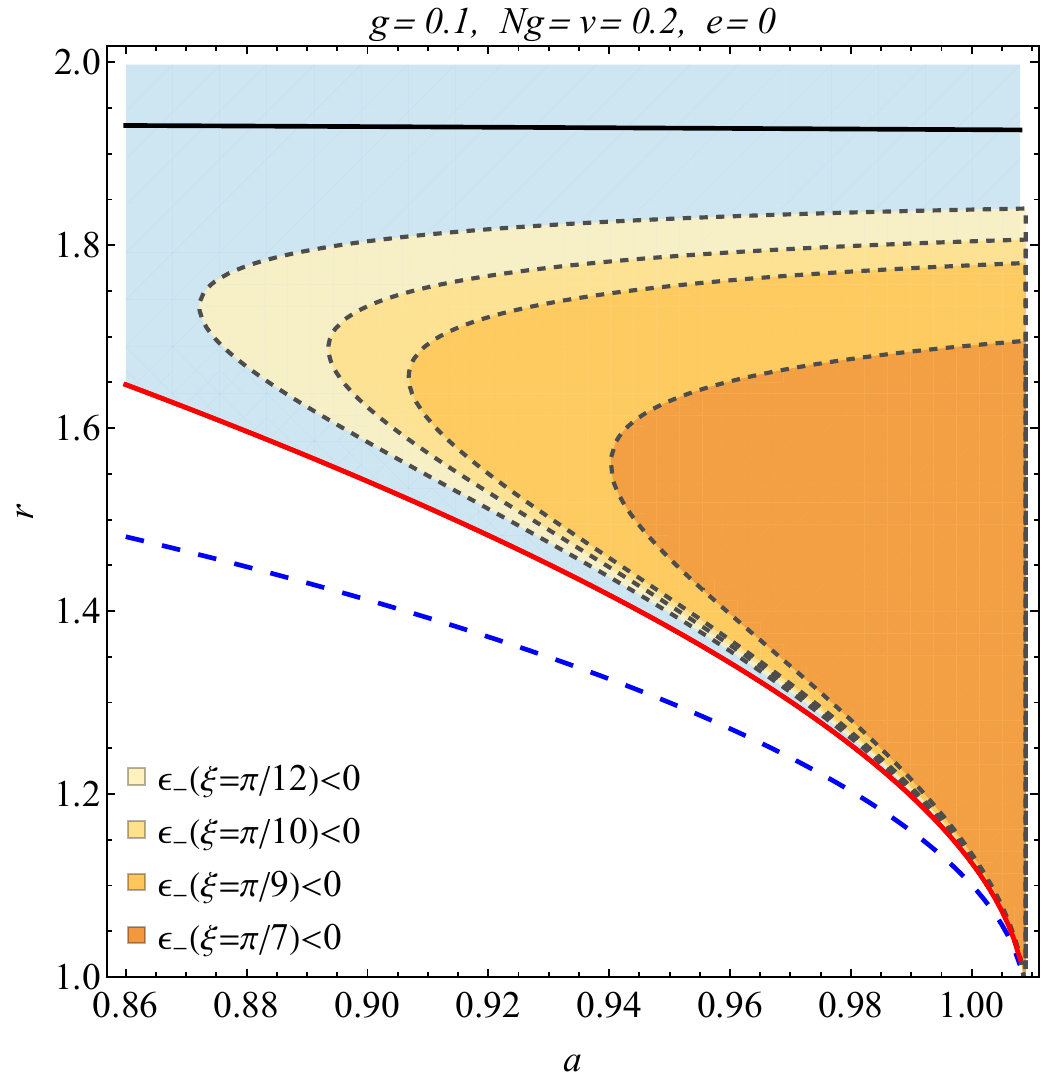}}
{\includegraphics[width=0.32\textwidth]{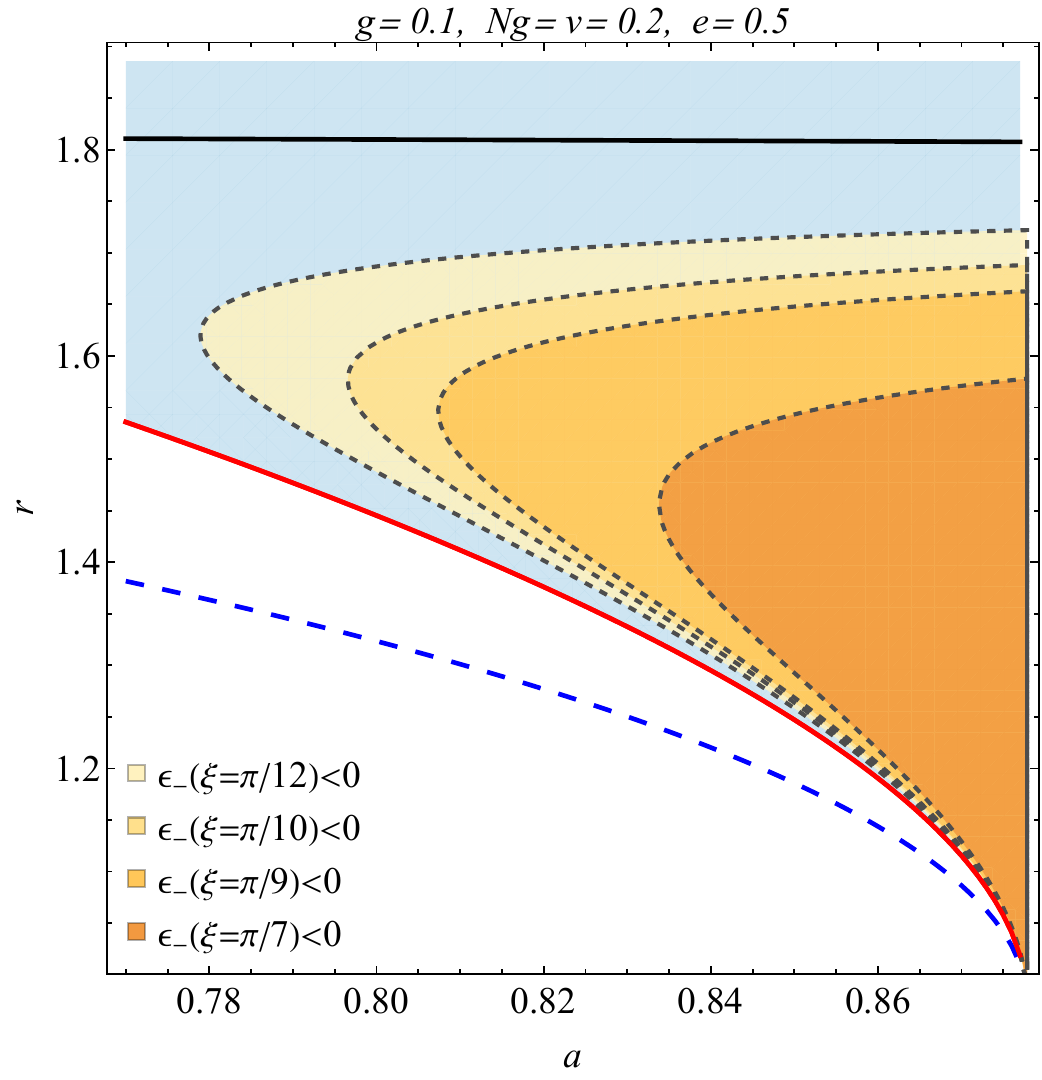}}
{\includegraphics[width=0.32\textwidth]{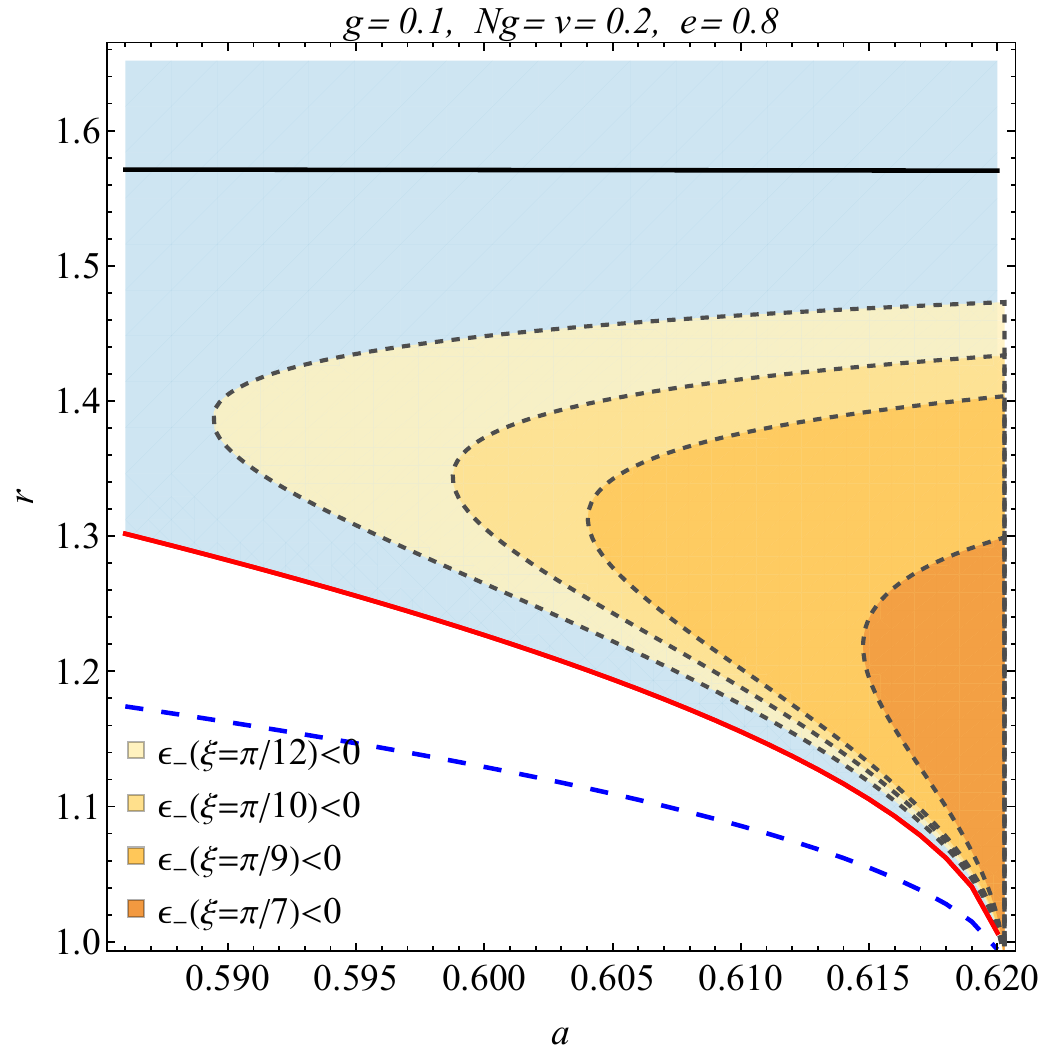}}
\caption{      \justifying{Regions acquiring negative $\epsilon^{\infty}_{-}$ mapped in $(a,r)$ plane for distinct $\xi$ values. The configuration expresses the dependence on $(g, N_g,v,e)$,  where dashed(Blue), thick(black), and thick(Red) lines show $r_E$, $r_{SLS}$ and $r_{ph}$ respectively, whereas $M=1$ and $\sigma_0=100$.}}
\label{Region Plot: spin vs r}
\end{figure*}

\section{Energy Extraction via the Comisso-Ansenjo Process}
\label{sec: Reconnection}

Magnetic reconnection occurs when a current sheet forms (i.e., where the magnetic field reverses in the accretion flow) and undergoes plasmoid instabilities, converting stored magnetic energy into plasma kinetic energy  \cite{Lyubarsky2005, Liu2023MagnetizedNonlinear, Xu2023}.  In this process, part of the plasma can be accelerated outward, while another part can be decelerated inward. The inward-directed plasma acquires negative energy as measured at infinity. In contrast, the outward-directed plasma carries positive energy; the net effect is an extraction of rotational energy and reconnection energy \cite{Comisso2021, ComissoAsenjo2018}. We therefore analyze the conditions under which reconnection yields such negative-energy inflows. We adopt the zero-angular-momentum observer frame (ZAMO), which is locally non-rotating with respect to the dragging of inertial frames and helps determine the plasma energy density. In Boyer–Lindquist coordinates,  $(t,x^i; i\in(1,3))$ is related to the ZAMO frame coordinates $(\hat{t},\hat{x}^i)$ via a transformation
\begin{equation}
    d\hat t = \alpha\,dt,\qquad
d\hat x^i = \sqrt{g_{ii}}\,dx^i - \alpha\,\beta^i\,dt,
\end{equation}
Where \(\alpha\) is the lapse function and \(\beta^i\) the shift vector \cite{Simpson2023, Wu2022KerrNewmanReconnection}.  In the equatorial plane, one explicitly finds 
\begin{figure*}[htbp]
\centering
\vspace{0.05 cm}
{\includegraphics[width=0.32\textwidth, height=4.4cm, keepaspectratio=false]{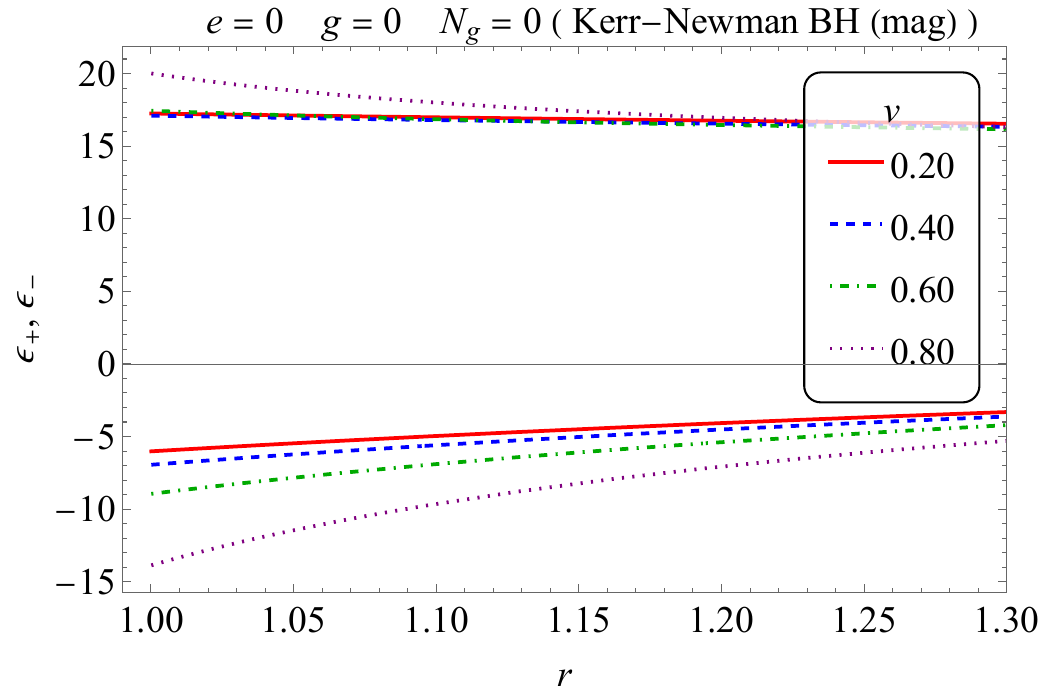}}
{\includegraphics[width=0.32\textwidth, height=4.4cm, keepaspectratio=false]{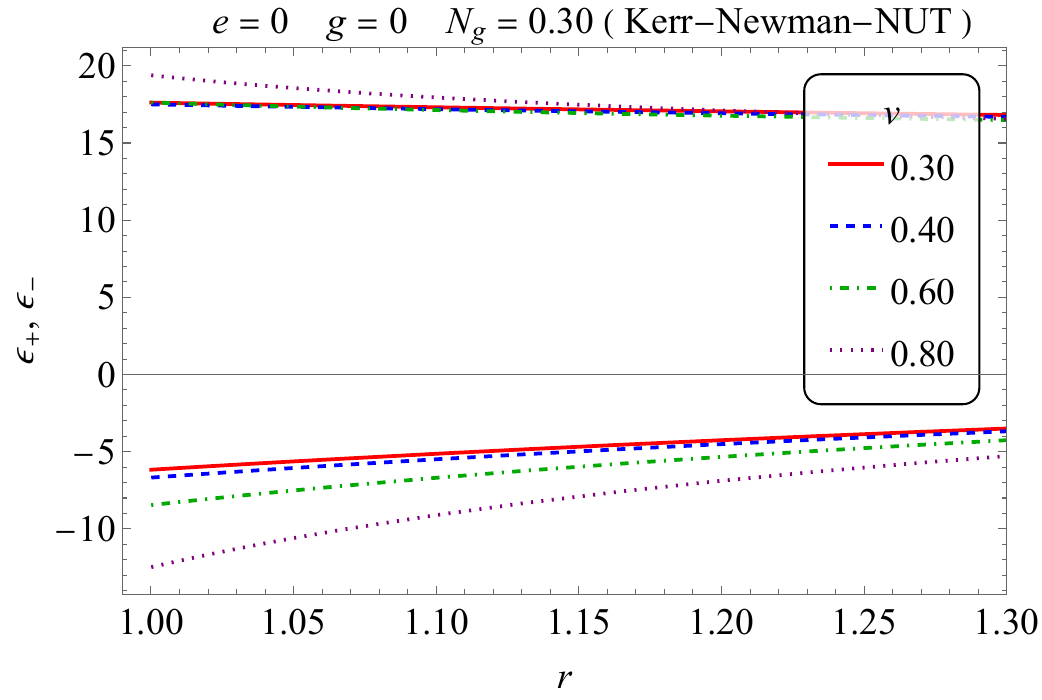}}
{\includegraphics[width=0.32\textwidth, height=4.4cm, keepaspectratio=false]{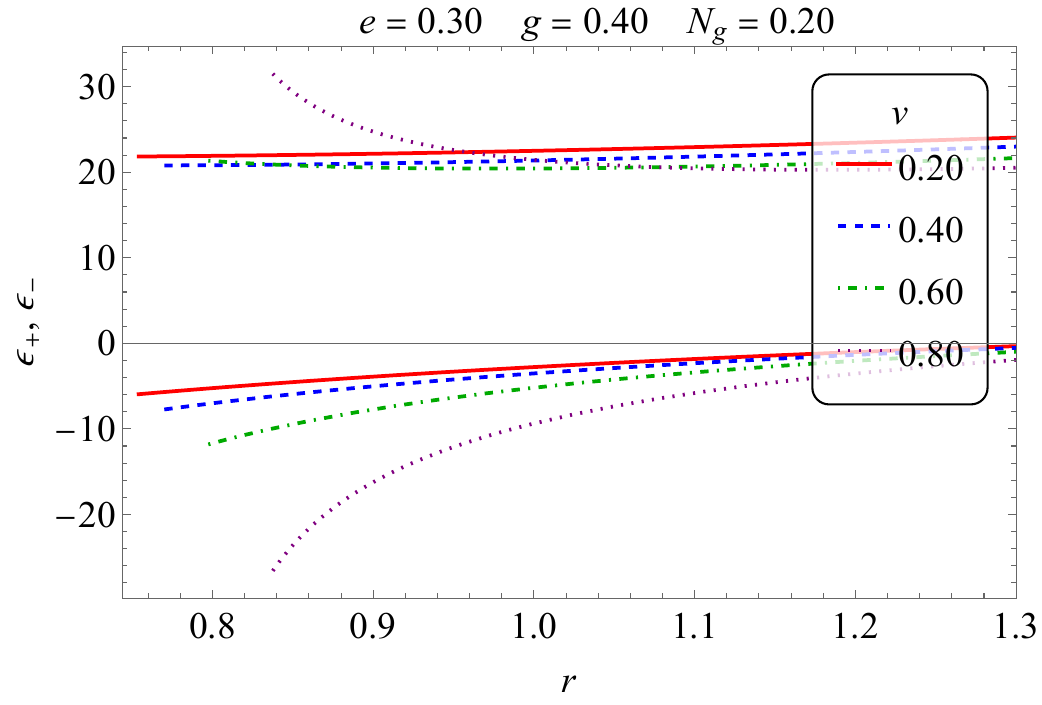}}
\vspace{0.05 cm}

{\includegraphics[width=0.32\textwidth, height=4.4cm, keepaspectratio=false]{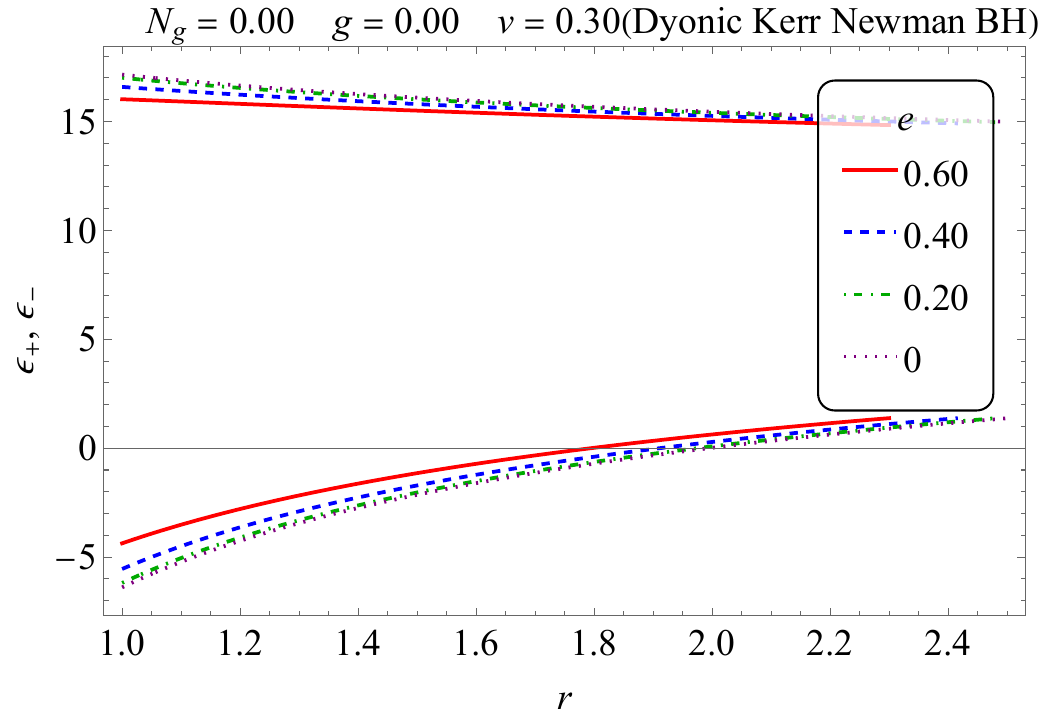}}
{\includegraphics[width=0.32\textwidth, height=4.4cm, keepaspectratio=false]{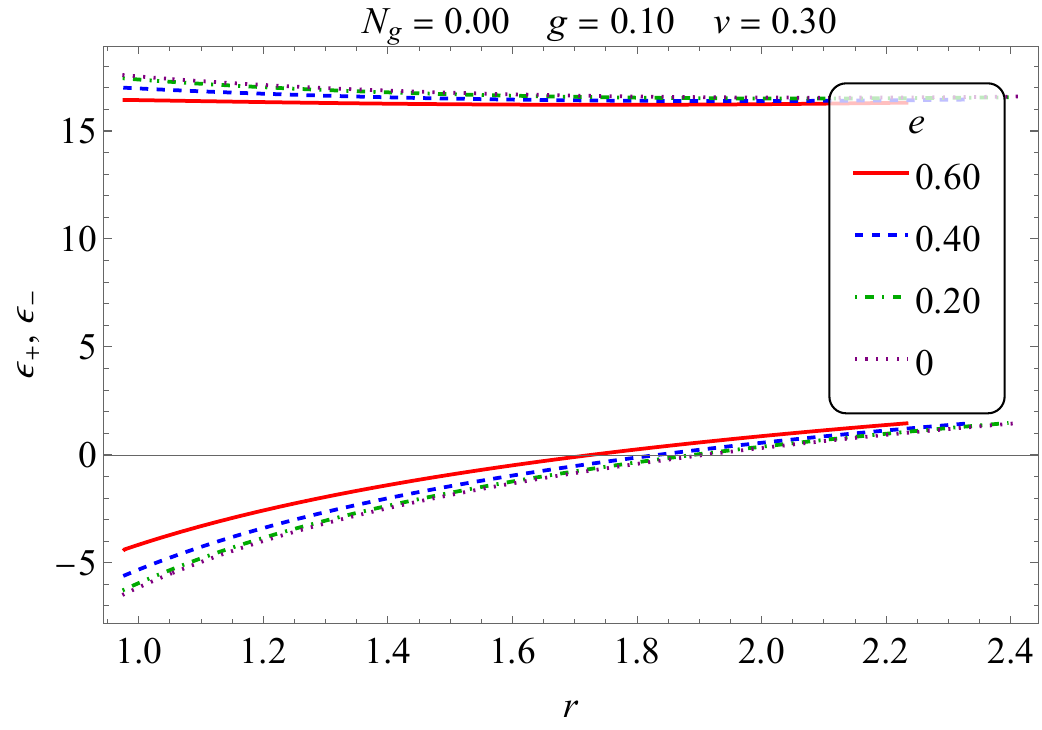}}
{\includegraphics[width=0.32\textwidth, height=4.4cm, keepaspectratio=false]{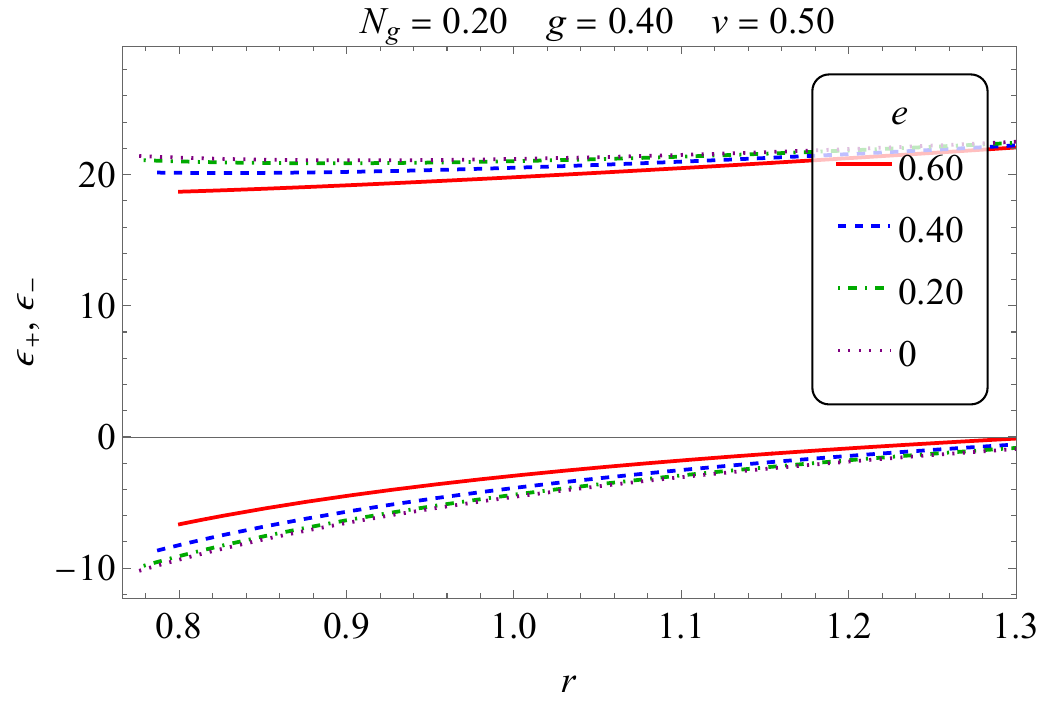}}

\vspace{0.05 cm}

{\includegraphics[width=0.32\textwidth, height=4.4cm, keepaspectratio=false]{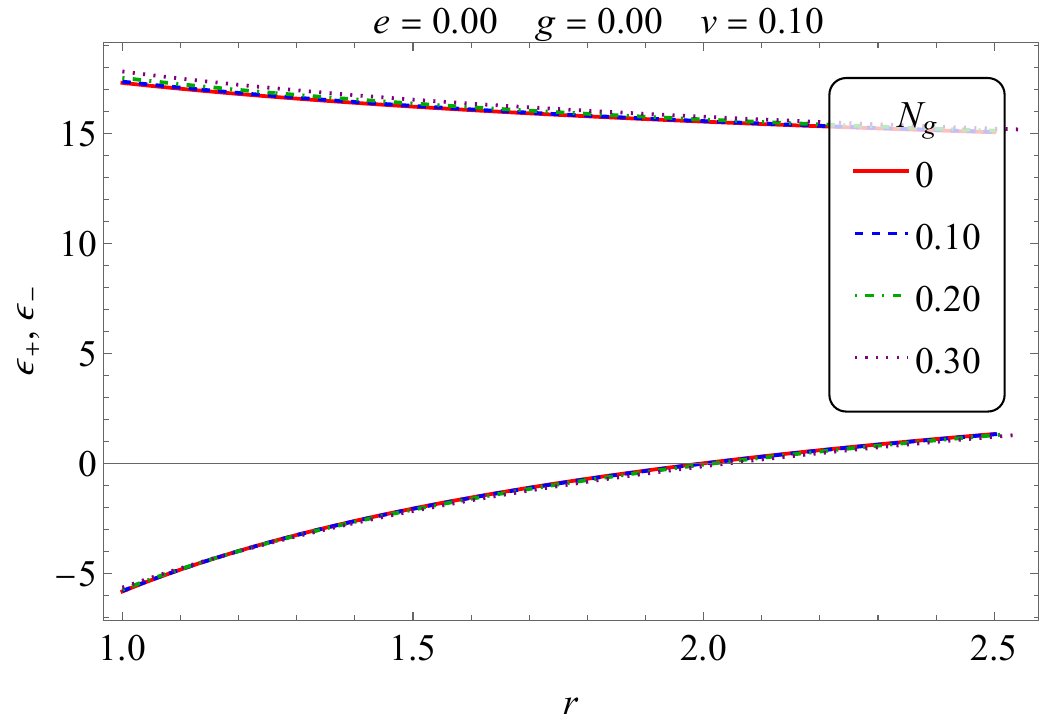}}
{\includegraphics[width=0.32\textwidth, height=4.4cm, keepaspectratio=false]{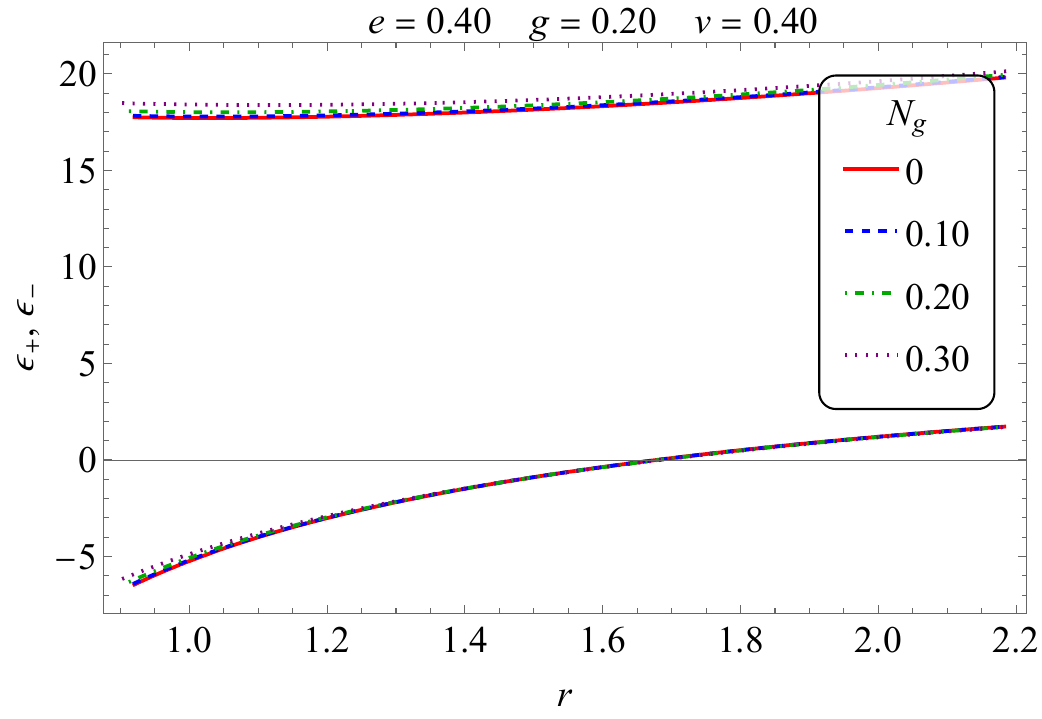}}
{\includegraphics[width=0.32\textwidth, height=4.4cm, keepaspectratio=false]{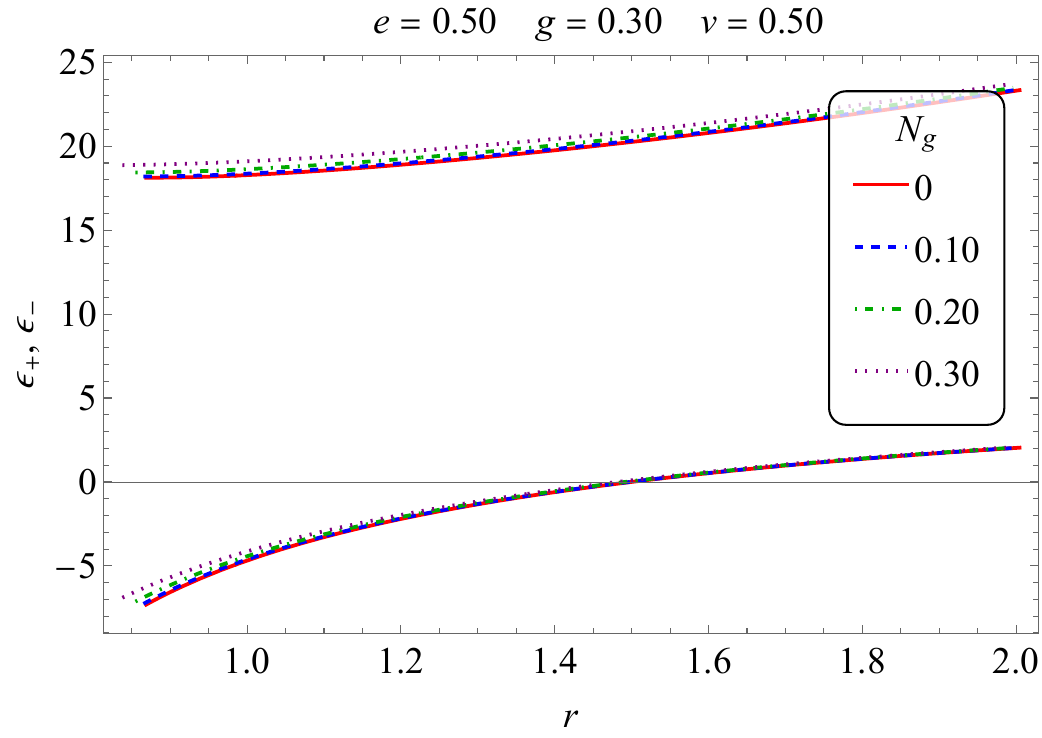}}

\vspace{0.05 cm}

{\includegraphics[width=0.32\textwidth, height=4.4cm, keepaspectratio=false]{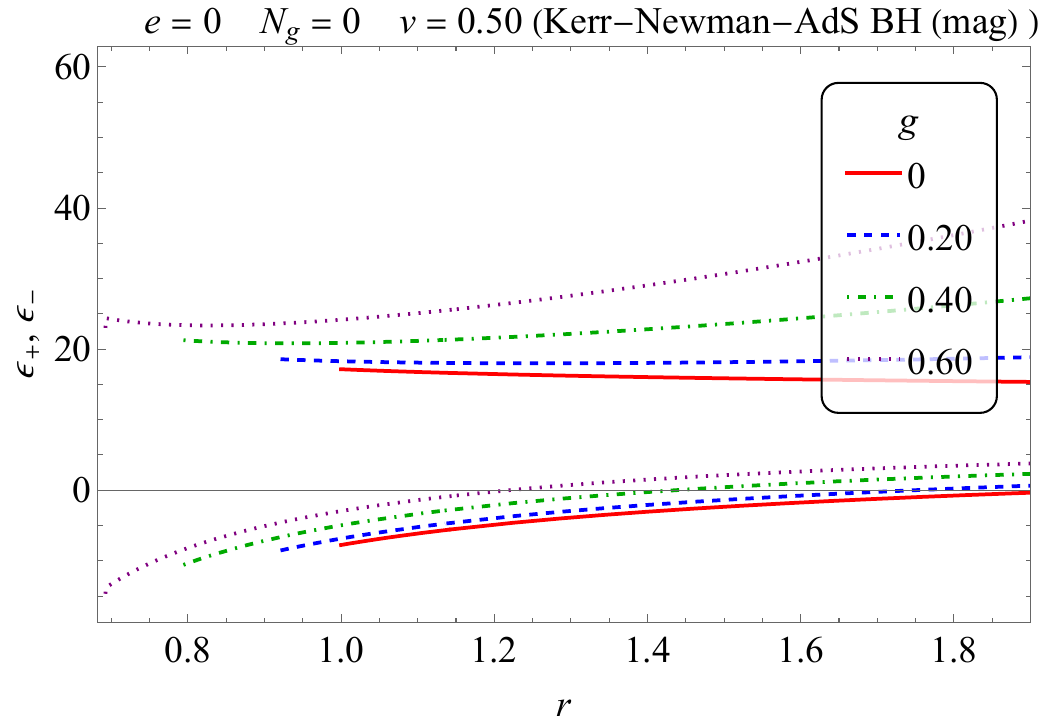}}
{\includegraphics[width=0.32\textwidth, height=4.4cm, keepaspectratio=false]{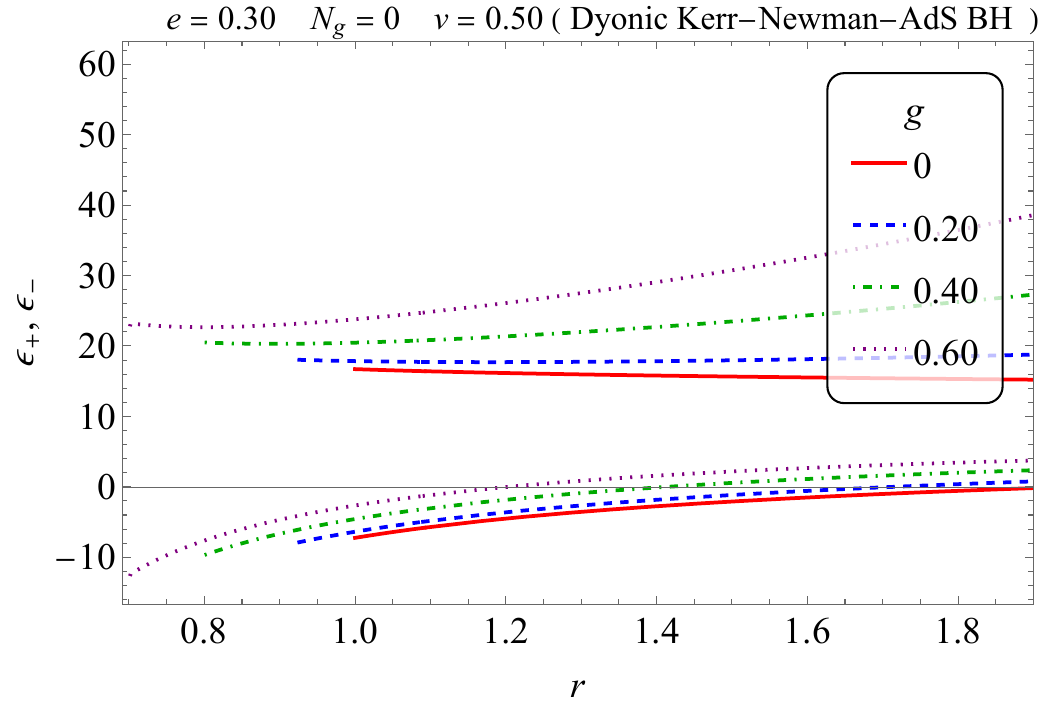}}
{\includegraphics[width=0.32\textwidth, height=4.4cm, keepaspectratio=false]{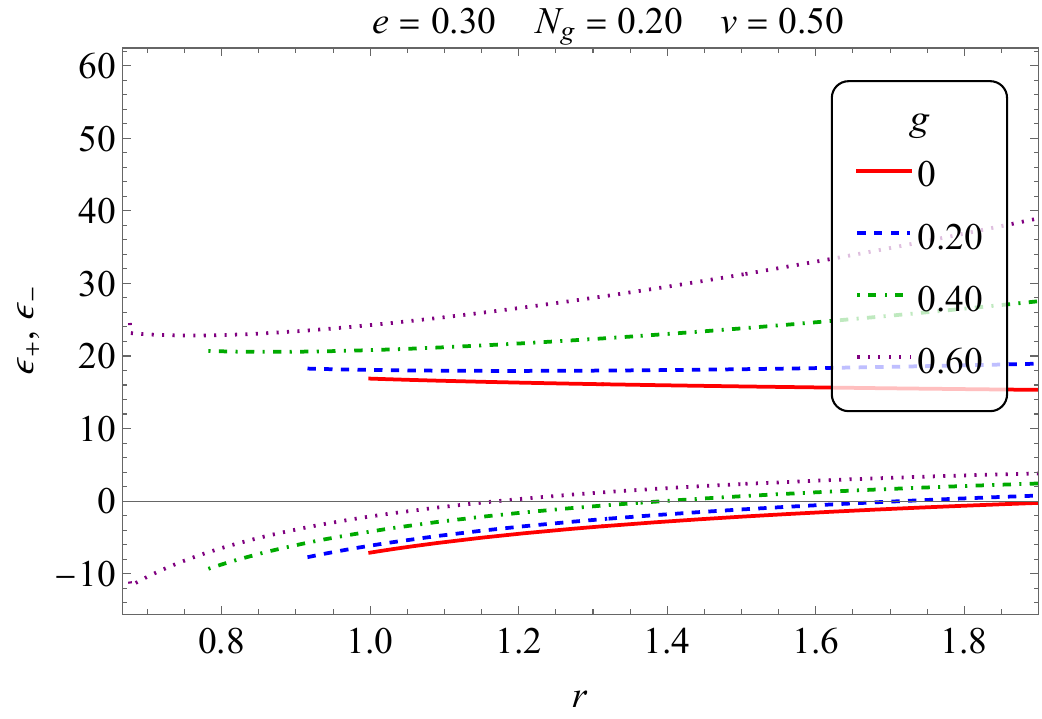}}

\caption{      \justifying
 Radial dependence of  $\epsilon_{\pm}$, each line corresponds to a different $a_{ext}$ value, and same color lines represent two different branches ($\epsilon_{+}$ and $\epsilon_{-}$) of the same equation as defined by Eq.~(\ref{Eneergy at infinity}). $(\sigma_0,\xi
)\equiv$ (100, 0) respectively, whereas $M=1$.}
\label{Plot: energy_pm vs r}
\end{figure*}

\begin{equation}\label{alpha and beta}
    \alpha=\sqrt{-g_{tt} + g_{t\phi}^2/g_{\phi\phi}} \qquad \mbox{and}
\qquad \beta^\phi = (\sqrt{g_{\phi\phi}}/\alpha)\,\omega^\phi\, ,
\end{equation}
 where \(\omega^\phi = -g_{t\phi}/g_{\phi\phi}\) refers to the frame-dragging angular velocity. We then obtain  the lapse function, shift vector, and angular velocity, respectively, as 
 \begin{equation}
 \label{alpha val and beta val}
 \begin{aligned}
       \qquad & \alpha(r,\theta) =
\sqrt{
\frac{(aA - \mathcal{B})\, R_g\, \Theta_g  \sin^2\theta}{A^2 R_g - \mathcal{B}^2 \Theta_g  \sin^2\theta}
}
,\\
& \beta^\phi(r,\theta) =
\frac{
A R_g - a \mathcal{B} \,\Theta_g \sin^2\theta
}{
(aA - \mathcal{B})\,\sqrt{R_g\, \Theta_g}\,\sin\theta
}\,\\
&
\omega(r,\theta) = -\Xi \, \frac{R_g A - \Theta_g a \mathcal{B} \sin^2\theta}{-R_g A^2 + \Theta_g \mathcal{B}^2 \sin^2\theta}.
,
 \end{aligned}
\end{equation}

so that the following line element in the ZAMO frame can be written as
 \begin{equation}
   ds^2= -d\hat{t}^2+ \Sigma_{i=1}^{3} (d\hat{x})^2= \eta_{\mu\nu} dx^{\mu}dx^{\nu}\, .
\end{equation}
It should be emphasized that the hat-labeled quantities are observed in the ZAMO frame and connected to the Boyer-Lindquist coordinates by using the transformation \cite{Khodadi2022, Simpson2023}
 \begin{equation}
 \begin{aligned}
\hat{l_{0}} & = \frac{l_{0}}{\alpha} + \sum_{i=1}^{3} 
\frac{\beta^{i}}{\sqrt{g_{ii}}} {l}_{i}, 
\quad 
\hat{l_{i}} = \frac{{l}_{i}}{\sqrt{g_{ii}}},\quad (Covariant form)\\
\\  
\hat{l}^{0} &= \alpha l^{0}, 
\,\,\, 
\hat{l}^{i} = \sqrt{g_{ii}}\, l^{i} - \alpha \beta^{i} l^{0}\, .\,\,(Contravariant form)
 \end{aligned}
\end{equation}

Having established ZAMO transformations, we now evaluate the Comisso-Asenjo (CA) process's ability to extract energy from the BH. One of the main conditions for this extraction mechanism to be physically viable is that the decelerated plasmoid must acquire a negative energy as measured by an asymptotic observer. To derive the exact conditions under which this occurs, we employ the one-fluid approximation, defining the stress-energy tensor of the magnetized plasma as: 
\begin{equation}
T^{\mu\nu} = p\,g^{\mu\nu} + \omega\,U^\mu U^\nu + F^{\mu\sigma}F^\nu{}_{\sigma} - \tfrac{1}{4}g^{\mu\nu}F_{\alpha\beta}F^{\alpha\beta},
\end{equation}
where $p$ is the pressure due to plasma, $\omega$ is the total enthalpy density (rest-mass plus thermal energy), $U^\mu$ is the plasma four-velocity, and $F^{\mu\nu}$ is the electromagnetic field tensor \cite{Wei2022Braneworld, Simpson2023, Liu2023MagnetizedNonlinear}. The energy-at-infinity density of the plasma as seen by a distant observer is 
\begin{equation}
e_\infty = -\,\alpha\,g_{\mu 0}\,T^{\mu0} = \alpha\,\hat e + \alpha\,\beta^\phi\,\hat P_\phi\, ,
\end{equation}
where \(\hat e\) and \(\hat P_\phi\) are the total energy density and azimuthal momentum densities.  One can decompose \(e_\infty\) using the components as given in \cite{Comisso2021} into a hydrodynamic part \(e_\infty^\text{hyd}\) and an electromagnetic part \(e_\infty^\text{em}\) as
\begin{equation}
\begin{aligned}
    e^{\infty} =& \underbrace{\alpha ( \omega \hat{\gamma}^{2} - p ) + \alpha \beta^{\phi} \omega \hat{\gamma}^{2} \hat{v}^{\phi}}_{\textstyle e^{\infty}_{\mathrm{hyd}}} \\
\quad &+\quad \underbrace{\frac{\alpha}{2} ( \hat{B}^{2} + \hat{E}^{2} ) + \alpha \beta^{\phi} ( \hat{B} \times \hat{E} )_{\phi}}_{\textstyle e^{\infty}_{\mathrm{em}} }\, .
\end{aligned}
\end{equation}

In the reconnection event, the electromagnetic energy is quickly converted into plasma kinetic energy, causing \(e^\infty_\text{em}\) to be radiated away. Consequently, the contribution of the EM energy can be neglected relative to the hydrodynamic energy. Thus, the net extracted energy is carried by the plasma (hydrodynamic) component, and we take $e_\infty\simeq e^{\infty}_{hyd}$ at a large distance. The observed energy at infinity is therefore given by  
\begin{equation}\label{ehydfinal}
    e^{\infty}= e^{\infty}_{hyd}=   \alpha\,\omega\,\hat\gamma_K\,(1 + \beta^\phi \hat v^\phi) - \frac{\alpha\,p}{\hat\gamma_K}\, ,
\end{equation}
where \(\hat\gamma_K=1/\sqrt{1-\hat v_K^2}\) is the Lorentz factor in the ZAMO frame. We further consider a plasma element in the local rest frame ${x'^{\mu}} = (x'^{0},x'^{1},x'^{2},x'^{3})$ with a Keplerian angular velocity \(\Omega_K\) in the equatorial plane. Directions $x'^{1},x'^{3}$ are then chosen to be parallel to the radial and azimuthal directions $r,\phi$, respectively. Using the transformation relation relative to a ZAMO frame, its local azimuthal velocity is given as 
\begin{equation}\label{vk}
\hat v_K \;=\; \frac{d\hat x^\phi}{d\hat t}
     \;=\;\frac{\sqrt{g_{\phi\phi}}}{\alpha}\,\bigl(\Omega_K\bigr) - \beta^\phi
\;=\;\frac{\sqrt{g_{\phi\phi}}}{\alpha}\,(\Omega_K - \omega^\phi)\,,
\end{equation}
where we have used \(\beta^\phi=(\sqrt{g_{\phi\phi}}/\alpha)\,\omega^\phi\). Thus, the resulted form of $\hat{v}_K$ becomes 
\begin{equation}
   \hat{v}_K= \frac{A R_g (A \Omega_K - \Xi) + \mathcal{B} \Theta_g (a \Xi - \mathcal{B} \Omega_K) \sin^2\theta}{(aA - \mathcal{B}) \Xi \sin\theta \sqrt{R_g \Theta_g}}\, .
\end{equation}

 We acquire the final form for $\hat{v_K}$ by putting $\Omega_K$ in Eqs.~(\ref{ehydfinal}) and (\ref{vk}). Now, $\hat{v}^{\phi}$ as measured in the ZAMO frame is the outflow velocity, and reads as~\cite{Comisso2021}
\begin{equation}\label{vout}
\hat{v}^{\phi}
=
\frac{\hat{v}_{K} \pm v_{{out}} \cos \xi}
{1 \pm \hat{v}_{K} v_{{out}} \cos \xi},
\quad
v_{{out}}
=
\sqrt{\frac{\sigma_{0}}{1+\sigma_{0}}}\, ,
\quad \sigma_0=\frac{{B}^2}{4\pi\omega},
\end{equation}
where $v_{out}$ is the relativistic Alfv\'en velocity and the parameters ($\sigma_0$, $\xi$) are the plasma magnetization and orientation angle between the magnetic field lines and the azimuthal direction in the equatorial plane, respectively. By taking (\ref{vk}) and (\ref{vout}) into (\ref{ehydfinal}) and defining the specific energy (per enthalpy) \(\epsilon^\infty = e^\infty/\omega\), one obtains the final form for hydrodynamic energy at infinity per plasma enthalpy as follows (see details in \cite{Comisso2021})
\begin{equation} 
\begin{aligned}
\epsilon^{\infty}_{\pm} =  \alpha\, \hat{\gamma}_K \Bigg[&
(1 + \beta^\phi \hat{v}_K) \sqrt{1 + \sigma_0}
\pm \cos \xi\, (\beta^\phi + \hat{v}_K) \sqrt{\sigma_0} \\ 
& \qquad - \frac{ \sqrt{1 + \sigma_0} \mp \cos \xi\, \hat{v}_K \sqrt{\sigma_0} }
{4 \hat{v}_K^2 \left( 1 + \sigma_0 - \cos^2 \xi\, \hat{v}_K^2 \sigma_0 \right)}
\Bigg]\, ,
\end{aligned}
\label{Eneergy at infinity}
\end{equation}
where $\pm$ corresponds to the accelerated and decelerated plasma. 
As clearly observed in the formula, the $\epsilon^{\infty}_{\pm}$ is dependent on the parameters of the BH spacetime and the configuration in the reconnection $(\sigma_0,\xi)$ . The accelerated plasma acquires more positive energy compared to its rest mass energy, while the decelerated plasma acquires negative energy, measured at infinity. Using (\ref{metric-main}), and taking the orientation angle ($\xi \rightarrow0$) for all the configurations, the resulting rotating energy at the equatorial plane ($\theta=\pi/2$) is thus obtained as
\begin{widetext}
\begin{equation}
 \begin{aligned}
\epsilon_{\pm}^{\infty}& = \left\{ \frac{Z^2 R_g \Theta_g \Xi^2}{X \big(-X \Omega_K ^2 + 2 A \Omega_K  R_g \Xi - 2 a \mathcal{B} \Omega_K  \Theta_g \Xi + (-R_g + a^2 \Theta_g) \Xi^2\big)} \right\}^{\frac{1}{2}} \\
& \times \sqrt{\frac{-Z R_g \Theta_g}{X}} \Bigg[ \pm \frac{\Omega_K  \sqrt{\frac{X}{-Z \Xi^2}} \sqrt{\sigma_0}}{\sqrt{\frac{-Z R_g \Theta_g}{X}}} \, + \,\frac{X (A \Omega_K  R_g - a \mathcal{B} \Omega_K  \Theta_g - R_g \Xi + a^2 \Theta_g \Xi) \sqrt{1 + \sigma_0}}{Z^2 R_g \Theta_g \Xi} \\
& \quad \pm \frac{1}{4} \Big( Z^2 R_g \Theta_g \Xi^2 - (A^2 \Omega_K  R_g - A R_g \Xi + \mathcal{B} \Theta_g (-\mathcal{B} \Omega_K  + a \Xi))^2 \Big)\,\, ( X \sqrt{\frac{-Z R_g \Theta_g}{X}} )^{-1} \\
& \quad \times \Big( Z^2 R_g \Theta_g \Xi^2 + Z^2 R_g \Theta_g \Xi^2 \sigma_0 - (A^2 \Omega_K  R_g - A R_g \Xi + \mathcal{B} \Theta_g (-\mathcal{B} \Omega_K  + a \Xi))^2 \sigma_0 \Big)^{-1} \\
& \quad \times \Bigg( \sqrt{\frac{X}{-Z \Xi^2}} (A^2 \Omega_K  R_g - A R_g \Xi + \mathcal{B} \Theta_g (-\mathcal{B} \Omega_K  + a \Xi)) \sqrt{\sigma_0}  - X \sqrt{\frac{-Z R_g \Theta_g}{X}} \sqrt{1 + \sigma_0} \Bigg) \Bigg]\, , 
\end{aligned}
\end{equation}
\end{widetext}
\noindent where we have used $X = A^2 R_g - \mathcal{B}^2 \Theta_g$ and $Z = \mathcal{B} - a A$, respectively. The corresponding conditions for the allowed energy extraction are then given as follows 
\begin{equation}\label{eqn: energy condition}
\begin{aligned}
   & \epsilon_{-}^{\infty} < 0,\qquad
   \Delta \epsilon_{+}^{\infty}
= \epsilon_{+}^{\infty}
- \left( 1 - \frac{\Gamma}{\Gamma - 1}\,\frac{p}{\omega} \right)
= \epsilon_{+}^{\infty} > 0\, .
\end{aligned}
\end{equation}
 Due to these assumptions, the accelerated part of the plasma acquires the energy greater than the rest mass energy and thermal energy, and, because the plasma is assumed relativistically hot, we choose a polytropic index $\Gamma= 4/3$ and $\mathcal{\omega}  =p/4$ in Eq.~(\ref{eqn: energy condition})~\cite{Simpson2023, Comisso2021, Zhang2023ChargedAdS}.
\begin{figure*}[htbp]
\centering

\begin{subfigure}{0.32\textwidth}
    \centering
    \includegraphics[width=\linewidth, height=4.5cm, keepaspectratio=false]{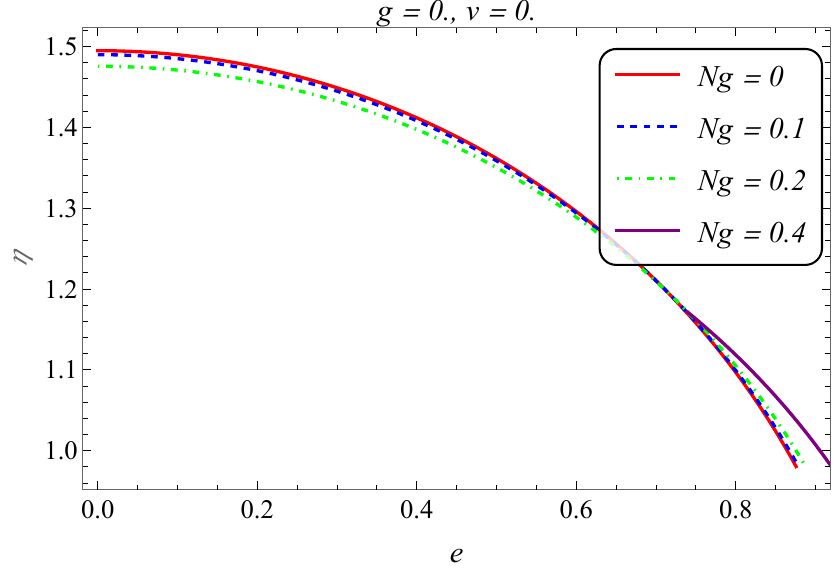}

\end{subfigure}
\hfill
\begin{subfigure}{0.32\textwidth}
    \centering
    \includegraphics[width=\linewidth, height=4.5cm, keepaspectratio=false]{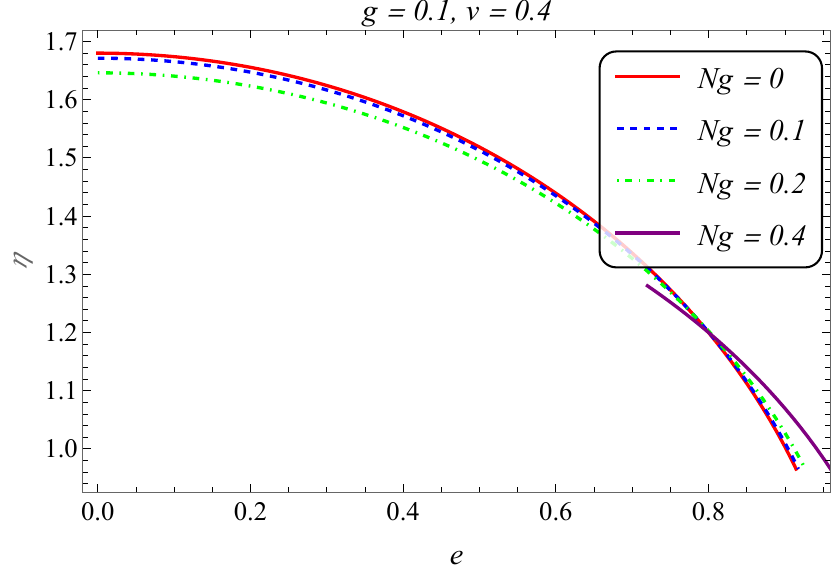}

\end{subfigure}
\hfill
\begin{subfigure}{0.32\textwidth}
    \centering
    \includegraphics[width=\linewidth, height=4.5cm, keepaspectratio=false]{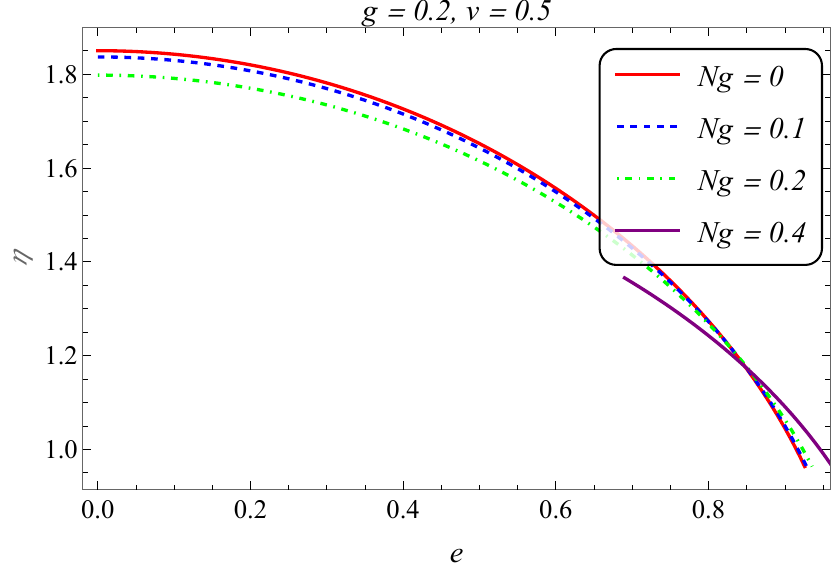}

\end{subfigure}

\begin{subfigure}{0.32\textwidth}
    \centering
    \includegraphics[width=\linewidth, height=4.5cm, keepaspectratio=false]{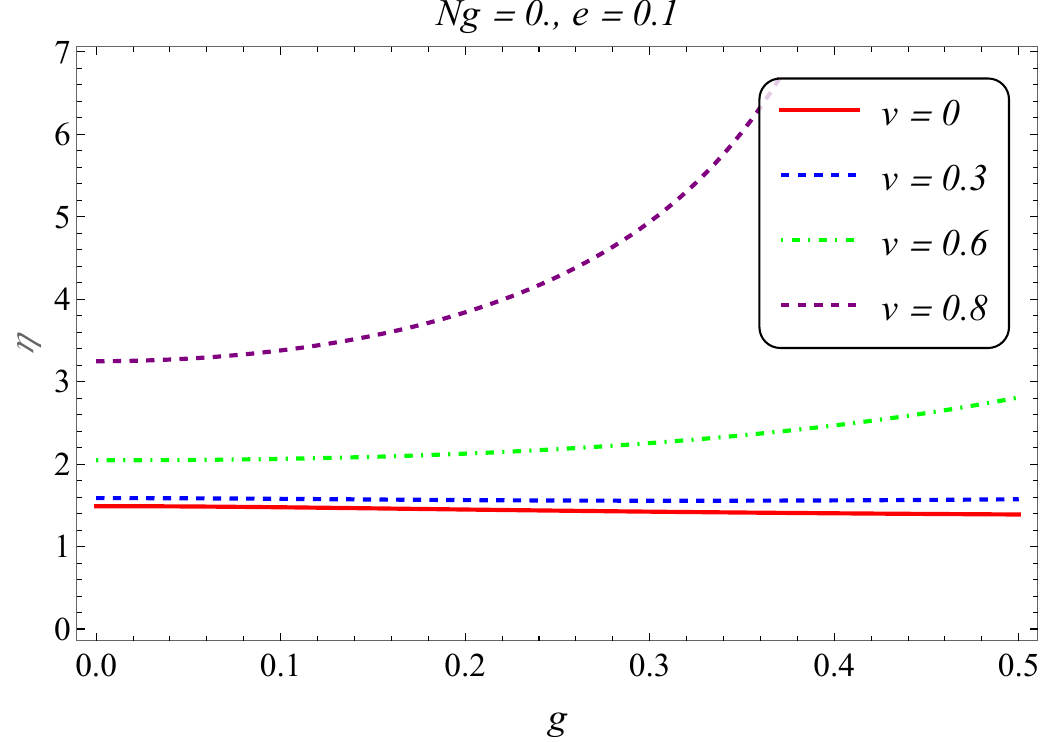}

\end{subfigure}
\hfill
\begin{subfigure}{0.32\textwidth}
    \centering
    \includegraphics[width=\linewidth, height=4.5cm, keepaspectratio=false]{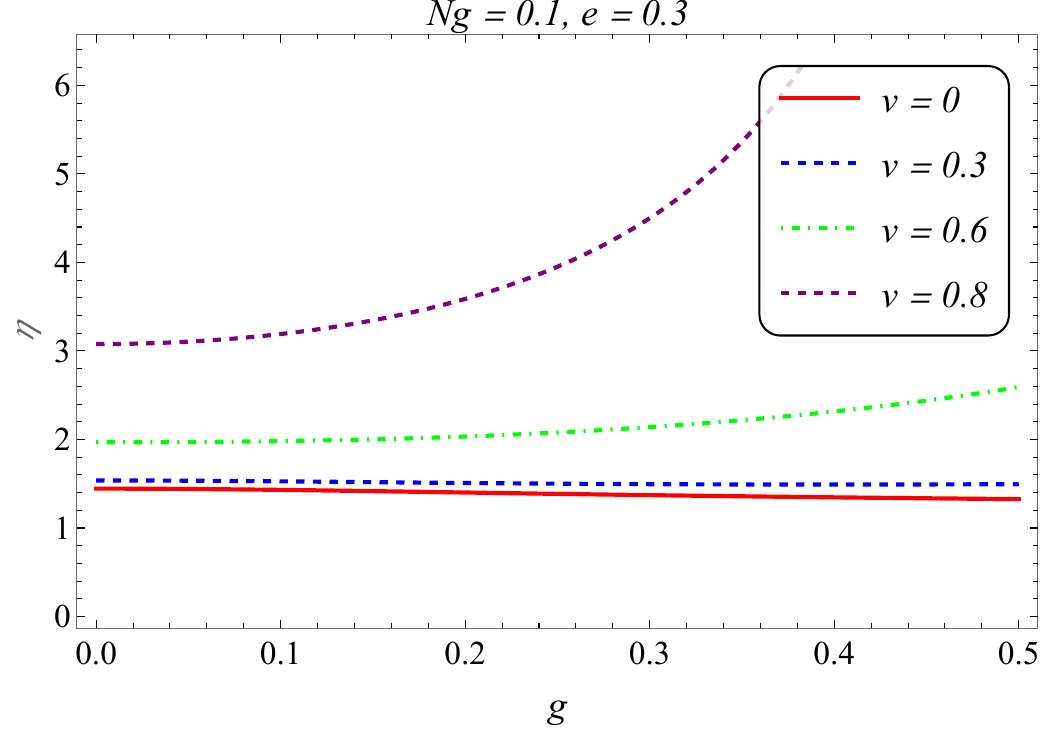}

\end{subfigure}
\hfill
\begin{subfigure}{0.32\textwidth}
    \centering
    \includegraphics[width=\linewidth, height=4.5cm, keepaspectratio=false]{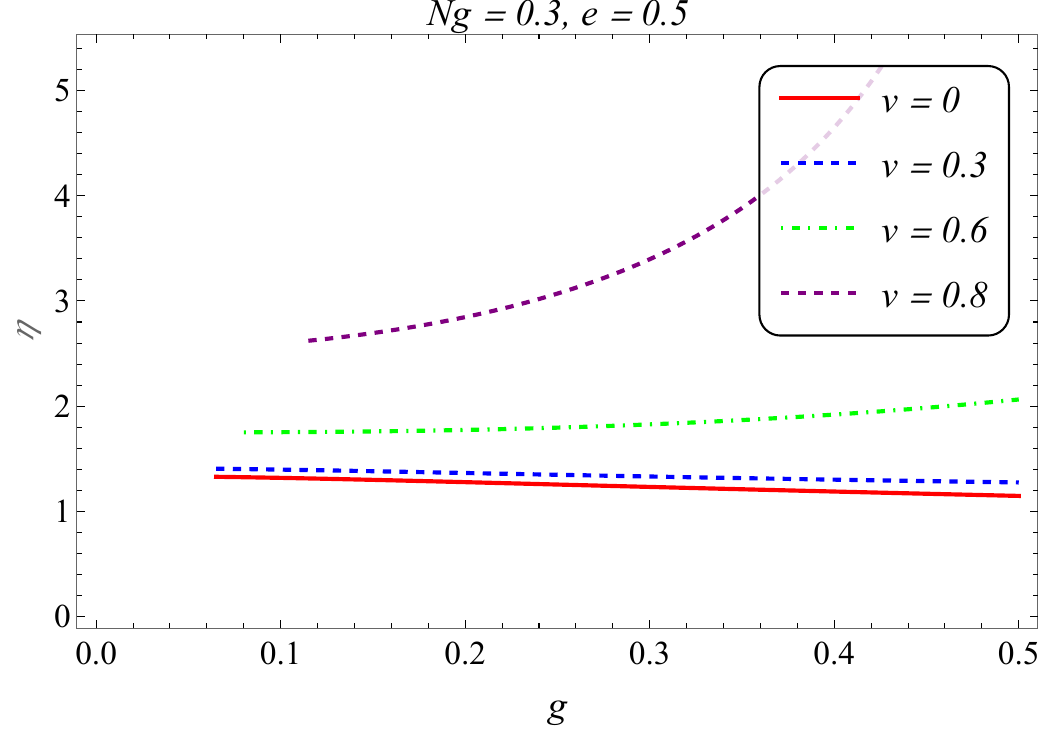}

\end{subfigure}

\begin{subfigure}{0.32\textwidth}
    \centering
    \includegraphics[width=\linewidth, height=4.5cm, keepaspectratio=false]{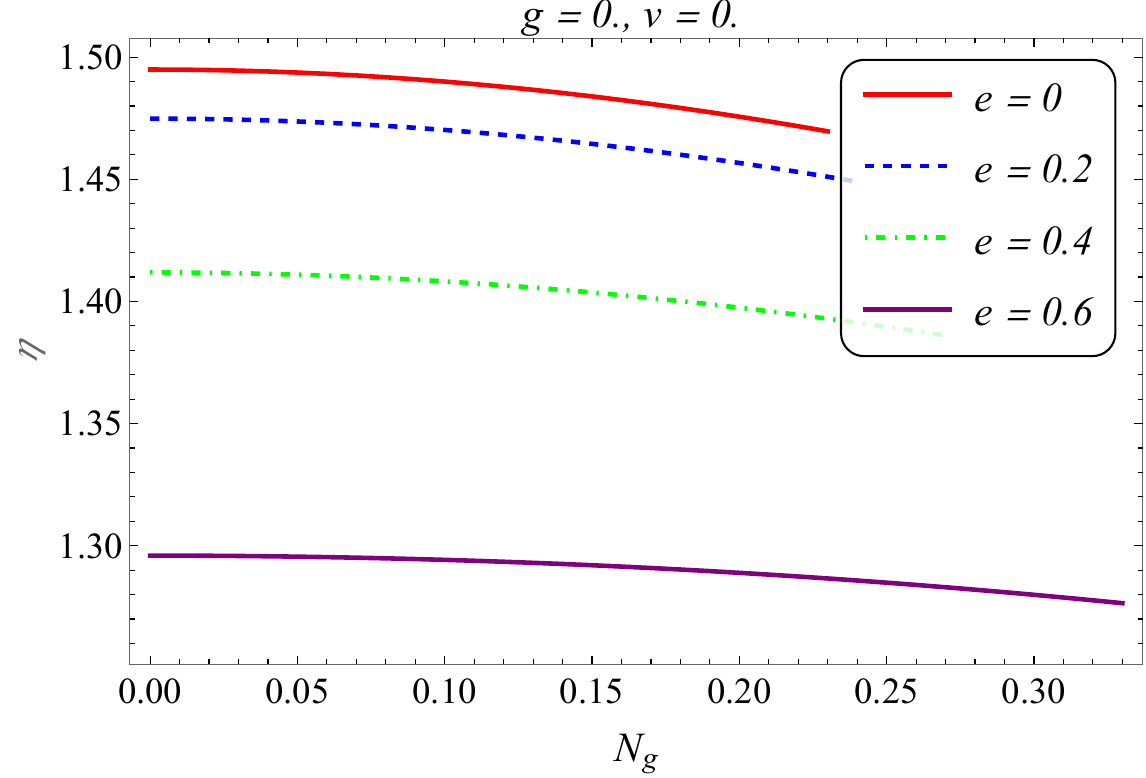}

\end{subfigure}
\hfill
\begin{subfigure}{0.32\textwidth}
    \centering
    \includegraphics[width=\linewidth, height=4.5cm, keepaspectratio=false]{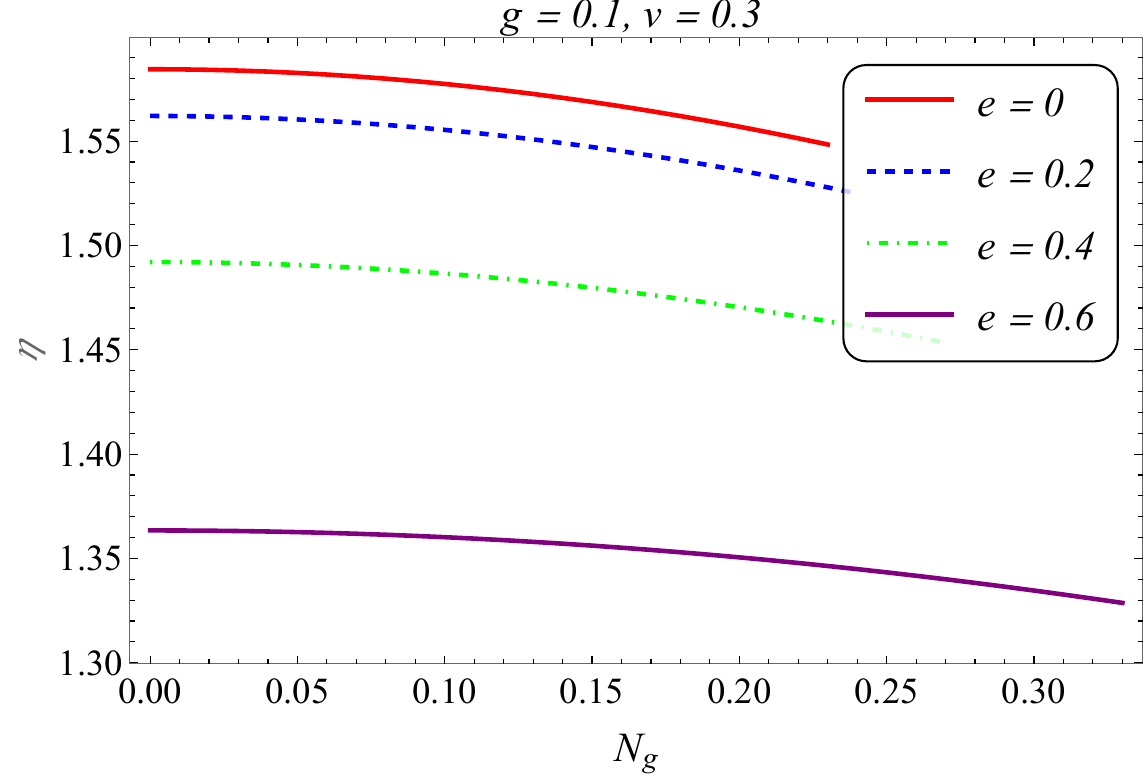}

\end{subfigure}
\hfill
\begin{subfigure}{0.32\textwidth}
    \centering
    \includegraphics[width=\linewidth, height=4.5cm, keepaspectratio=false]{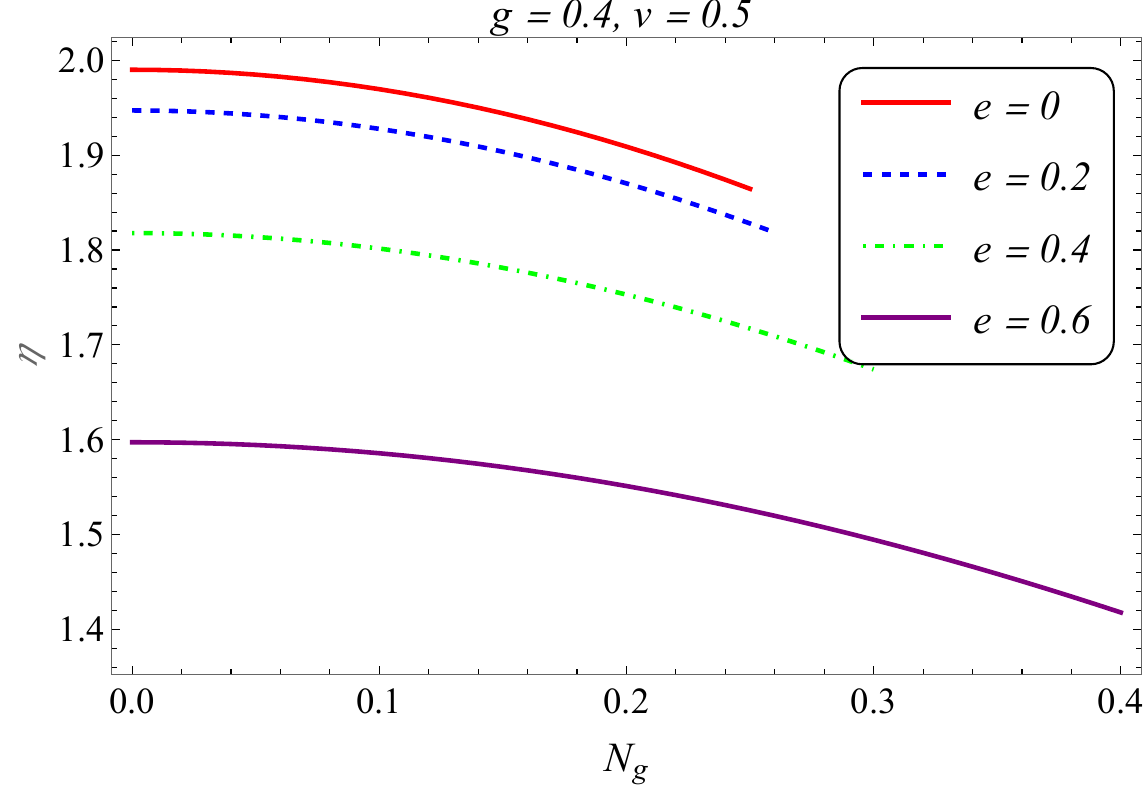}

\end{subfigure}

\begin{subfigure}{0.32\textwidth}
    \centering
    \includegraphics[width=\linewidth, height=4.5cm, keepaspectratio=false]{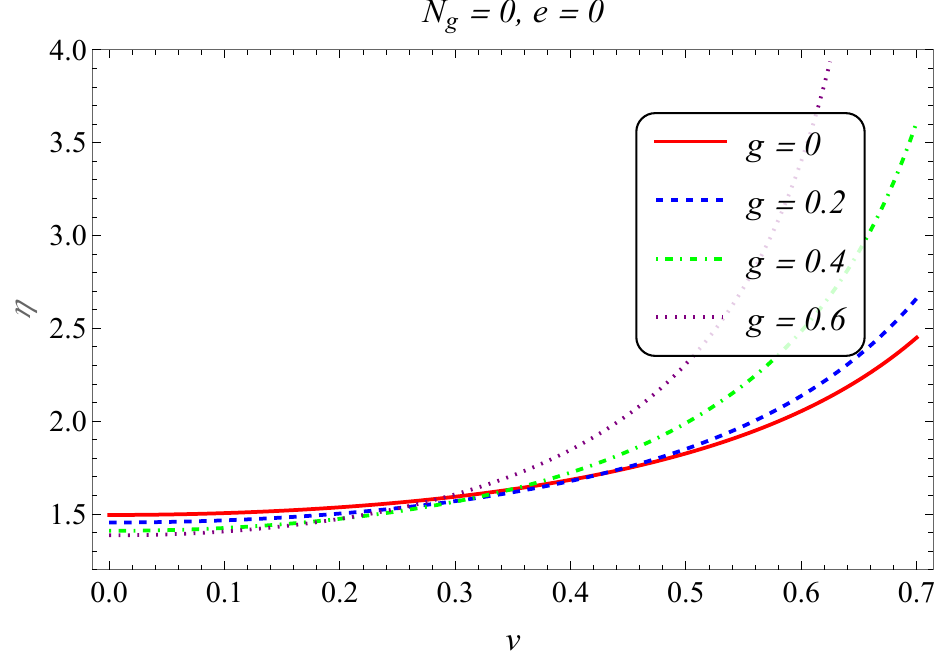}

\end{subfigure}
\hfill
\begin{subfigure}{0.32\textwidth}
    \centering
    \includegraphics[width=\linewidth, height=4.5cm, keepaspectratio=false]{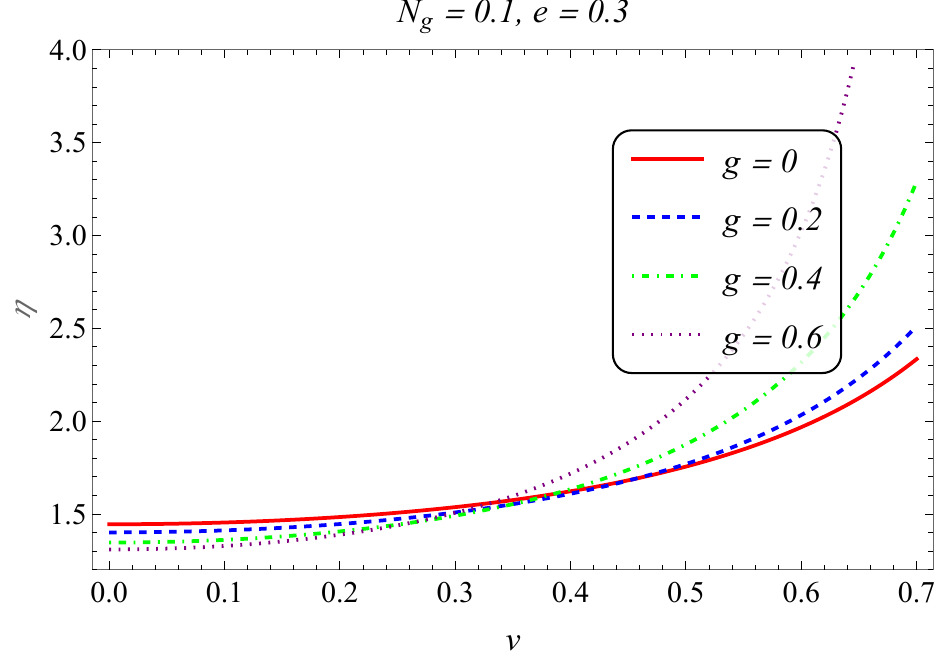}

\end{subfigure}
\hfill
\begin{subfigure}{0.32\textwidth}
    \centering
    \includegraphics[width=\linewidth, height=4.5cm, keepaspectratio=false]{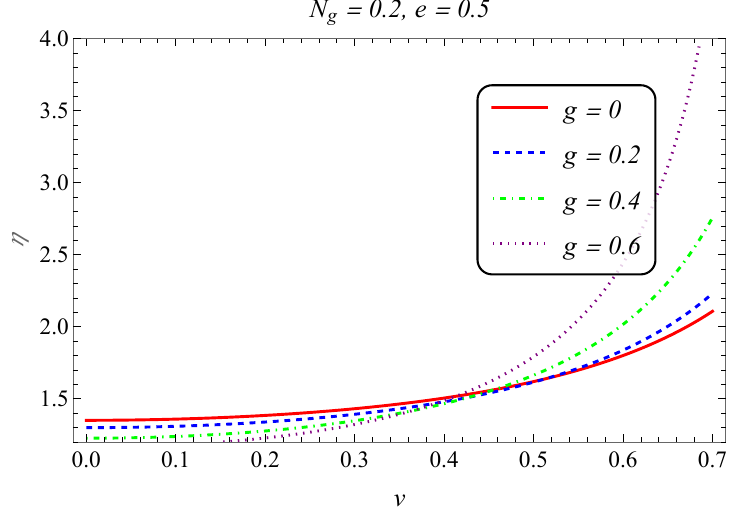}

\end{subfigure}

\caption{      \justifying{ Parametric dependence of efficiency $\eta$. Each parameter in the set $(e, g, N_g, v)$ varied individually, keeping other parameters constant; consequently, $(r_E, a_{ext})$ pair varies continuously at each point along each curve. The solid line represents the lowest parameter value for the particular plot.  Parameters $(\sigma_0,\xi)$ $\equiv$ (100, 0) respectively, whereas $M=1$.}}
\label{Plot: eff_vs_diff_parameters}
\end{figure*}

\begin{figure*}[htbp]
\centering
{\includegraphics[width=0.32\textwidth, height=4.5cm, keepaspectratio=false]{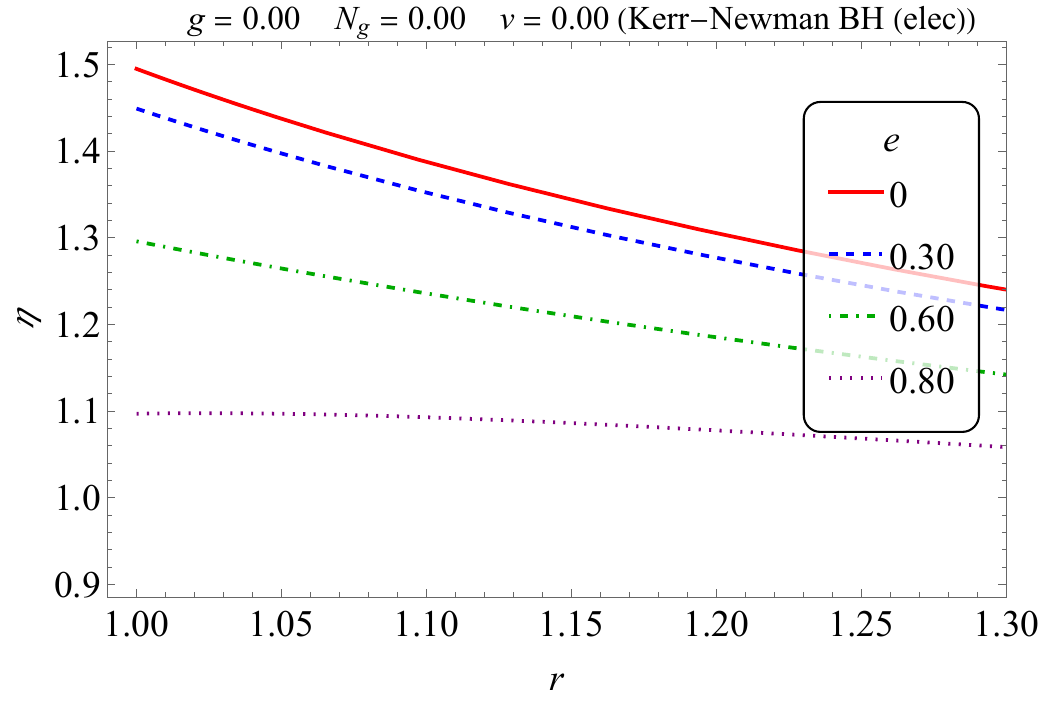}}
{\includegraphics[width=0.32\textwidth, height=4.5cm, keepaspectratio=false]{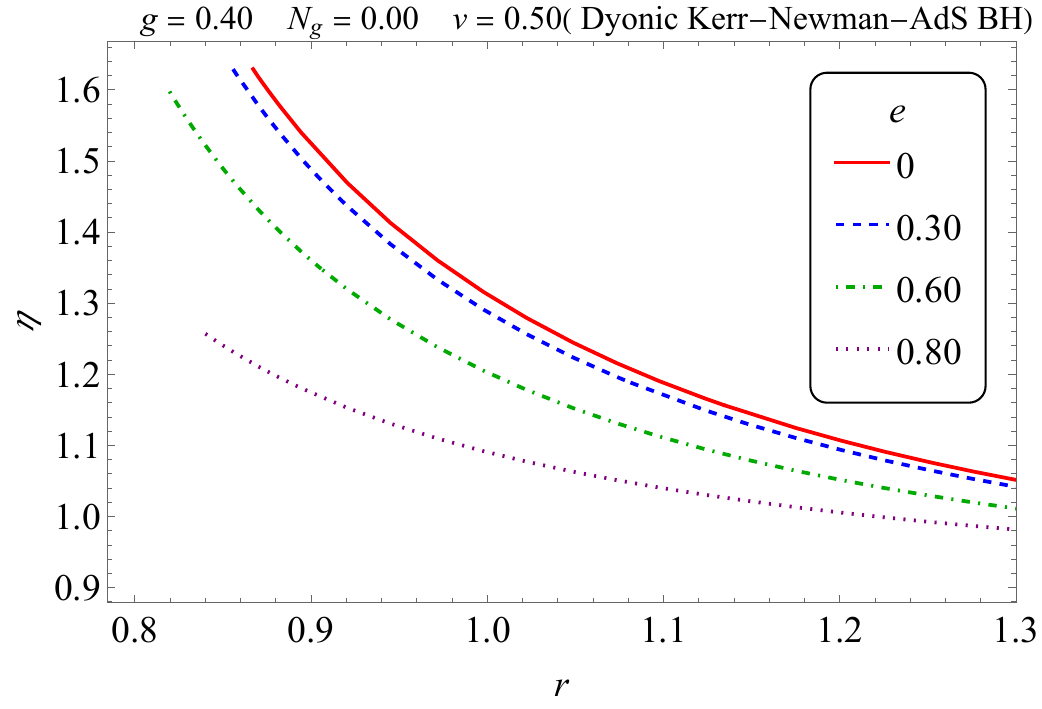}}
{\includegraphics[width=0.32\textwidth, height=4.5cm, keepaspectratio=false]{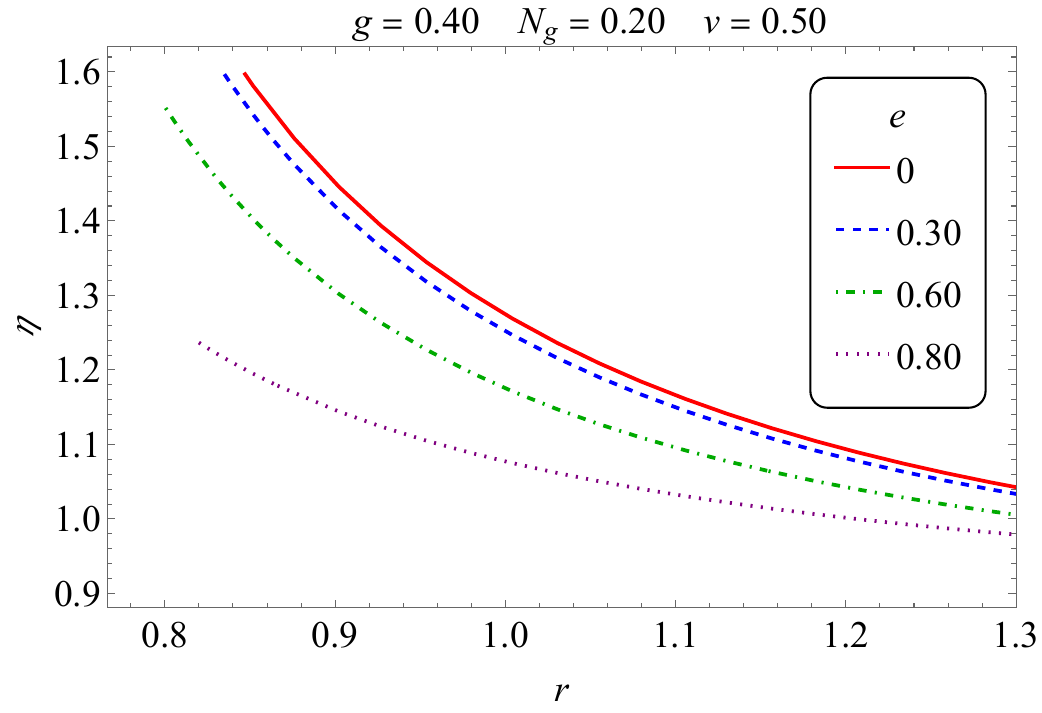}}
\vspace{0.05 cm}

{\includegraphics[width=0.32\textwidth, height=4.5cm, keepaspectratio=false]{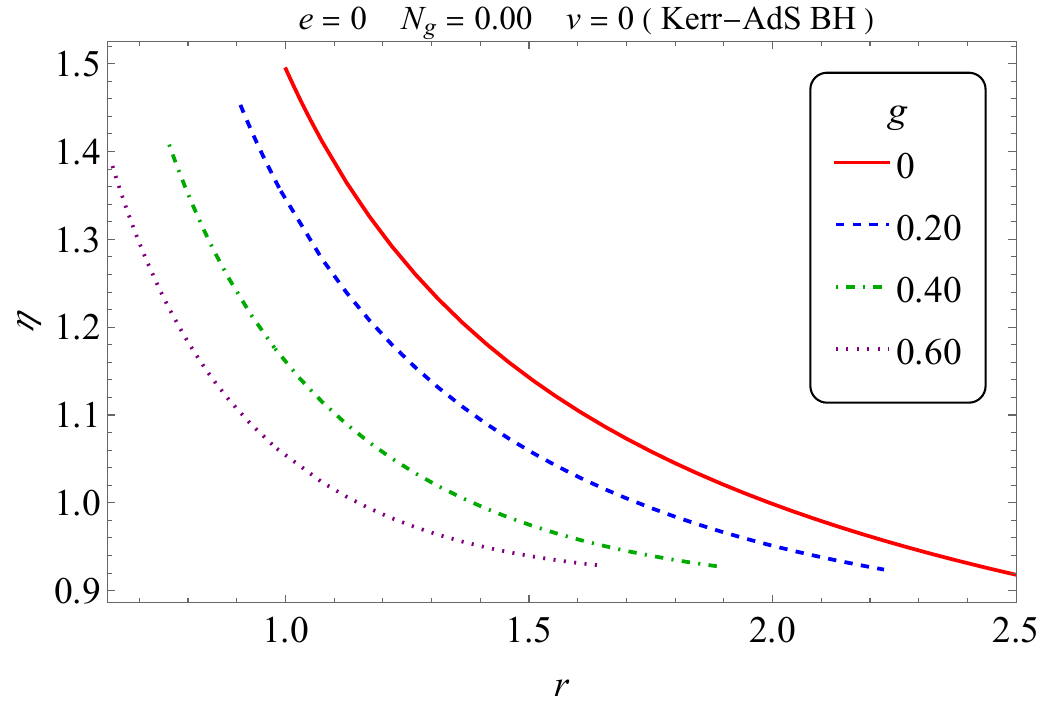}}
{\includegraphics[width=0.32\textwidth, height=4.5cm, keepaspectratio=false]{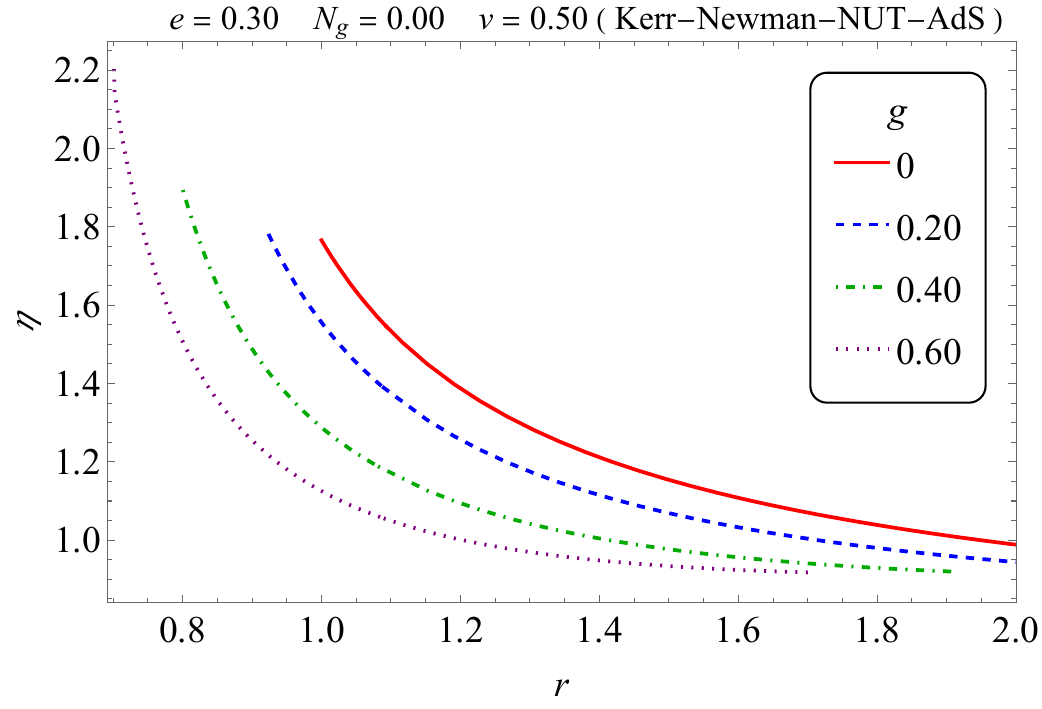}}
{\includegraphics[width=0.32\textwidth, height=4.5cm, keepaspectratio=false]{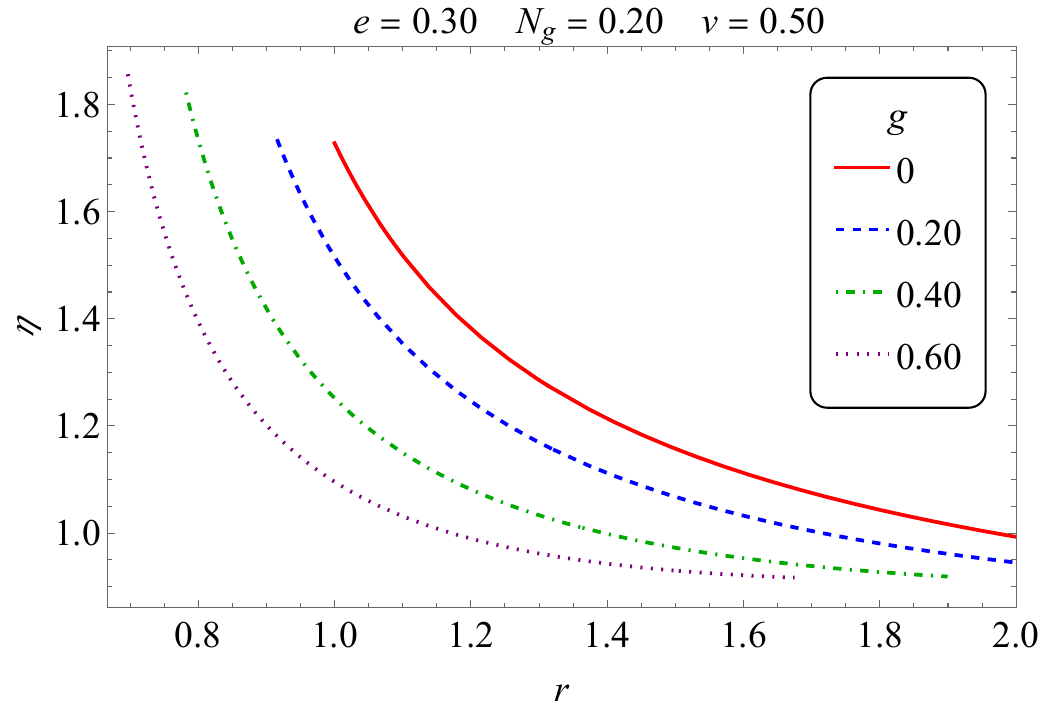}}

\vspace{0.05 cm}

{\includegraphics[width=0.32\textwidth, height=4.5cm, keepaspectratio=false]{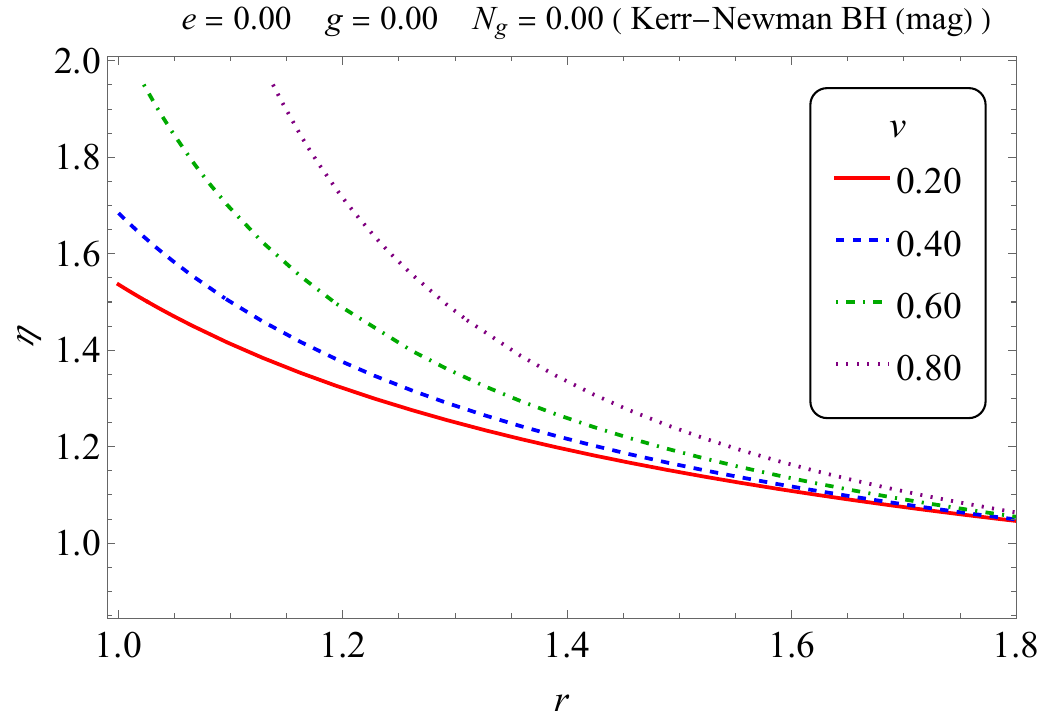}}
{\includegraphics[width=0.32\textwidth, height=4.5cm, keepaspectratio=false]{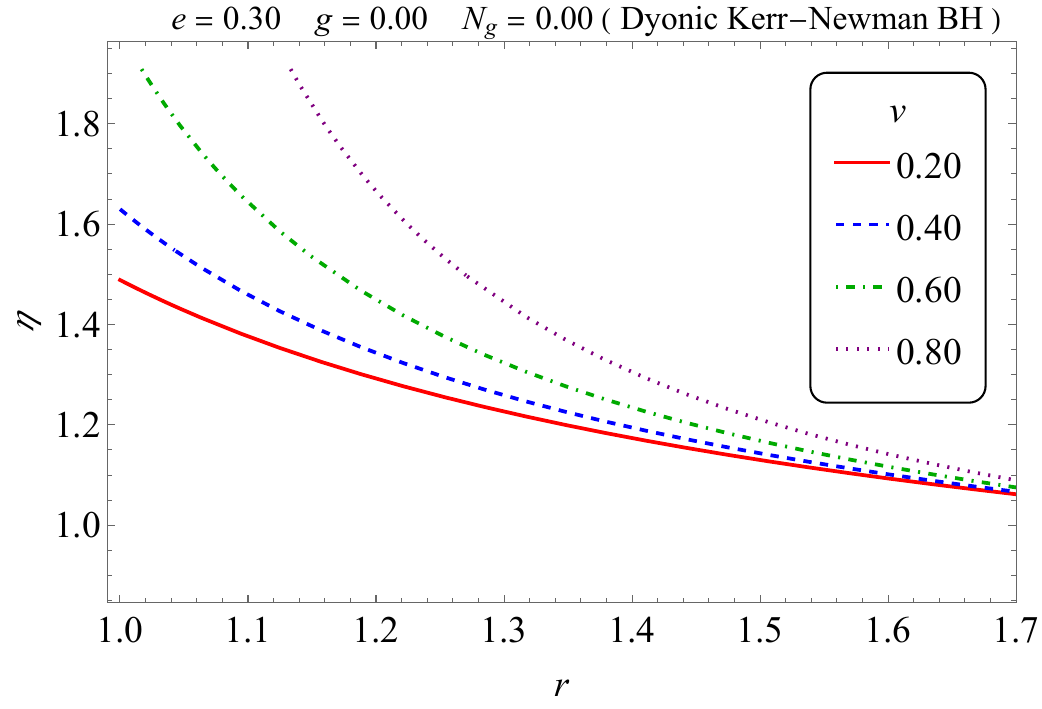}}
{\includegraphics[width=0.32\textwidth, height=4.5cm, keepaspectratio=false]{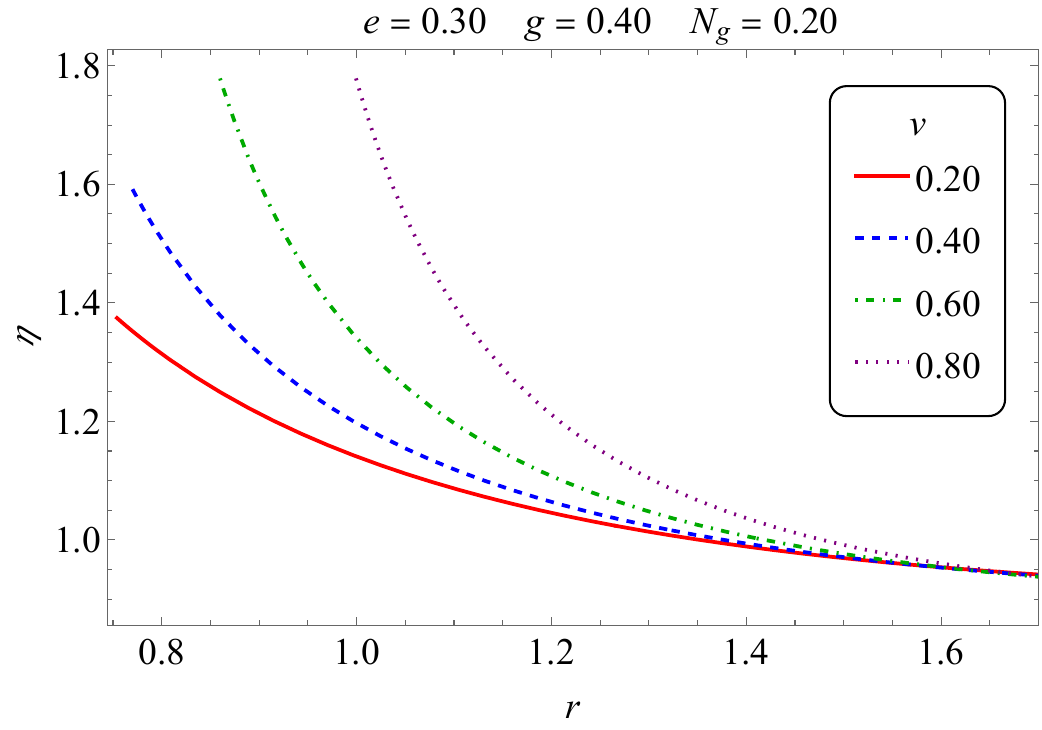}}

\vspace{0.05 cm}

{\includegraphics[width=0.32\textwidth, height=4.5cm, keepaspectratio=false]{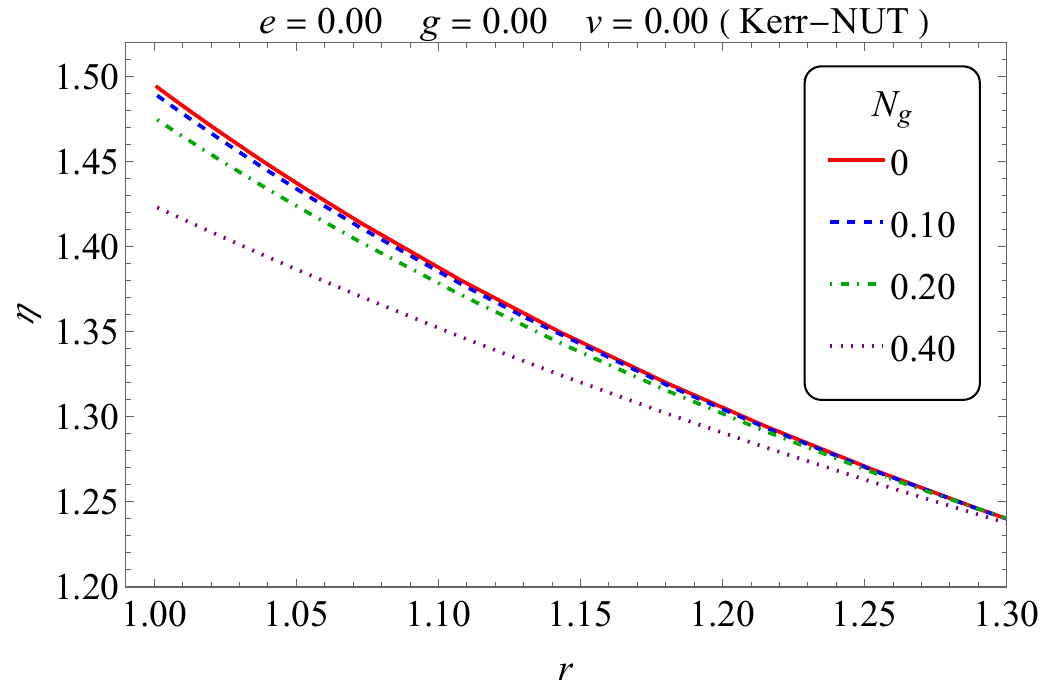}}
{\includegraphics[width=0.32\textwidth, height=4.5cm, keepaspectratio=false]{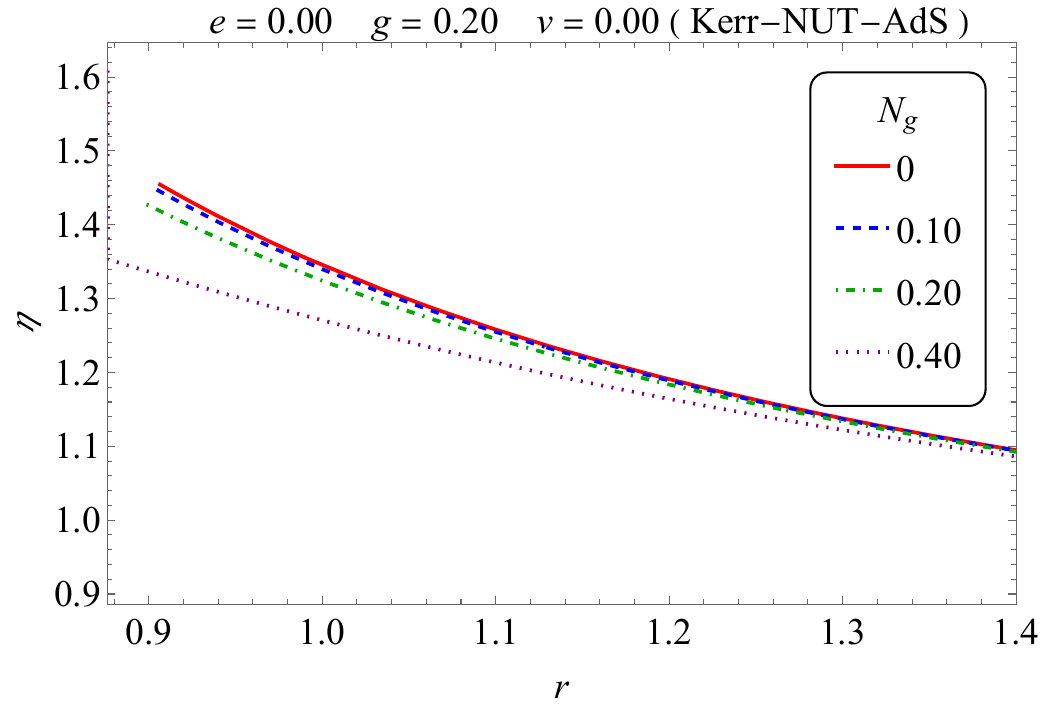}}
{\includegraphics[width=0.32\textwidth, height=4.5cm, keepaspectratio=false]{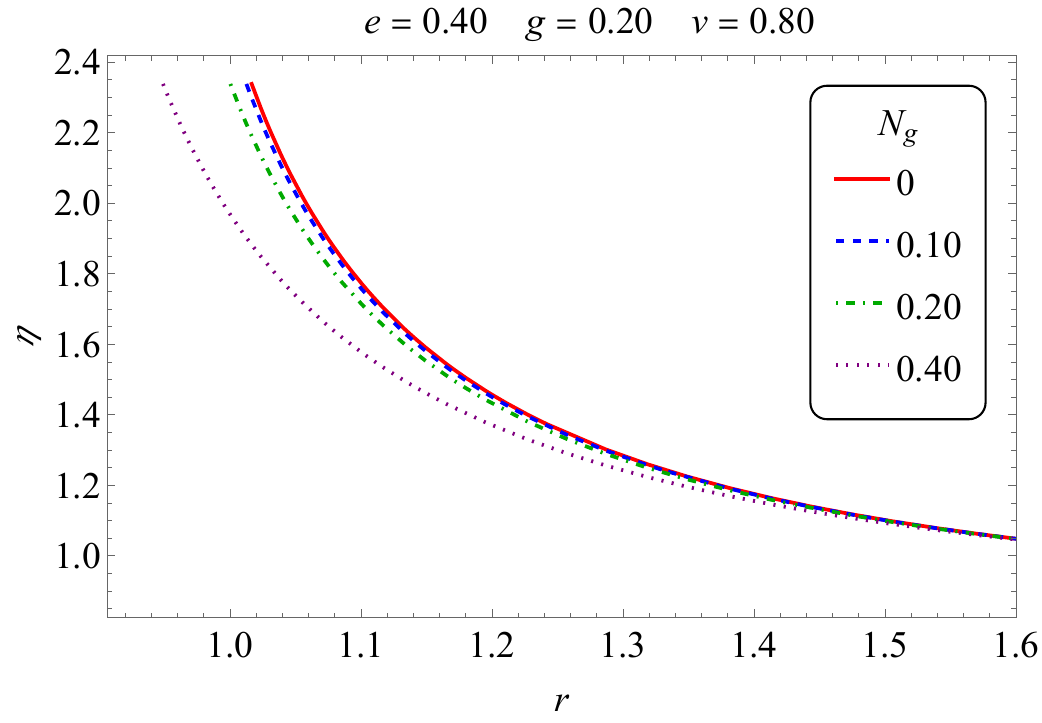}}

\caption{      \justifying Radial dependence of efficiency $\eta$, each parameter in the set $(e, g, N_g, v)$ varied individually, keeping other parameters constant. The leftmost origin of each curve represents the extremal horizon $r_E$, and $a_{ext}$, and is different for each line in the plot. $\eta$ attains a maximum at each ($r_E$, $a_{ext}$), solid lines denoting the lowest value of the parameter in the plot. Parameters $(\sigma_0,\xi
)$ $\equiv$ (100, 0) respectively, whereas $M=1$.}
\label{Plot: eff vs r different parameter values}
\end{figure*}

Before we analyze the spacetime metric, we fix some parameters to analyze the spacetime metric via the CA Process. We assume that magnetic reconnection occurs at the equatorial plane ($\theta=\pi/2$) throughout the analysis. Hereafter, we examine the conditions for energy extraction within the Dyonic Kerr-Newman-NUT-AdS spacetime. The analysis will be conducted at various reconnection points. Unlike the Kerr BH, which has an extremal at $(r, a)=(1, 1)$, each parameter set has different extremal values $(r_E, a_{ext})$, as seen in Table~\ref{Table: Various Parameters and different conditions}.
We construct region plots in the $(a,r)$ phase space for various parameter combinations, as illustrated in Fig.~\ref{Region Plot: spin vs r}, to identify the negative energy regions that permit the energy extraction via the magnetic reconnection. By explicitly mapping the locations of the degenerate event horizon ($r_E$) and the photon sphere ($r_{ph}$), we properly constrain the relevant geometric region between $r_{ph}$ and $r_{SLS}$. 
Within this region, we observe the decelerated plasma gaining the negative energy required for energy extraction as measured by the observer at infinity. Consequently, inside this region,  both positive and negative energy branches ($\epsilon_{\pm}$) show radial dependence, as seen in Fig.~\ref{Plot: energy_pm vs r}. By varying one parameter while keeping the others fixed, and simultaneously satisfying the causality conditions in Eq.~(\ref{singularity}), we can determine the difference between the positive and negative-energy states ($\epsilon_+,\,\epsilon_-$). The difference between $\epsilon_+$ and $\epsilon_-$ increases as we inch towards the extremal case, i.e., ($r\rightarrow r_E$). In addition, we can observe various radii where we don't get the negative energy, there we see that the efficiency ($\eta<1$), as we will see in section \ref{sec: Efficiency and Power}.
\begin{figure*}[htbp]
    \centering

    \includegraphics[width=0.83\textwidth, height=5.25cm, keepaspectratio=false]{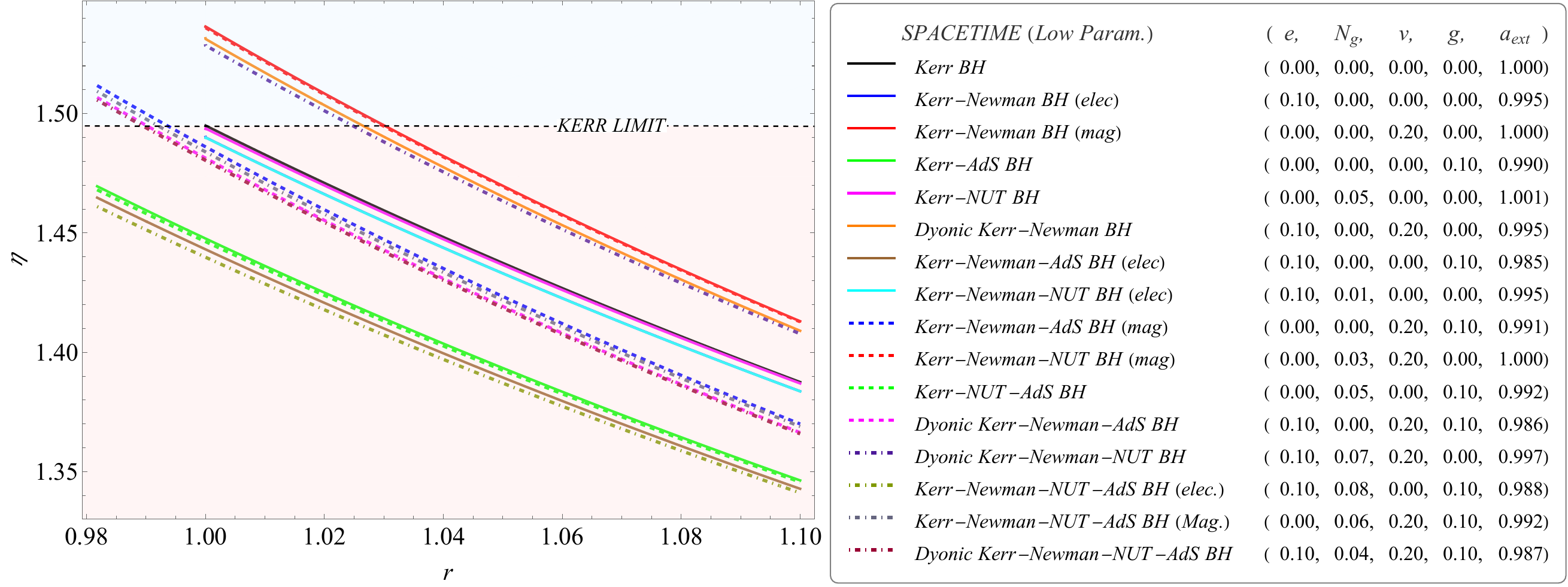}
    
    \vspace{0.0cm} 

    \includegraphics[width=0.83\textwidth, height=5.25cm, keepaspectratio=false]{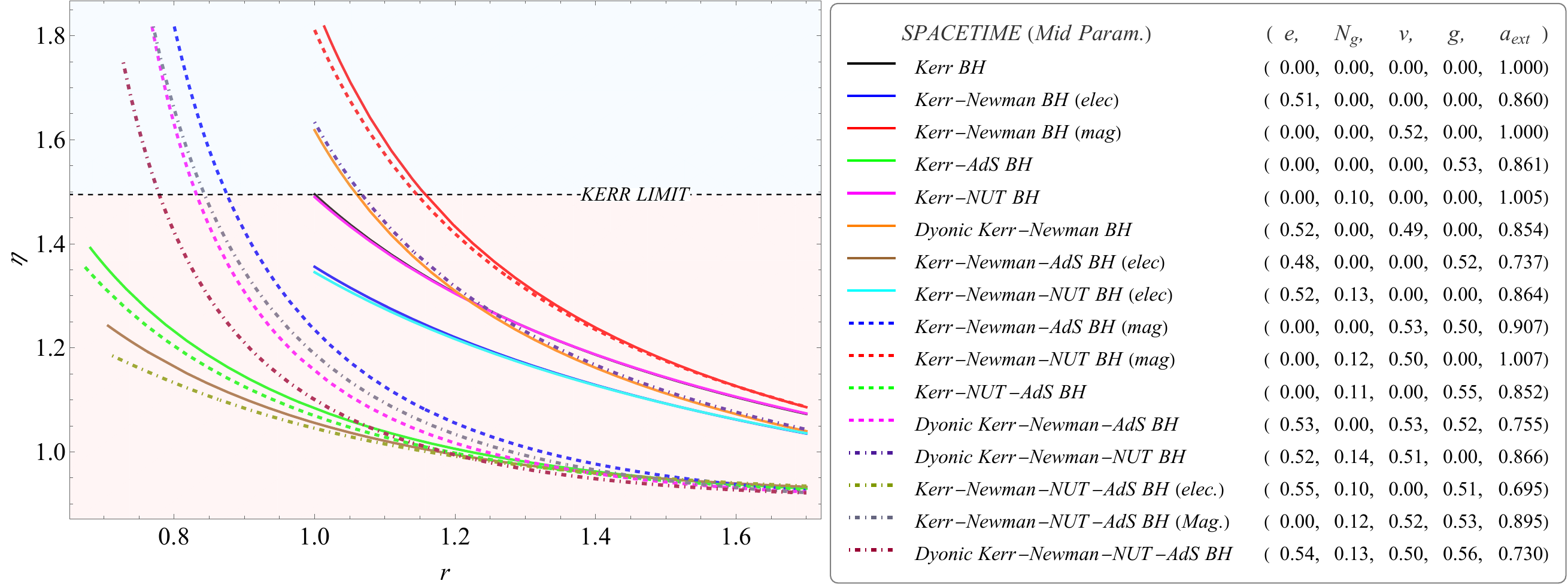}
    
    \vspace{0.0cm}

    \includegraphics[width=0.83\textwidth, height=5.25cm, keepaspectratio=false]{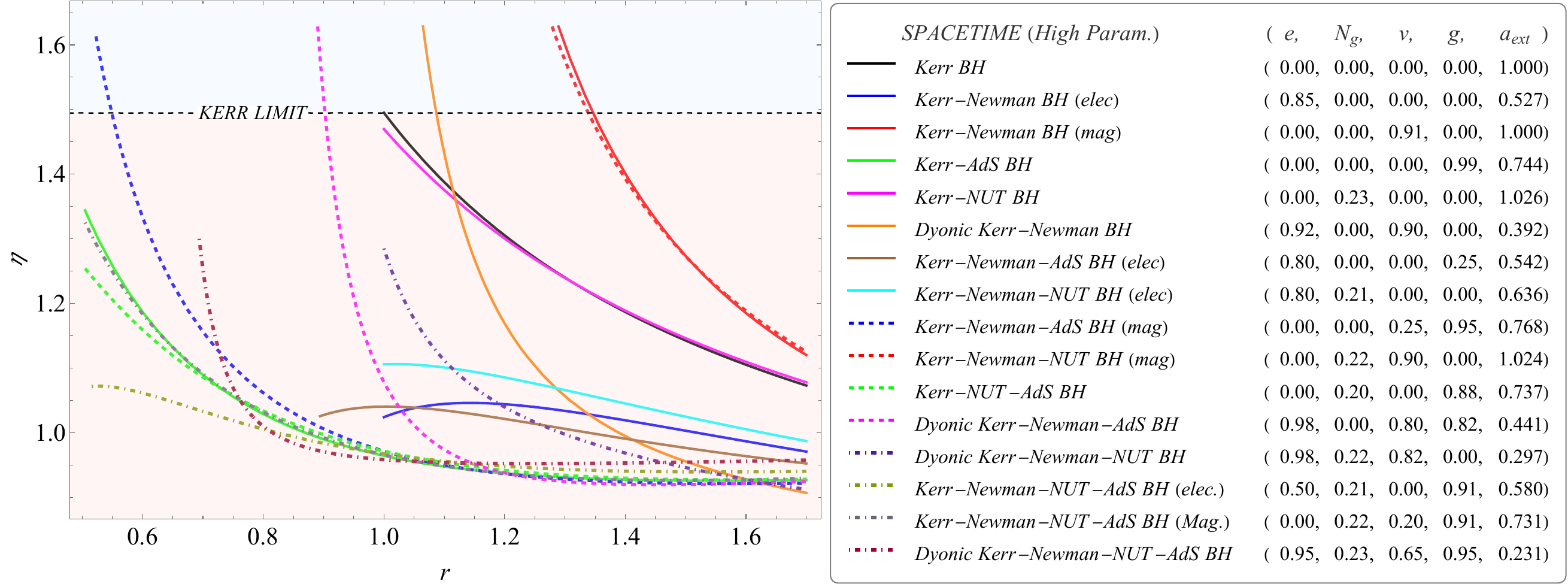}

  \vspace{0.0 cm}
    
    \includegraphics[width=0.83\textwidth, height=5.25cm, keepaspectratio=false]{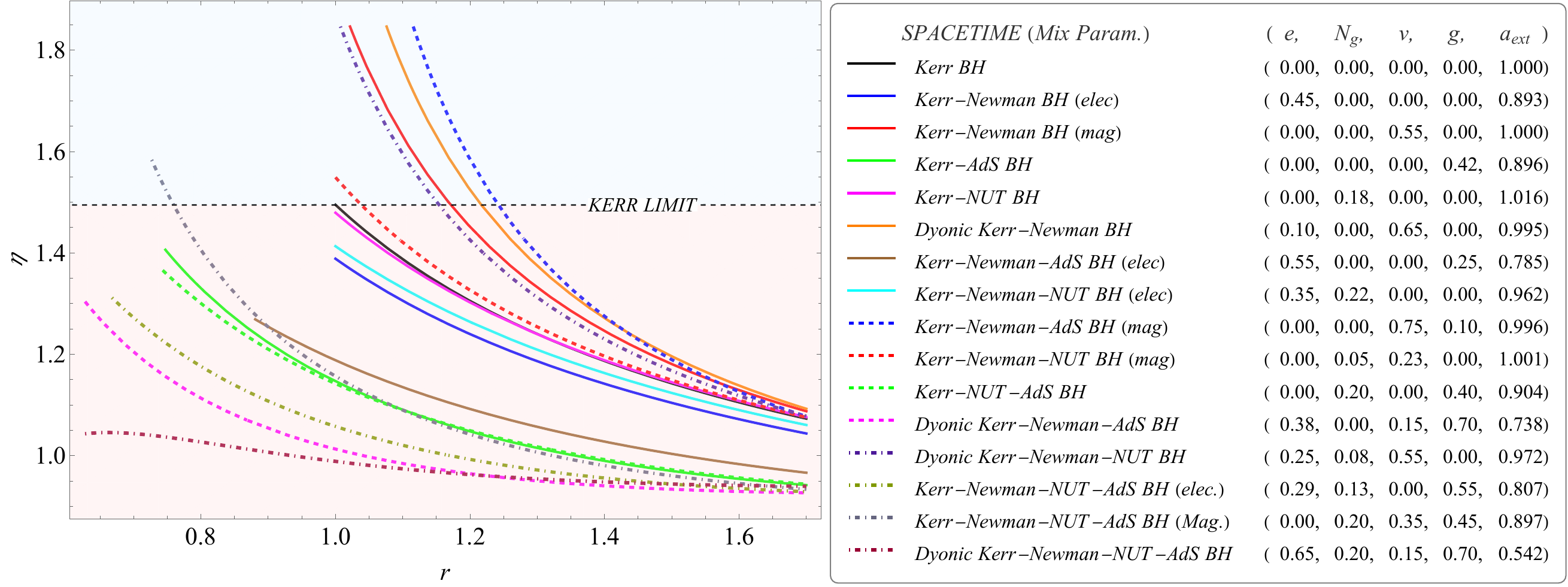}

    \caption{\justifying{ Radial Dependence of efficiency $\eta$; various spacetimes compared with KBH. Horizontal dashed line is the Kerr efficiency limit, dividing two regions(Blue and pink), where the blue region is above the Kerr efficiency limit, and the pink region is below the Kerr efficiency limit, the parameters ($\sigma_0,\xi$) $\equiv$ (100,0) and M=1.}}
{
    \centering
    \begin{tabular*}{0.9\textwidth}{@{\extracolsep{\fill}} ||c|c|c|c|| }
        \hline\hline
        \textbf{Low} & \textbf{Mid} & \textbf{High} & \textbf{Mix} \\
        \hline
        $\,\forall p_i < 0.2 \land N_g < 0.08\,$ & $\,\exists p_i \geq 0.5 \land N_g \in (0.08,0.15)\,$ & $\,\exists p_i > 0.7 \land N_g \in (0.16, 0.23)\,$ & $\,\forall p_i \in (0,1) \land N_g \in (0,0.23)\,$ \\
        \hline\hline
    \end{tabular*}
}
    \label{eff_vs_r_diff_g}
\end{figure*}

\section{Efficiency and Power Analysis}
\label{sec: Efficiency and Power}

As the reconnection occurs, the magnetic energy in the plasma inside the ergoregion of a rapidly rotating BH is redistributed again. The plasma carrying positive energy escapes to infinity, while the remaining decelerated plasma falls into the horizon. As a consequence, higher energy extraction efficiency is achieved. The higher negative energy corresponds to the greater amount of energy extracted from the BH. Therefore, an important factor for assessing the CA process is the efficiency of the plasma energization via magnetic reconnection (see details \cite{Comisso2021, Xu2023, Simpson2023, ComissoAsenjo2018}), given by
\begin{equation}
    \eta= \frac{\epsilon^{\infty}_{+}}{\epsilon^{\infty}_{+}+\epsilon^{\infty}_{-}}\, .
\end{equation}
One can then deduce that the more negative the decelerated plasma energy, the higher the energy extraction efficiency. Energy extraction from the black hole occurs when satisfying $\eta>1$. To analyze the efficiency, we shall, for simplicity, fix the reconnection parameters such that we measure the maximum efficiency, i.e. at $(\sigma_0=100,\, \xi=0)$ which will be kept same throughout the paper, and the spacetime parameter ($N_g,e,v,g$) vary from low $[\,\forall p_i < 0.2 \land N_g < 0.08\,]$, mid $[\,\exists p_i  \sim 0.5 \land N_g \in (0.08,0.15)\,]$, high $[\,\exists p_i > 0.7 \land N_g \in (0.16, 0.23)\,]$, and mixed $[\,\forall p_i \in (0,1) \land N_g \in (0,0.23)\,]$ values of parameter sets and the analyse the radial dependency of efficiency $\eta$ for different spin parameters $a$ in Fig.~\ref{Plot: eff vs r different parameter values}. We can see that different parameter sets produce distinct values of ($a_{\rm{ext}}, r_+$) for each curve. From the first row in Fig.~\ref{Plot: eff vs r different parameter values} to the fourth row, we vary $(e,g,v, Ng)$ in each column while keeping the others fixed. We can then observe the behavior for certain parameter combinations, where the efficiency exceeds the Kerr BH case (i.e., $\eta>1.495$) as the extremality is approached, which is also reflected in Table~\ref{tab:extraction_bounds_combined}. The red line in Fig.~\ref{Plot: eff vs r different parameter values} shows the lowest values of that particular parameter.  In Fig.~\ref{Plot: eff_vs_diff_parameters}, we present the efficiency with each parameter from the first to the fourth row varying $e,g, N_g,v$, respectively. Each point on the plot line is the extremum case where we obtain ($a_{\rm{ext}},r_E$), and we observe that at each point we obtain efficiency for the extremal case, where we obtain efficiency higher than that of the Kerr BH case for most of the cases. From Fig.~\ref{Plot: eff_vs_diff_parameters}, we can obtain a rough estimate of the parameter choices for the efficiency comparison with a Kerr BH. Furthermore, Fig.~\ref{Plot: eff_vs_diff_parameters} clearly shows that satisfying the causality conditions does not necessarily require the parameters to vanish, which is particularly evident in the first and second rows.
 Similarly, we compare efficiency results for various spacetimes with the Kerr BH efficiency, as shown in Fig.~\ref{eff_vs_r_diff_g}. We see that the efficiency $\eta$ for various spacetimes clearly breaches the Kerr limit as they approach the extremal limits over similar parameter ranges.\\
 Another quantity that depends on the negative energy is the power absorbed by the BH at infinity in a unit time. According to energy conservation, a high absorbing power potentially results in a high energy extraction rate by escaping plasmas, the power $\mathcal{P}_{CA}$, which is the $P_{ext}$ per unit enthalpy, and can be written as \cite{Simpson2023, Khodadi2022}
\begin{equation}\label{eqn: Power extracted}
  \mathcal{P_{CA}}= \frac{ P_{extr}}{\mathcal{\omega}}= -\epsilon^{\infty}_{-} A_{in} U_{in}\, ,
\end{equation}
where $U_{in}$ is the constant parameter that depends on the plasma density, which is typically $ \mathcal{O}(10^{-1})$ and $\mathcal{O}(10^{-2})$ for the collisionless and collisional regimes, respectively. \cite{Xu2023, Khodadi2022}, and $A_{in}$ is the cross-sectional area of the inflowing plasma and is given as

\begin{equation}
A_{\mathrm{in}} \sim 
\begin{cases}
\left(r_{\mathrm{SLS}}\right)^2 - r_{\mathrm{photon}}^2, & \text{with photon orbit}, \\[6pt]
\left(r_{\mathrm{SLS}}\right)^2, & \text{without photon orbit}.
\end{cases}
\end{equation}

\begin{table*}[htpb]
\centering
\caption{      \justifying Strong field efficiency ($\eta$) and Extracted Power ($P_{CA}$) for low, mixed, and high spacetime parameter sets. The efficiency ratio ($R_{\eta}$) and power ratio ($R_{P}$) are evaluated against the standard Kerr black hole baseline. Plasma magnetization and orientation angle are taken ($\sigma_0\rightarrow100,\xi\rightarrow0)$ respectively{, whereas mass $M=1$}  \textit{(Note: Negative power indicates energy extraction is suppressed. $\mathcal{P}_{ext}$ and $\eta$ are approximated to four decimal places.)} }
\vspace{-0.05cm}


\textbf{Table A: Low Parameter Values $[\,\forall p_i < 0.2 \land N_g < 0.08\,]$} \\
\vspace{0.05cm}
\resizebox{0.9\textwidth}{!}{
\begin{tabular}{|llccccccc|}
\hline
\textbf{Spacetime} & \textbf{$(e,\quad \,\, v,\,\,\quad g,\,\,\quad N_g)$} & $a_{\it{ext}}$ & $r_+$ & $r_{\it{erg}}$ & $\mathbf{\eta}$ & $\mathcal{P}_{CA}$ & $R_\eta$ & $R_P$ \\
\hline
Kerr BH & $(0.00, 0.00, 0.00, 0.00)$ & $1.0000$ & $1.0000$ & $2.0000$ & $1.4950$ & $0.2349$ & $1.0000$ & $1.0000$ \\
Kerr-Newman BH (elec) & $(0.10, 0.00, 0.00, 0.00)$ & $0.9950$ & $1.0000$ & $1.9950$ & $1.4900$ & $0.2283$ & $0.9967$ & $0.9720$ \\

\rowcolor{lightgray}
\multicolumn{9}{|l|}{Kerr-Newman BH (mag) \qquad \qquad \quad\,$(0.00, 0.20, 0.00, 0.00)$ \,$1.0000$ \,$1.0000$ \,$2.0000$ \,$1.5365$ \,$0.2554$ \,$1.0277$ \,$1.0872$} \\

Kerr-AdS BH & $(0.00, 0.00, 0.10, 0.00)$ & $0.9905$ & $0.9721$ & $1.9114$ & $1.4817$ & $0.1945$ & $0.9911$ & $0.8283$ \\
Kerr-NUT BH & $(0.00, 0.00, 0.00, 0.05)$ & $1.0012$ & $1.0000$ & $2.0012$ & $1.4938$ & $0.2352$ & $0.9992$ & $1.0015$ \\
Dyonic Kerr-Newman BH & $(0.10, 0.20, 0.00, 0.00)$ & $0.9950$ & $1.0000$ & $1.9950$ & $1.5313$ & $0.2489$ & $1.0243$ & $1.0596$ \\
Kerr-Newman-AdS BH (elec) & $(0.10, 0.00, 0.10, 0.00)$ & $0.9855$ & $0.9722$ & $1.9069$ & $1.4767$ & $0.1879$ & $0.9877$ & $0.8002$ \\
Kerr-Newman-NUT BH (elec) & $(0.10, 0.00, 0.00, 0.01)$ & $0.9950$ & $1.0000$ & $1.9950$ & $1.4900$ & $0.2283$ & $0.9967$ & $0.9720$ \\
Kerr-Newman-AdS BH (mag) & $(0.00, 0.20, 0.10, 0.00)$ & $0.9908$ & $0.9728$ & $1.9128$ & $1.5250$ & $0.2169$ & $1.0200$ & $0.9235$ \\

\rowcolor{orange!10}
\multicolumn{9}{|l|}{Kerr-Newman-NUT BH (mag) \quad \quad\,\,\,$(0.00, 0.20, 0.00, 0.03)$ \,$1.0004$ \,$1.0000$ \,$2.0004$ \,$1.5360$ \,$0.2555$ \,$1.0274$ \,$1.0877$} \\

Kerr-NUT-AdS BH & $(0.00, 0.00, 0.10, 0.05)$ & $0.9916$ & $0.9719$ & $1.9122$ & $1.4804$ & $0.1947$ & $0.9902$ & $0.8290$ \\
Dyonic Kerr-Newman-AdS BH & $(0.10, 0.20, 0.10, 0.00)$ & $0.9858$ & $0.9729$ & $1.9083$ & $1.5197$ & $0.2104$ & $1.0165$ & $0.8958$ \\
Dyonic Kerr-Newman-NUT BH & $(0.10, 0.20, 0.00, 0.07)$ & $0.9974$ & $1.0000$ & $1.9974$ & $1.5286$ & $0.2495$ & $1.0225$ & $1.0624$ \\
Kerr-Newman-NUT-AdS BH (elec.) & $(0.10, 0.00, 0.10, 0.08)$ & $0.9884$ & $0.9718$ & $1.9091$ & $1.4733$ & $0.1883$ & $0.9855$ & $0.8018$ \\
Kerr-Newman-NUT-AdS BH (Mag.) & $(0.00, 0.20, 0.10, 0.06)$ & $0.9925$ & $0.9726$ & $1.9140$ & $1.5228$ & $0.2171$ & $1.0186$ & $0.9242$ \\
General Dyonic BH & $(0.10, 0.20, 0.10, 0.04)$ & $0.9866$ & $0.9728$ & $1.9088$ & $1.5188$ & $0.2105$ & $1.0159$ & $0.8962$ \\
\hline
\end{tabular}
} 

\vspace{0.2 cm}
\textbf{Table B: Mix Parameter Values $[\,\forall p_i \in (0,1) \land N_g \in (0,0.23)\,]$} \\
\vspace{0.05cm}
\resizebox{0.9\textwidth}{!}{
\begin{tabular}{|llccccccc|}
\hline
\textbf{Spacetime} & \textbf{$(e,\quad \,\, v,\,\,\quad g,\,\,\quad N_g)$} & $a_{\it{ext}}$ & $r_+$ & $r_{\it{erg}}$ & $\mathbf{\eta}$ & $\mathcal{P}_{CA}$ & $R_\eta$ & $R_P$ \\
\hline
Kerr BH & $(0.00, 0.00, 0.00, 0.00)$ & $1.0000$ & $1.0000$ & $2.0000$ & $1.4950$ & $0.2349$ & $1.0000$ & $1.0000$ \\
Kerr-Newman BH (elec) & $(0.45, 0.00, 0.00, 0.00)$ & $0.8930$ & $1.0000$ & $1.8930$ & $1.3886$ & $0.0838$ & $0.9288$ & $0.3569$ \\
Kerr-Newman BH (mag) & $(0.00, 0.55, 0.00, 0.00)$ & $1.0000$ & $1.0000$ & $2.0000$ & $1.9265$ & $0.3947$ & $1.2886$ & $1.6804$ \\
Kerr-AdS BH & $(0.00, 0.00, 0.42, 0.00)$ & $0.8956$ & $0.7471$ & $1.3618$ & $1.4062$ & $-0.0812$ & $0.9406$ & $-0.3457$ \\
Kerr-NUT BH & $(0.00, 0.00, 0.00, 0.18)$ & $1.0161$ & $1.0000$ & $2.0161$ & $1.4793$ & $0.2394$ & $0.9895$ & $1.0192$ \\
Dyonic Kerr-Newman BH & $(0.10, 0.65, 0.00, 0.00)$ & $0.9950$ & $1.0000$ & $1.9950$ & $2.2143$ & $0.4592$ & $1.4811$ & $1.9552$ \\
Kerr-Newman-AdS BH (elec) & $(0.55, 0.00, 0.25, 0.00)$ & $0.7847$ & $0.8807$ & $1.5222$ & $1.2690$ & $-0.2029$ & $0.8488$ & $-0.8639$ \\
Kerr-Newman-NUT BH (elec) & $(0.35, 0.00, 0.00, 0.22)$ & $0.9622$ & $1.0000$ & $1.9622$ & $1.4132$ & $0.1584$ & $0.9453$ & $0.6744$ \\

\rowcolor{lightgray}
\multicolumn{9}{|l|}{Kerr-Newman-AdS BH (mag) \qquad \quad $(0.00, 0.75, 0.10, 0.00)$ \,$0.9958$ \,$0.9823$ \,$1.9306$ \,\,$2.8526$ \,\,$0.5418$ \,\,$1.9081$ \,\,\,$2.3069$} \\

Kerr-Newman-NUT BH (mag) & $(0.00, 0.23, 0.00, 0.05)$ & $1.0012$ & $1.0000$ & $2.0013$ & $1.5491$ & $0.2623$ & $1.0362$ & $1.1167$ \\
Kerr-NUT-AdS BH & $(0.00, 0.00, 0.40, 0.20)$ & $0.9038$ & $0.7431$ & $1.3725$ & $1.3658$ & $-0.0949$ & $0.9135$ & $-0.4042$ \\
Dyonic Kerr-Newman-AdS BH & $(0.38, 0.15, 0.70, 0.00)$ & $0.7380$ & $0.6177$ & $1.0457$ & $1.3211$ & $-0.3682$ & $0.8837$ & $-1.5675$ \\
Dyonic Kerr-Newman-NUT BH & $(0.25, 0.55, 0.00, 0.08)$ & $0.9715$ & $1.0000$ & $1.9715$ & $1.8734$ & $0.3567$ & $1.2531$ & $1.5185$ \\
Kerr-Newman-NUT-AdS BH (elec.) & $(0.29, 0.00, 0.55, 0.13)$ & $0.8069$ & $0.6678$ & $1.1756$ & $1.3116$ & $-0.2705$ & $0.8773$ & $-1.1516$ \\
Kerr-Newman-NUT-AdS BH (Mag.) & $(0.00, 0.35, 0.45, 0.20)$ & $0.8971$ & $0.7268$ & $1.3275$ & $1.5837$ & $0.0003$ & $1.0593$ & $0.0012$ \\
General Dyonic BH & $(0.65, 0.15, 0.70, 0.20)$ & $0.5420$ & $0.6191$ & $0.9424$ & $1.0415$ & $-1.0827$ & $0.6966$ & $-4.6098$ \\
\hline
\end{tabular}
}

\vspace{0.2 cm}
\textbf{Table C: High Parameter Values $[\,\exists p_i > 0.7 \land N_g \in (0.16, 0.23)\,]$} \\
\vspace{0.05cm}
\resizebox{0.9\textwidth}{!}{
\begin{tabular}{|llccccccc|}
\hline
\textbf{Spacetime} & \textbf{$(e,\quad \,\, v,\,\,\quad g,\,\,\quad N_g)$} & $a_{\it{ext}}$ & $r_+$ & $r_{\it{erg}}$ & $\mathbf{\eta}$ & $\mathcal{P}_{CA}$ & $R_\eta$ & $R_P$ \\
\hline
Kerr BH & $(0.00, 0.00, 0.00, 0.00)$ & $1.0000$ & $1.0000$ & $2.0000$ & $1.4950$ & $0.2349$ & $1.0000$ & $1.0000$ \\
Kerr-Newman BH (elec) & $(0.85, 0.00, 0.00, 0.00)$ & $0.5268$ & $1.0000$ & $1.5268$ & $1.0242$ & $-0.9948$ & $0.6851$ & $-4.2350$ \\
\rowcolor{lightgray}
\multicolumn{9}{|l|}{Kerr-Newman BH (mag) \qquad \qquad \quad $(0.00, 0.91, 0.00, 0.00)$ \,$1.0000$ \,$1.0000$ \,$2.0000$ \,$6.2873$ \,\,\,$0.8205$ \,\,$4.2056$ \,\,\,\,$3.4930$} \\
Kerr-AdS BH & $(0.00, 0.00, 0.99, 0.00)$ & $0.7438$ & $0.4946$ & $0.8734$ & $1.3667$ & $-0.3976$ & $0.9142$ & $-1.6926$ \\
Kerr-NUT BH & $(0.00, 0.00, 0.00, 0.23)$ & $1.0261$ & $1.0000$ & $2.0261$ & $1.4697$ & $0.2421$ & $0.9831$ & $1.0307$ \\
Dyonic Kerr-Newman BH & $(0.92, 0.90, 0.00, 0.00)$ & $0.3919$ & $1.0000$ & $1.3919$ & $2.5522$ & $-0.1851$ & $1.7072$ & $-0.7880$ \\
Kerr-Newman-AdS BH (elec) & $(0.80, 0.00, 0.25, 0.00)$ & $0.5424$ & $0.8942$ & $1.3474$ & $1.0257$ & $-1.0375$ & $0.6861$ & $-4.4168$ \\
Kerr-Newman-NUT BH (elec) & $(0.80, 0.00, 0.00, 0.21)$ & $0.6357$ & $1.0000$ & $1.6357$ & $1.1058$ & $-0.4994$ & $0.7397$ & $-2.1260$ \\
Kerr-Newman-AdS BH (mag) & $(0.00, 0.25, 0.95, 0.00)$ & $0.7682$ & $0.5227$ & $0.9176$ & $1.6134$ & $-0.2344$ & $1.0792$ & $-0.9979$ \\
Kerr-Newman-NUT BH (mag) & $(0.00, 0.90, 0.00, 0.22)$ & $1.0239$ & $1.0000$ & $2.0239$ & $4.7730$ & $0.7601$ & $3.1926$ & $3.2358$ \\
Kerr-NUT-AdS BH & $(0.00, 0.00, 0.88, 0.20)$ & $0.7370$ & $0.5011$ & $0.8948$ & $1.2580$ & $-0.4730$ & $0.8415$ & $-2.0136$ \\
Dyonic Kerr-Newman-AdS BH & $(0.98, 0.80, 0.82, 0.00)$ & $0.4410$ & $0.8323$ & $1.0640$ & $4.4602$ & $-0.1320$ & $2.9834$ & $-0.5620$ \\
Dyonic Kerr-Newman-NUT BH & $(0.98, 0.82, 0.00, 0.22)$ & $0.2966$ & $1.0000$ & $1.2967$ & $1.2850$ & $-0.8146$ & $0.8595$ & $-3.4679$ \\
Kerr-Newman-NUT-AdS BH (elec.) & $(0.50, 0.00, 0.91, 0.21)$ & $0.5801$ & $0.5159$ & $0.8288$ & $1.0716$ & $-0.9576$ & $0.7168$ & $-4.0766$ \\
Kerr-Newman-NUT-AdS BH (Mag.) & $(0.00, 0.20, 0.91, 0.22)$ & $0.7309$ & $0.4946$ & $0.8805$ & $1.3447$ & $-0.4124$ & $0.8995$ & $-1.7556$ \\
General Dyonic BH & $(0.95, 0.65, 0.95, 0.23)$ & $0.2312$ & $0.6939$ & $0.8174$ & $1.3000$ & $-1.2156$ & $0.8696$ & $-5.1750$ \\
\hline
\end{tabular}
}

\label{tab:extraction_bounds_combined}
\end{table*}

\begin{figure*}[htbp]
\centering
{\includegraphics[width=0.325\textwidth]{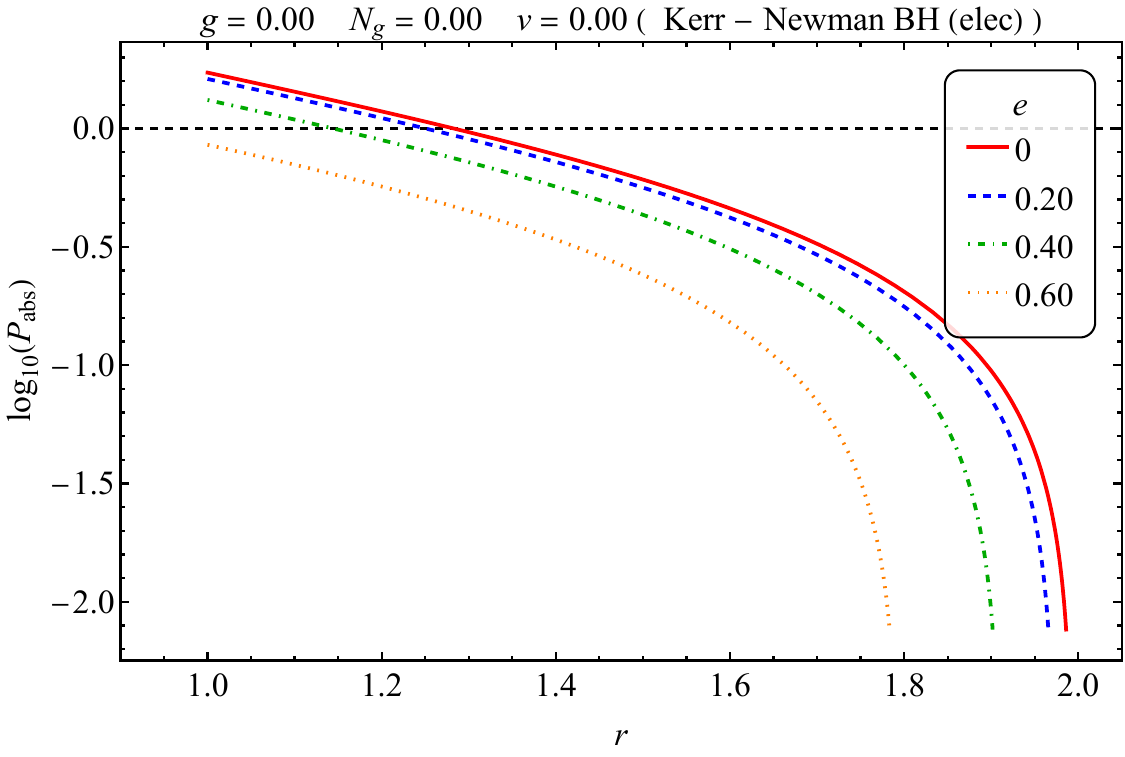}}
{\includegraphics[width=0.325\textwidth]{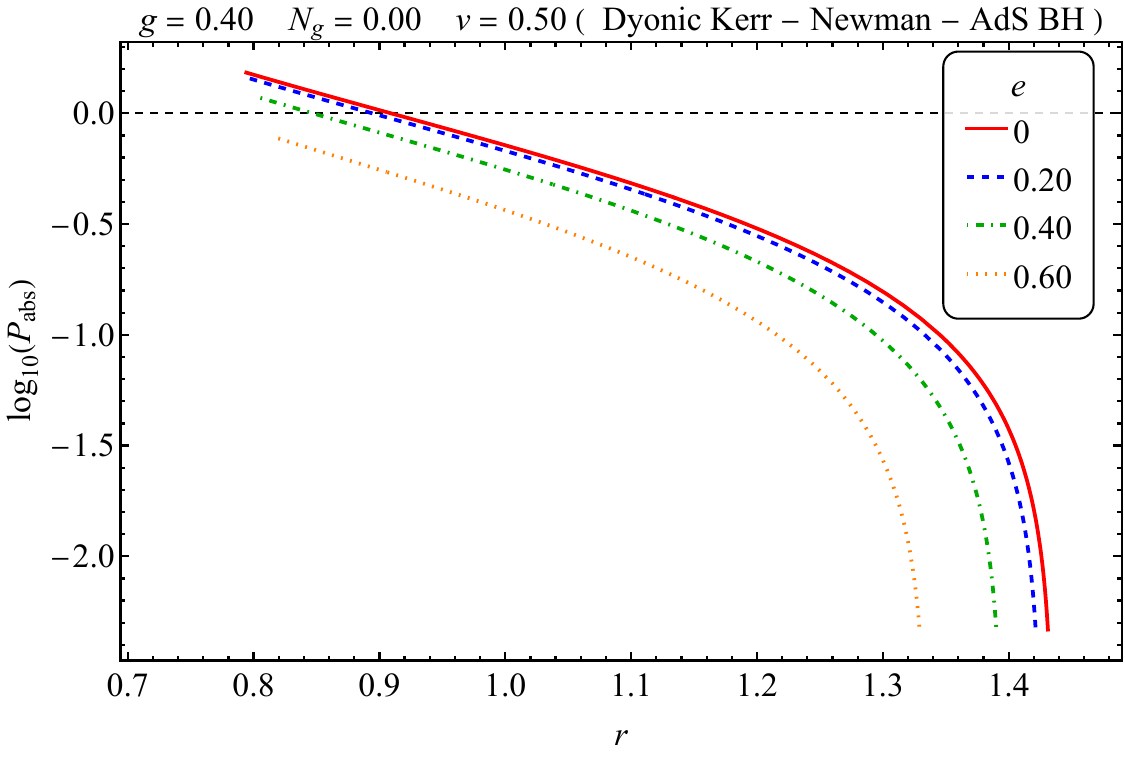}}
{\includegraphics[width=0.325\textwidth]{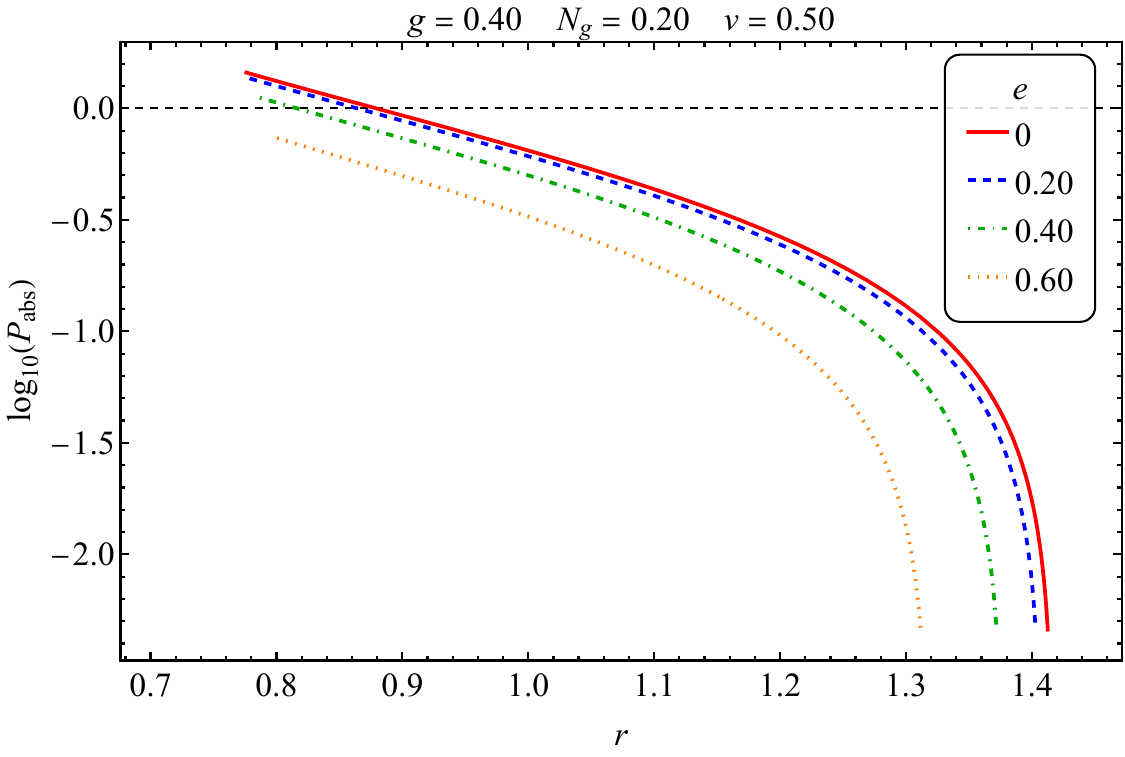}}
\vspace{0.2 cm}

{\includegraphics[width=0.325\textwidth]{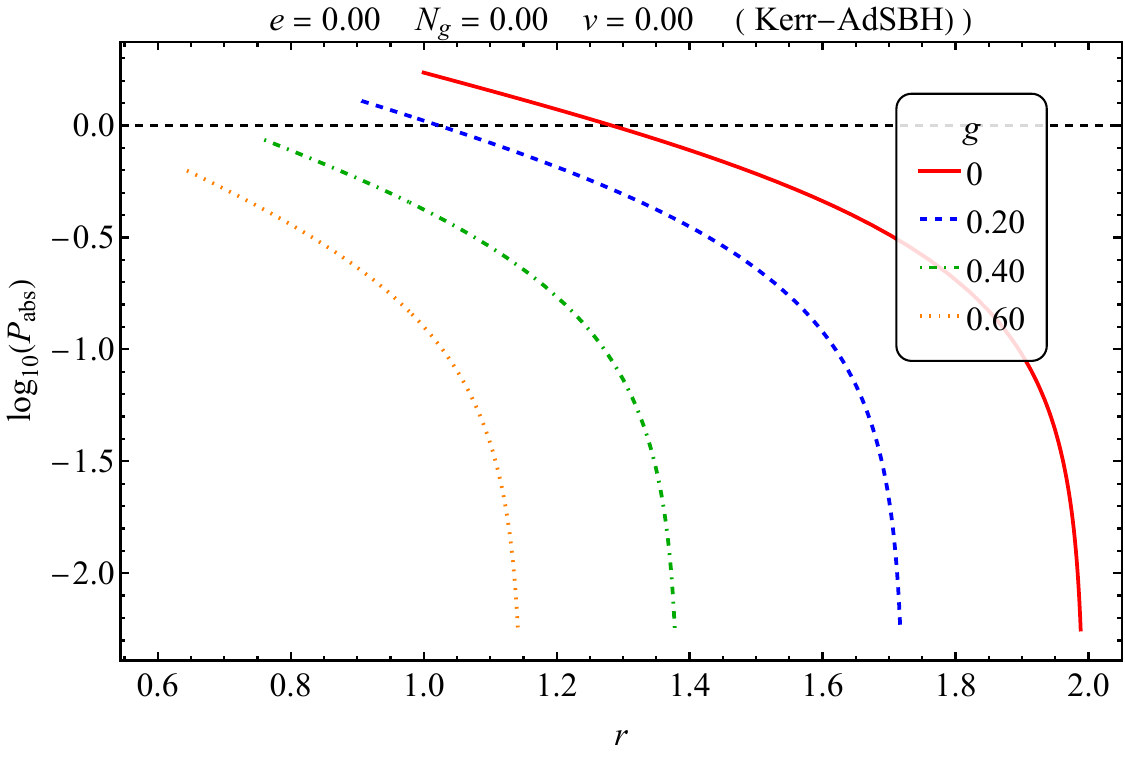}}
{\includegraphics[width=0.325\textwidth]{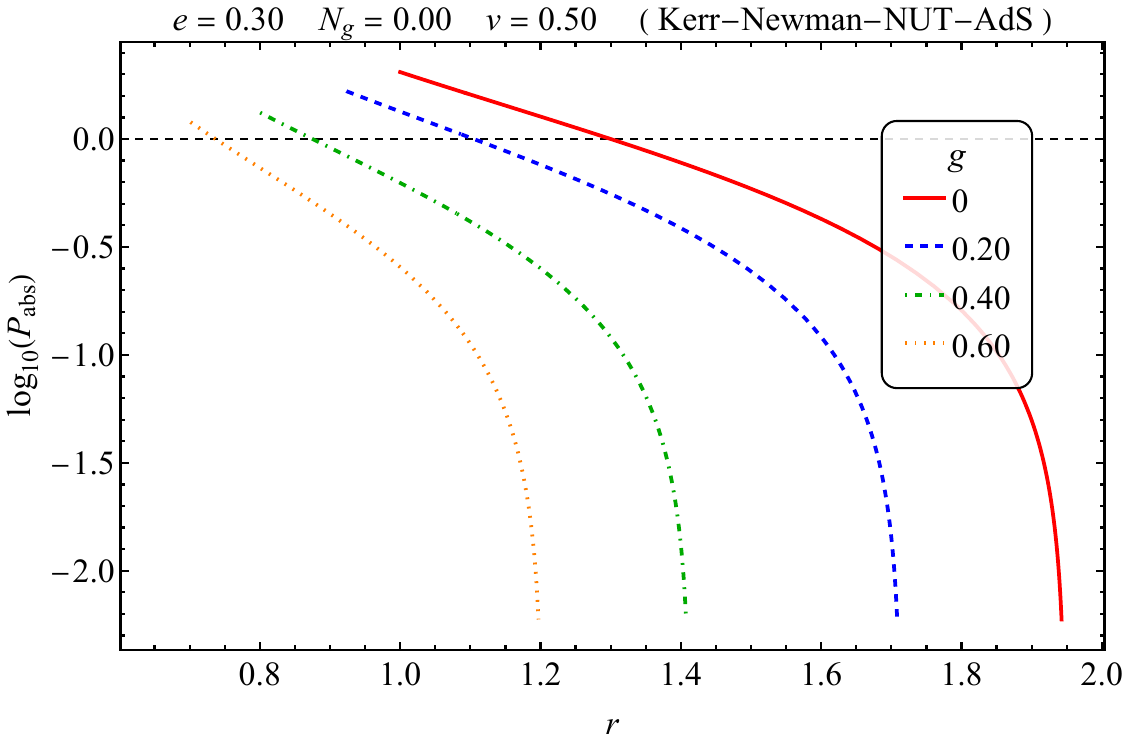}}
{\includegraphics[width=0.325\textwidth]{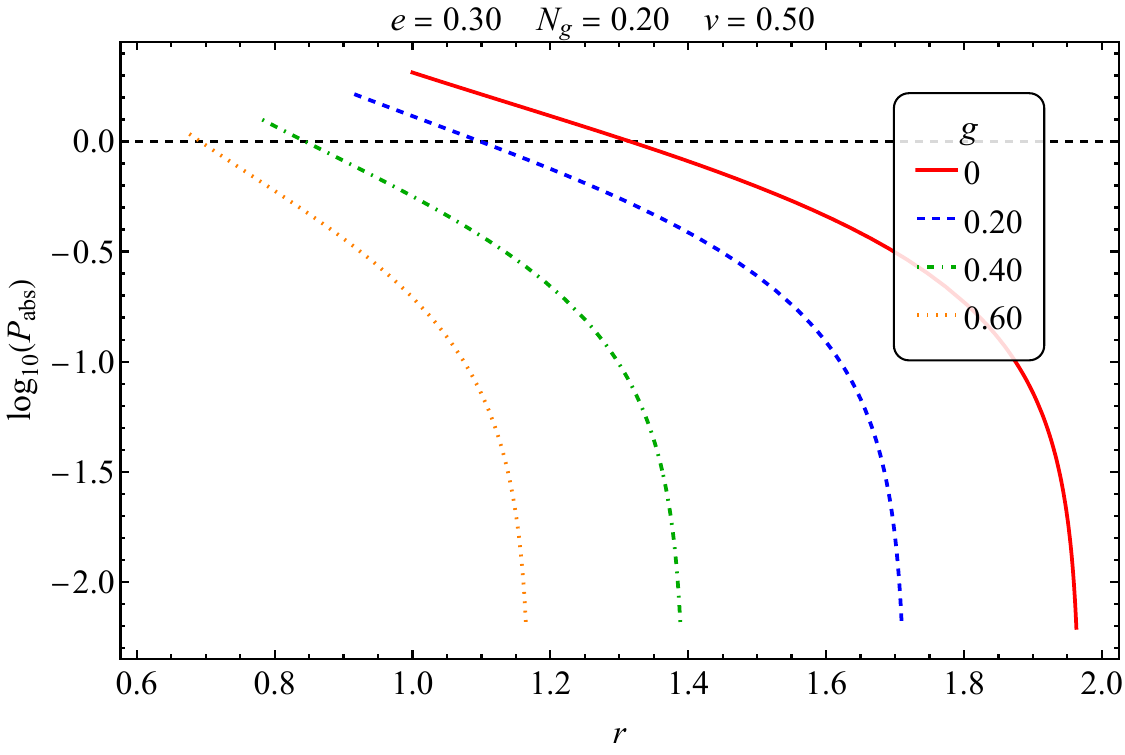}}

\vspace{0.2 cm}

{\includegraphics[width=0.325\textwidth]{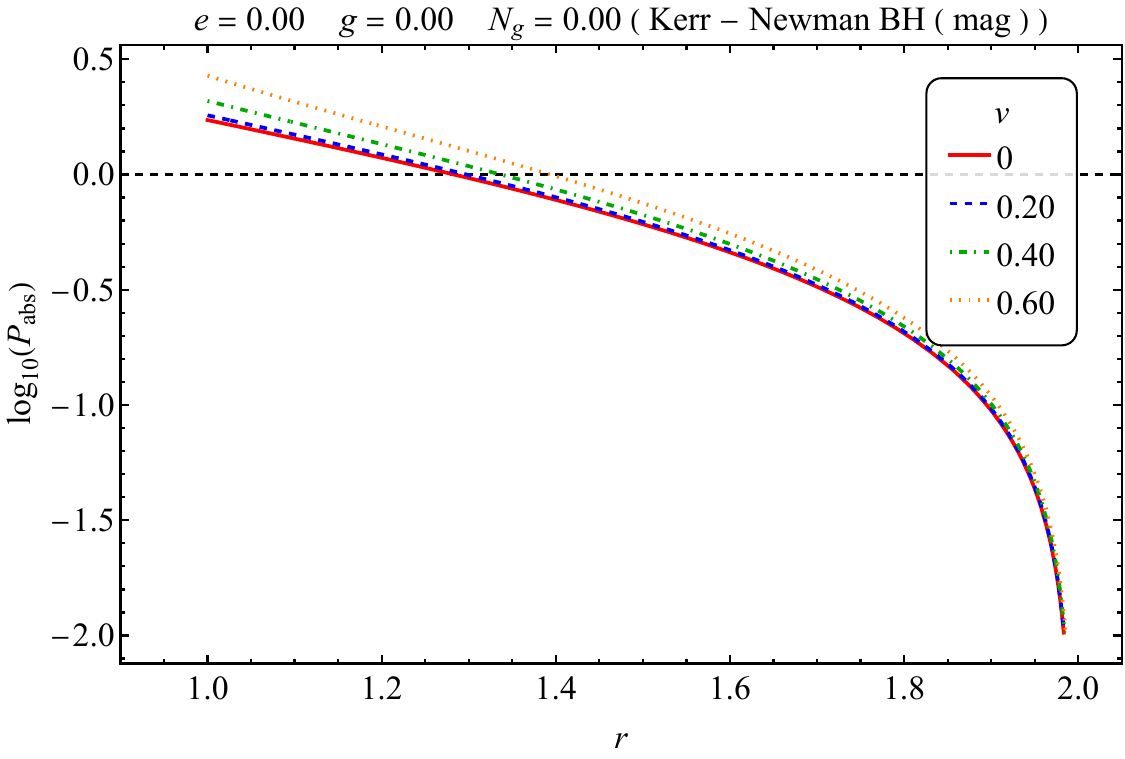}}
{\includegraphics[width=0.325\textwidth]{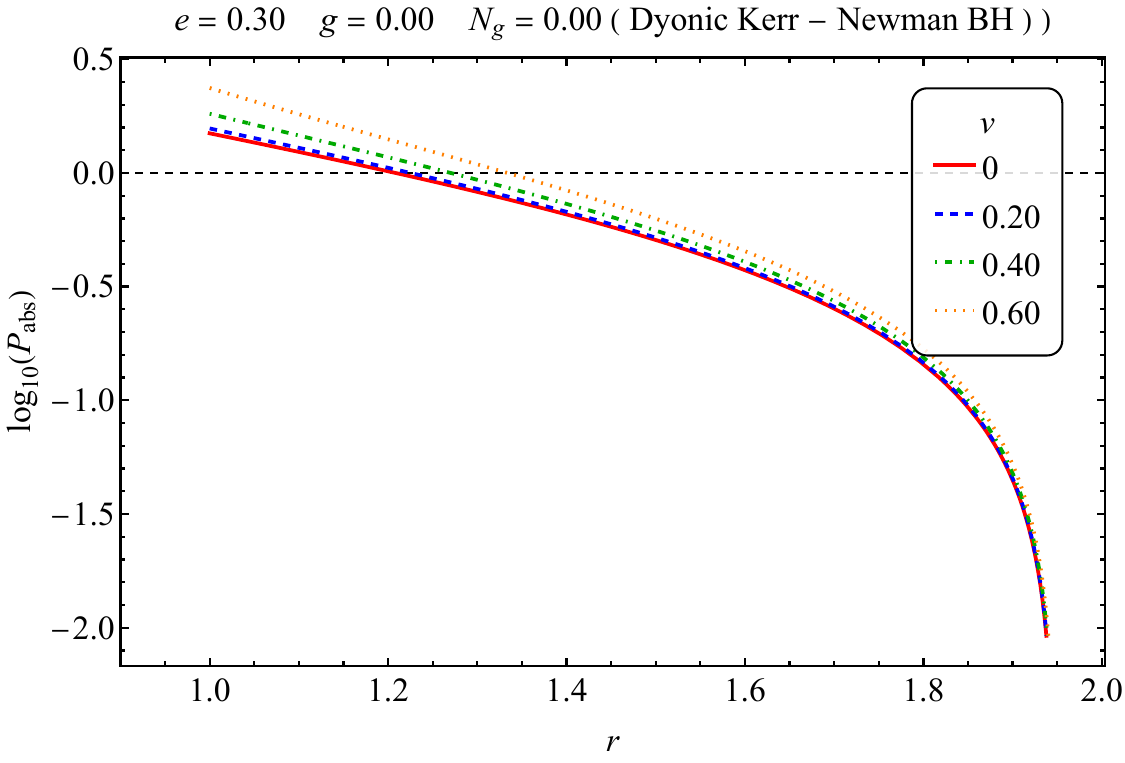}}
{\includegraphics[width=0.325\textwidth]{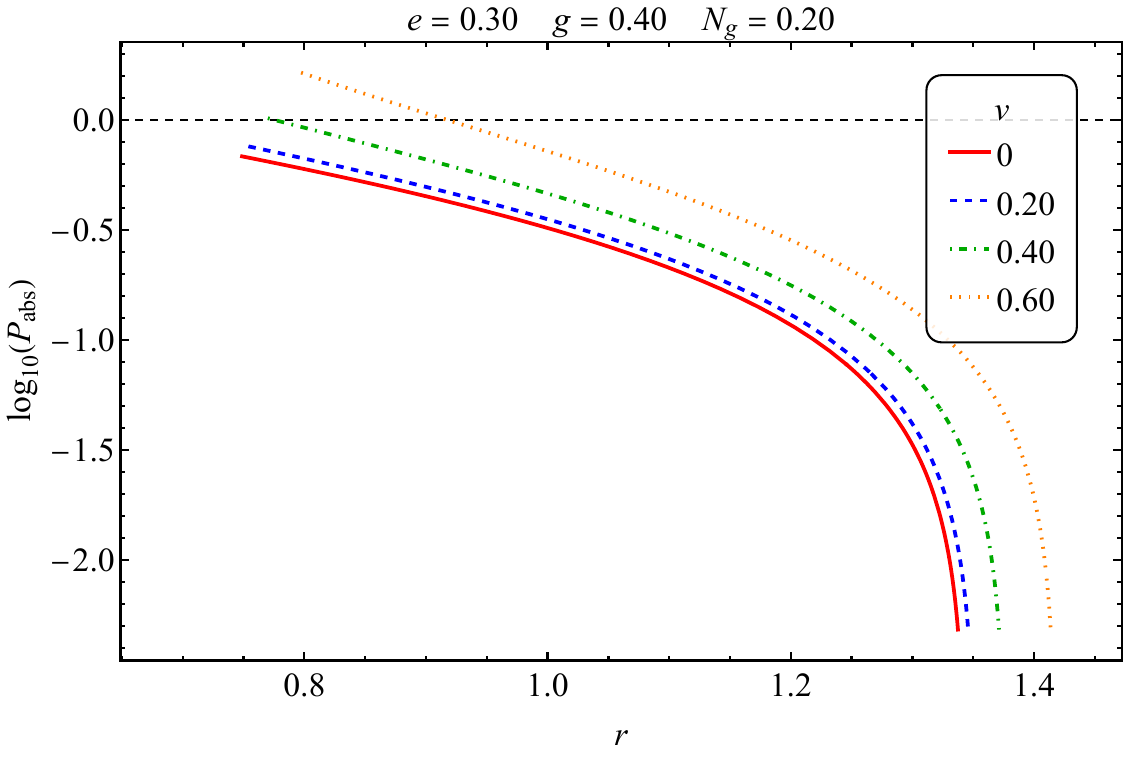}}

\vspace{0.2 cm}

{\includegraphics[width=0.325\textwidth]{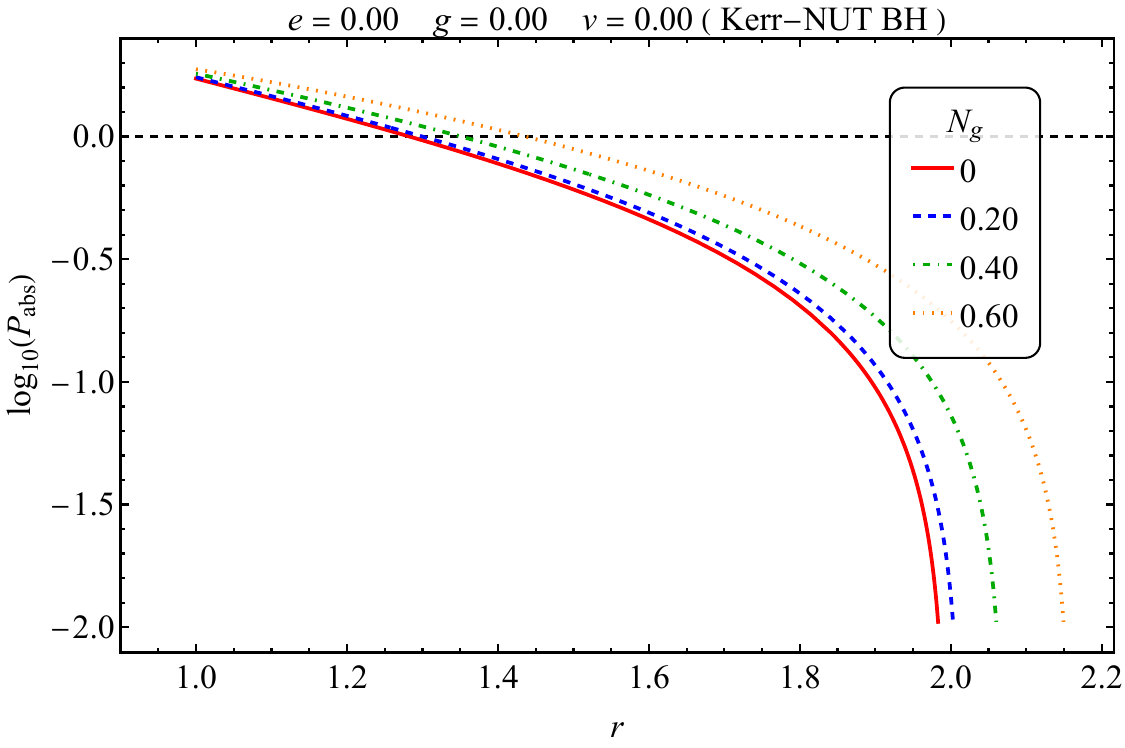}}
{\includegraphics[width=0.325\textwidth]{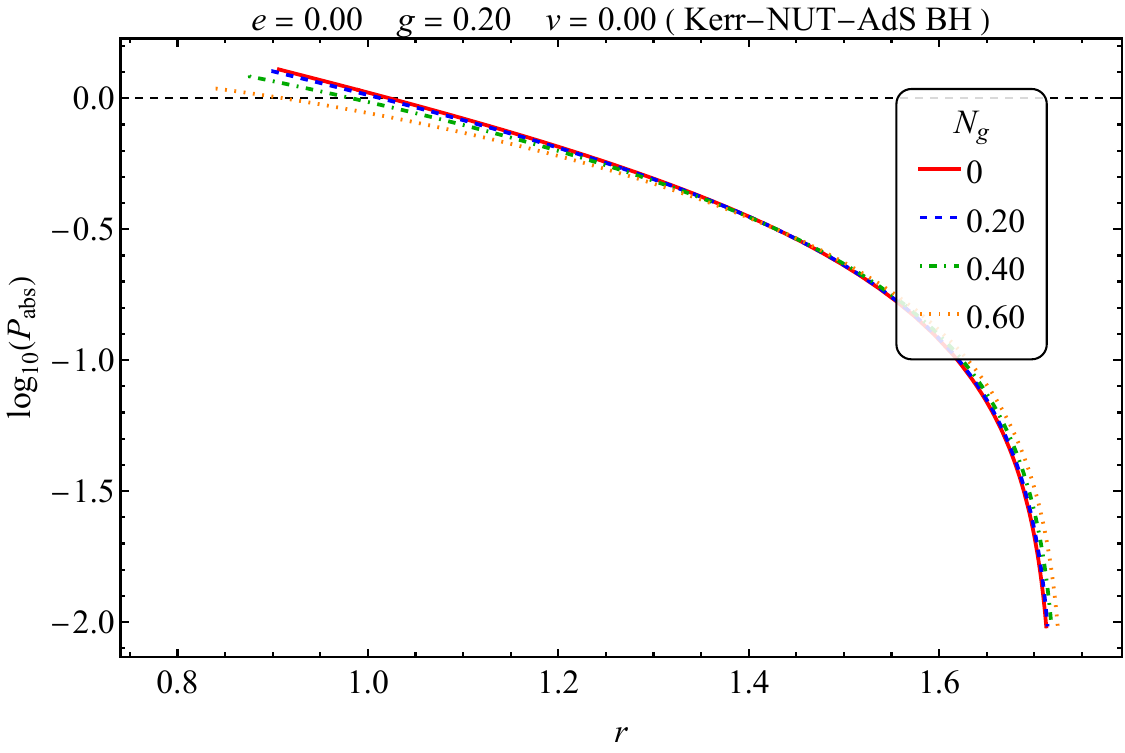}}
{\includegraphics[width=0.325\textwidth]{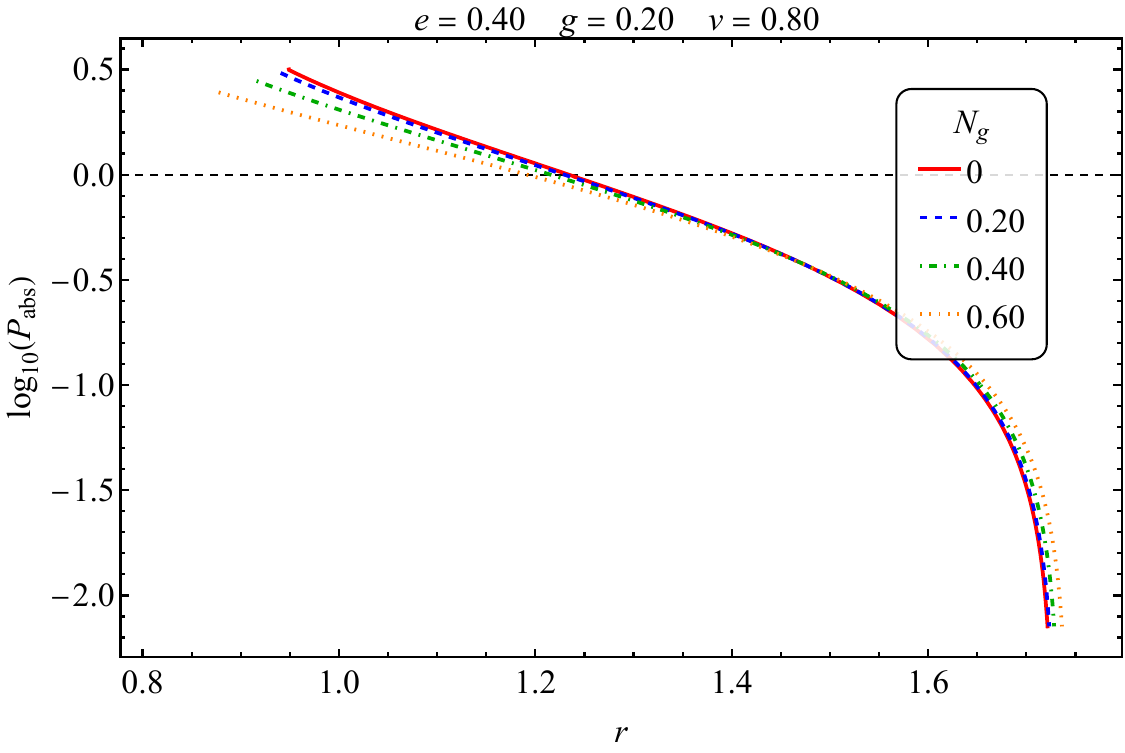}}

\caption{      \justifying
{Radial dependence of Extracted Power $log_{10}[\mathcal{P}_{CA}]$; where each parameter in the set $(e, g, N_g, v)$ varied individually keeping other parameters constant. $\mathcal{P}_{CA}\equiv\mathcal{P}_{CA}(r_{ergo},\sigma_0,\xi)$, depends on $r_{ergo}$ if parameters $(\sigma_0,\xi)$ $\equiv$ (100, 0) respectively, whereas $M=1$. $\mathcal{P}_{CA}$  attains a maximum at each ($r_{ergo}$, $a_{ext}$), solid lines denoting the lowest value of the varying parameter in the plot.}}
\label{plot: power vs r for diff a vals}
\end{figure*}

Since we prioritize extreme cases, our regime is collisionless ($\mathcal{O}(10^{-1})$), and the photon orbits all lie near the horizon ($r_{ph}\sim r_+$). So, our extracted power becomes
\begin{widetext}
\begin{equation}
  \begin{aligned}
&\mathcal{P}_{\mathcal{CA}} \sim -\mathcal{O}(10^{-1}) \times (r_{SLS}^2-r^2_+)  \times \left\{ \frac{Z^2 R_g \Theta_g \Xi^2}{X \big(-X \Omega_K ^2 + 2 A \Omega_K  R_g \Xi - 2 a \mathcal{B} \Omega_K  \Theta_g \Xi + (-R_g + a^2 \Theta_g) \Xi^2\big)} \right\}^{\frac{1}{2}} \\
& \times \sqrt{\frac{-Z R_g \Theta_g}{X}} \Bigg[ - \frac{\Omega_K  \sqrt{\frac{X}{-Z \Xi^2}} \sqrt{\sigma_0}}{\sqrt{\frac{-Z R_g \Theta_g}{X}}}  \, + \frac{X (A \Omega_K  R_g - a \mathcal{B} \Omega_K  \Theta_g - R_g \Xi + a^2 \Theta_g \Xi) \sqrt{1 + \sigma_0}}{Z^2 R_g \Theta_g \Xi} \\
& \ - \frac{1}{4} \Big( Z^2 R_g \Theta_g \Xi^2 - (A^2 \Omega_K  R_g - A R_g \Xi + \mathcal{B} \Theta_g (-\mathcal{B} \Omega_K  + a \Xi))^2 \Big) \\
& \times \left( X \sqrt{\frac{-Z R_g \Theta_g}{X}} \right)^{-1} \Big( Z^2 R_g \Theta_g \Xi^2 + Z^2 R_g \Theta_g \Xi^2 \sigma_0 - (A^2 \Omega_K  R_g - A R_g \Xi + \mathcal{B} \Theta_g (-\mathcal{B} \Omega_K  + a \Xi))^2 \sigma_0 \Big)^{-1} \\
&\times \Bigg( \sqrt{\frac{X}{-Z \Xi^2}} (A^2 \Omega_K  R_g - A R_g \Xi + \mathcal{B} \Theta_g (-\mathcal{B} \Omega_K  + a \Xi)) \sqrt{\sigma_0}  - X \sqrt{\frac{-Z R_g \Theta_g}{X}} \sqrt{1 + \sigma_0} \Bigg) \Bigg]\, ,
\end{aligned}
\end{equation}
\end{widetext}
\noindent where the terms $X,Z$ are  $X = A^2 R_g - \mathcal{B}^2 \Theta_g$ and $Z = \mathcal{B} - a A$. We vary the extracted power with respect to $r$ for parameter set ($N_g,g,v,e$) and plot it in a logarithmic scale $Log_{10}(\mathcal{P_{CA}})$  for various parameter combinations in Fig.~\ref{plot: power vs r for diff a vals}.  As we can see the extracted $Log[\mathcal{P}_{CA}]\rightarrow0$ as the negative energy ($\epsilon_{-}^{\infty}\rightarrow0$), and reaches to negative infinity as $r\rightarrow r_{erg}$ and the solid red line shows lowest parameter values in that plot in Fig.~\ref{plot: power vs r for diff a vals}. A significant result here is that both the efficiency and power reach their maximum result for the extremal Kerr BH as $(N_g,e,v,g)\rightarrow0$.

To investigate whether the spacetime considered here offers an advantage over the Kerr BH case, we define the following efficiency and power ratios 
\begin{equation}\label{Ratio formula}
R_{\eta} = \lim_{a,\, r \to M} \frac{\eta}{\eta_{\, \kappa_{\mathrm{err}}}}
\;, \qquad
R_{p} = \lim_{a,\, r \to M} \frac{P_{\mathrm{extr}}}{P_{\mathrm{Kerr}}}
\, ,
\end{equation}
where the denominators denote the efficiency and power of extremal Kerr black holes, respectively. If they are above one, the Kerr-Newman-NUT-AdS BH would be more efficient than the Kerr black hole to extract energy via the magnetic reconnection mechanism. Otherwise, the Kerr BH will dominate.
\begin{figure*}[htbp]
    \centering

    \includegraphics[width=0.82\textwidth, height=5.25cm, keepaspectratio=false]{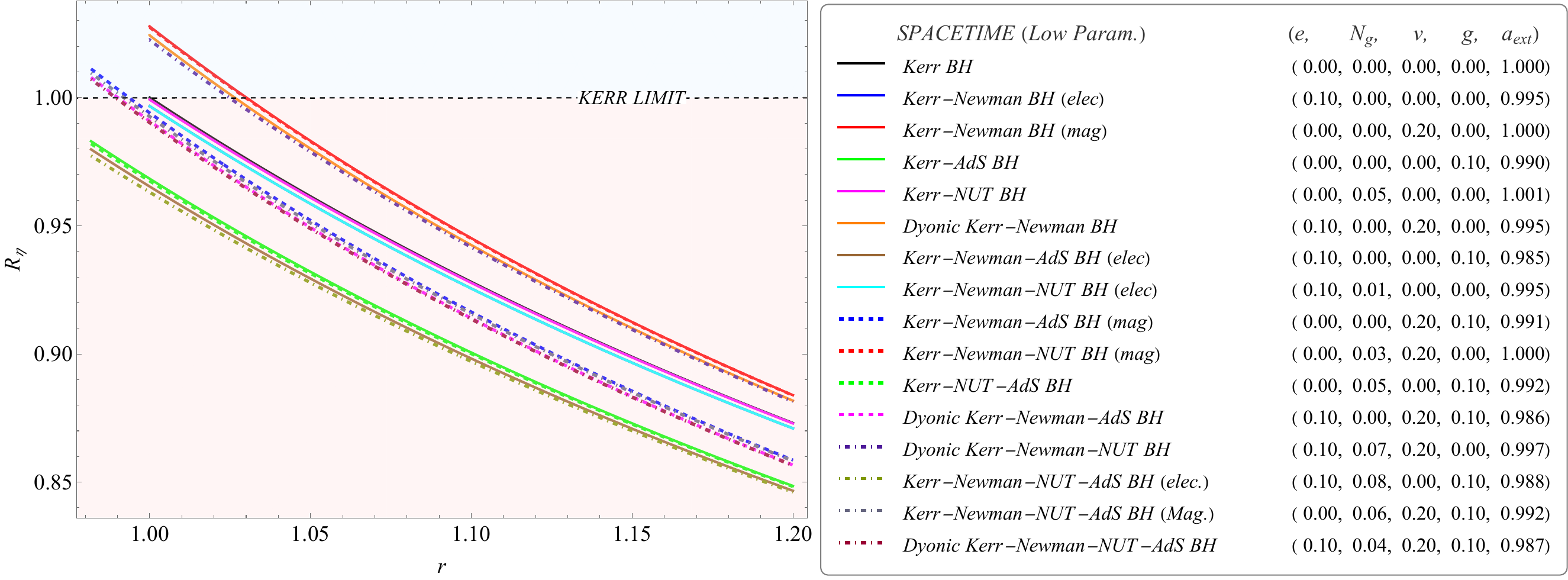}
    
    \vspace{0cm} 

    \includegraphics[width=0.82\textwidth, height=5.25cm, keepaspectratio=false]{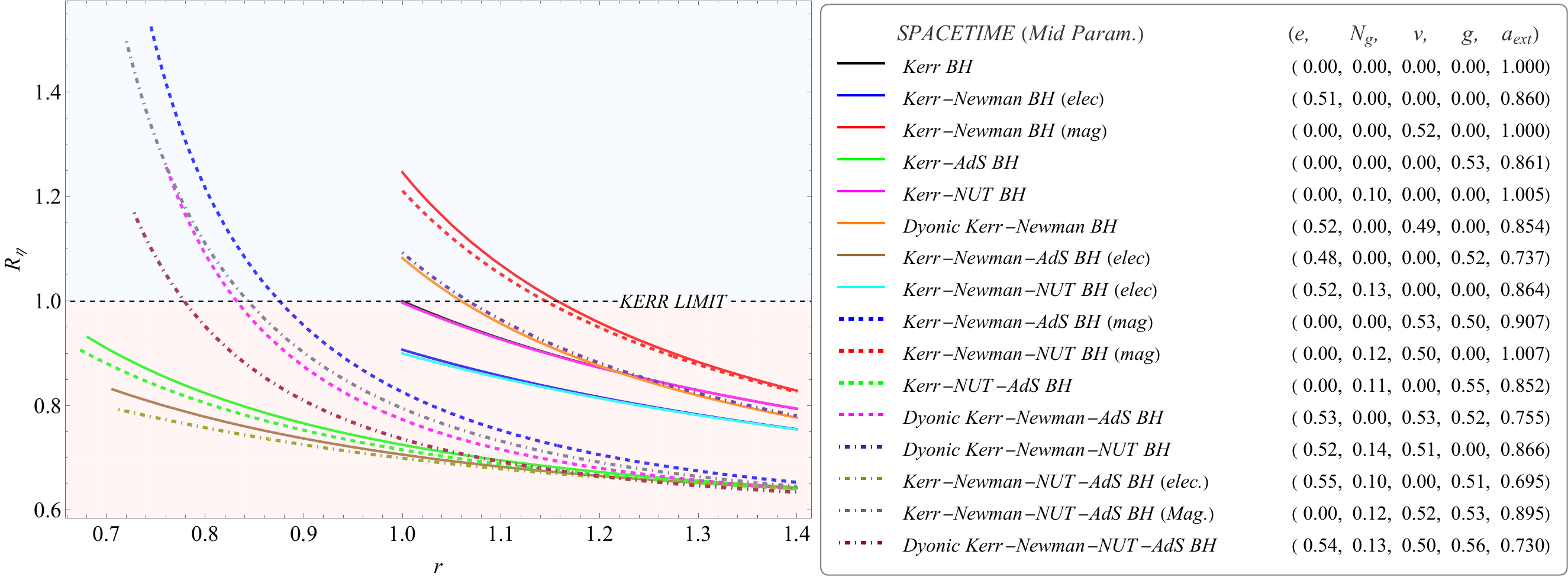}
    
    \vspace{0cm}

    \includegraphics[width=0.82\textwidth, height=5.25cm, keepaspectratio=false]{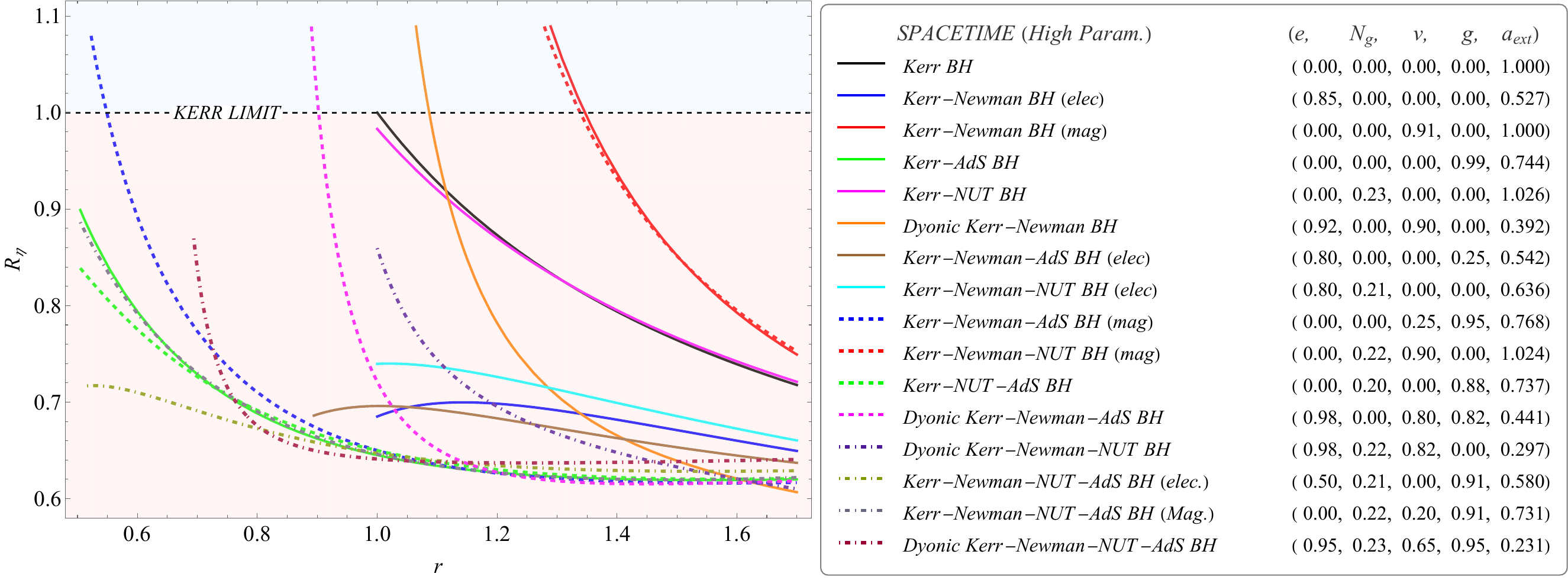}

     \vspace{0.0cm}

  \vspace{0.0 cm}
    
    \includegraphics[width=0.82\textwidth, height=5.25cm, keepaspectratio=false]{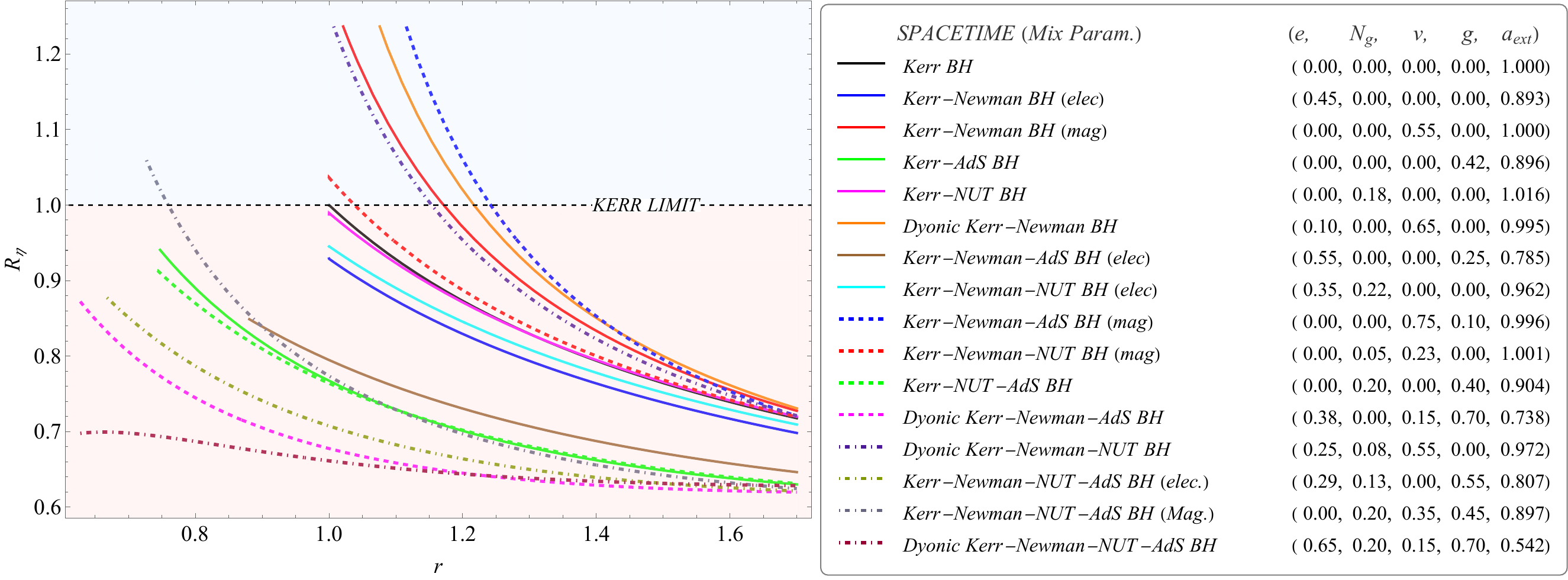}

    \caption{      \justifying { Radial Dependence of efficiency ratio ${R}_{\eta}$; Different spacetimes plotted and compared with KBH. The horizontal dashed line represents the Kerr efficiency ratio limit, dividing the two regions (blue and pink): the blue region is above the limit, and the pink region is below it. Whereas, $M=1$, $(\sigma_0,\xi)$ $\equiv$ (100, 0) and the parameter range are given in Fig.~\ref{eff_vs_r_diff_g}}}
    \label{Plot: eff_ratio_vs_r_diff_ST}.
\end{figure*}

The efficiency and power ratios are plotted as a function of $r$ in Figs.~(\ref{Plot: eff_ratio_vs_r_diff_ST}) and (\ref{Plot: power_ratio_vs_r_diff_parameters}) respectively, where the efficiency ratio is calculated using (\ref{Ratio formula}). The ratio $R_{\eta}$ of the efficiency exhibits a monotonously decreasing behavior for different parameter sets from low, middle, high, and mix from first to the fourth row, showing that for smaller $r$ values, i.e., near the horizon, Kerr BH is dominated by different spacetime combinations. As can be seen in both Figs.~\ref{Plot: eff_ratio_vs_r_diff_ST} and \ref{Plot: power_ratio_vs_r_diff_parameters}, the red region is the region where the efficiency and power are both dominated by the Kerr BH extremal case, whereas in the blue region, the Kerr BH is dominated by various spacetimes.

 To analyze and correlate the effect that independent spacetime parameters have on the geometric parameters ($r_+, r_{erg}, a_{ext}$) and efficiency parameters ($ \epsilon_{\pm},\eta,\mathcal{P}_{CA}$) and how all the parameters are interconnected with each other. We use the correlation map to correlate the datasets in the table~\ref{tab:extraction_bounds_combined}. Combining all three tables, we analyze the data points using Kendall's tau method \cite{kendall1938new,kendall1990rank} over other standard linear methods like Pearson's method \cite{pearson1896mathematical} or continuous ranking methods like Spearman's method \cite{spearman1904proof}, as it distinguishes the physical coupling from random fluctuations by detecting monotonic relationships by doing a rank based phase-space analysis, such that we can compensate for the loss of accuracy. It evaluates data pair by pair, making it a more reliable method for smaller datasets. 
  Analyzing the table~\ref{tab:extraction_bounds_combined}, we observe that the event horizon gets locked to $(a=1.000)$, as $g=0$, which can falsify the correlation for $r_+$,  as it is compared with other data points, averaging out most portions. To overcome this we separate the correlation into two regimes.
  \begin{itemize}
      \item Regime 1 ($g=0$): Correlation doesn't have the $(g,r_+)$ parameters as its constant throughout. 
      \item Regime 2 ($g\neq0$): Correlation has the $(g,r_+)$ parameters as its varying throughout the data points will have correct mapping.
  \end{itemize}
 The benefit of separating them into two regimes is to remove the aggregation bias, and the mixing of locked states (i.e., $g=0, r_+=1.00$) with the varying ones would have dragged the correlation coefficient towards zero. We thereby prevented the dormant variables from masking the strong parameter couplings in the AdS background. The obtained correlation map is given in Fig.~\ref{Plot: Correlation map}, and the following observations for the two regimes are:
 \begin{enumerate}[label=\Roman*.]
     \item Regime 1 ($g=0$) Asymptotic Flat Spacetime :
     \begin{description}
         \item[Booster Parameters] $v$ acts as a booster for both efficiency ($\eta, \tau=0.74$) and extracted power($\mathcal{P}_{CA}, \tau=0.46$), and remains neutral for $(a_{ext},\tau=-0.03), (r_{erg}, \tau=-0.03))$. 
         \item [Damping Parameters] $e$ damps the ergo region and extremal spin, with similar correlation ($a_{ext}; \tau=-0.88$), similarly damps extracted power faster than efficiency as it is dependent on ergoregion ($ \mathcal{P_{CA}}, \tau= -0.61$). Hence, $e$ acts as a great damper here.
         \item [Damper-cum-Booster] $N_g$ sits in both positions here, as it damps ($\eta, \tau=-0.19$) but it also increases ergoregion and extremal spin $(a_{ext},\eta,\tau=0.18)$. So, if we want higher extracted power, we can achieve the desired result by taking an input set where ($N_g(\uparrow),v(\uparrow),e(\downarrow)$).
     \end{description}
    \item Regime 2 ($g\neq 0$) AdS Spacetime:
    \begin{description}
         \item[Booster Parameters] $v$ increases the horizon radius ($r_+, \tau=0.14$), adn boost the efficiency($\eta, \tau=0.59$), extracted power($\mathcal{P}_{CA}, \tau=0.20$), and remains neutral for $(a_{ext},\tau=-0.02),(r_{erg},\tau=-0.01)$. 
         \item [Damping Parameters] $(g,e)$ damp the ergoregion($r_{erg}, \tau=-0.86$), horizon($r_+, \tau=-0.77$), and extremal spin($a_{ext}, \tau=-0.62$), similarly, it damps extracted power ($\mathcal{P}_{CA}, \tau=-0.59$) as it is dependent on ergoregion but efficiency ($\eta,\tau =-0.19$) decreases due to sharp decrease in horizon. Additionally, $e$ damps all parameters but remains neutral toward the horizon ($r_+,\tau =-0.00$). 
         \item [Degeneracy breaking] Parameter $g$ breaks the degeneracy between $a_{ext}$ and $r_{erg}$ ,$N_g$ here damps all the output parameters, especially $(r{+},\tau=-0.34),(r_{erg}, \tau=-0.30)$, and consequently horizon ($r_+,\tau=-0.25$) and extracted power($\mathcal{P}_{CA},\tau=-0.28$) gets damped as well.
     \end{description}
 \end{enumerate}

\section{General Relativistic Effects on the Observable Reconnection Rate}
\label{sec: Observable reconnection rate}

Magnetic reconnection in relativistic plasmas is described by the Sweet-Parker model, where the reconnection rate is governed by the Lundquist number $S$. While the previous section focuses on the local plasma dynamics near the BH, observable quantities are necessarily defined with respect to an external observer in curved spacetime. In strong gravitational field regimes around the rotating BHs, physical observables such as velocities, length scales, and electromagnetic fields become frame-dependent. Therefore, even if the intrinsic reconnection physics remains unchanged, observable reconnection properties can acquire non-trivial dependence on spacetime geometry. With this motivation, we study the GR effect on the observable reconnection rate.

A covariant formulation of magnetic reconnection in curved spacetime has been developed in ~\cite{Shen2024}. It is shown that the intrinsic Sweet-Parker scaling remains unchanged in the local plasma rest frame, namely
\begin{equation}
R = \frac{v_{\rm in}}{v_{out}} \sim S^{-1/2},
\label{eq:local_scaling}
\end{equation}
where $v_{\rm in}$ and $v_{out}$ is the inflow velocity and relativistic Alfv\'en velocity, respectively, and $S$ is the Lundquist number.
This implies that spacetime curvature does not alter the fundamental reconnection mechanism, while observable quantities differ due to frame transformations, as shown in ~\cite{Shen2024}.

To start, we write the intrinsic Lundquist number explicitly as

\begin{equation}
S = \frac{L_c v_{out}}{\eta_r},
\label{eqn: lundquist number}
\end{equation}
where $L_c$ is the characteristic length of the current sheet and $\eta_r$ is the resistivity. The relativistic Alfv\'en velocity is given in equation \ref{vout}~\cite{Lyubarsky2005, ComissoAsenjo2018}, and the relativistic enthalpy density is, $w = \rho + p + B^2/4\pi$, substituted in the expression \ref{eqn: lundquist number} , we obtain 

\begin{equation}
S =
\frac{L_c}{\eta_r}
\sqrt{
\frac{
{B^2}
}{
{4\pi w} + {B^2}
}
}.
\end{equation}
This expression shows that the intrinsic Lundquist number depends only on local plasma properties. In a curved spacetime, this local definition remains unchanged, while observable quantities are modified due to frame transformations, as described in the covariant formulation of magnetic reconnection in ~\cite{Shen2024}.

We now define the observable Lundquist number as
\begin{equation}
S_{\rm obs} = \frac{L_{\rm obs} \, v_{A,{\rm obs}}}{\eta_r}.
\end{equation}
For a stationary and axisymmetric spacetime, we adopt a locally non-rotating (ZAMO) observer with four-velocity
\begin{equation}
n^\mu = \frac{1}{\alpha}(1,0,0,\omega),
\end{equation}
where $\alpha$ and $\omega$ as given in (\ref{alpha and beta}).
Using the observer-frame projection formalism developed in ~\cite{Shen2024}, physical quantities measured by a local observer are obtained by projecting the fluid variables onto the observer’s frame. For a stationary and axisymmetric spacetime, choosing a locally non-rotating (ZAMO) observer leads to simple relations between locally defined and observed quantities. In particular, the observed velocity acquires corrections due to gravitational redshift through the lapse function $\alpha$ and frame dragging through the angular velocity $\omega$, while spatial lengths are determined by the induced spatial metric. Specializing to a current sheet aligned along the azimuthal direction, these relations simplify considerably.

Following this, the observed Alfv\'en velocity can be written as
\begin{equation}
v_{A,{\rm obs}} = \frac{v_{out}}{\alpha(1-\Omega \omega)},
\end{equation}

\begin{figure*}[htbp]
    \centering
\vspace{0 cm}
    \includegraphics[width=0.82\textwidth, height=5.25cm, keepaspectratio=false]{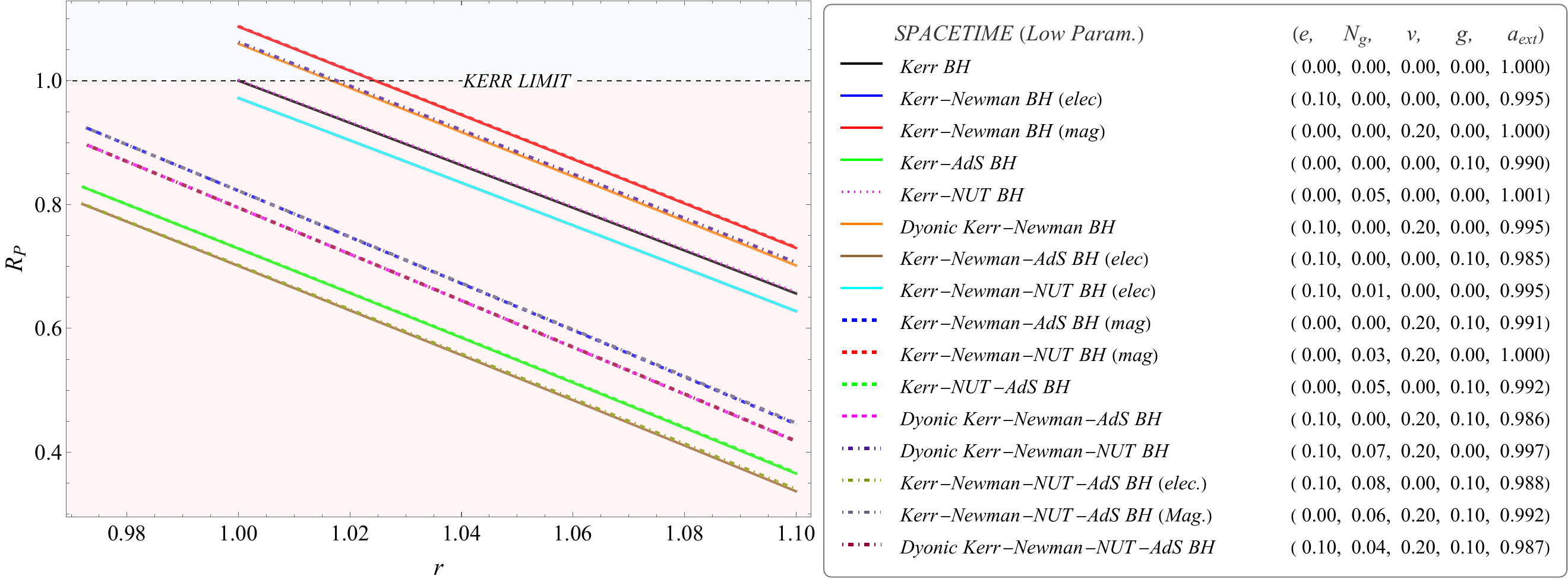}
    
    \vspace{0.0cm} 

    \includegraphics[width=0.82\textwidth, height=5.25cm, keepaspectratio=false]{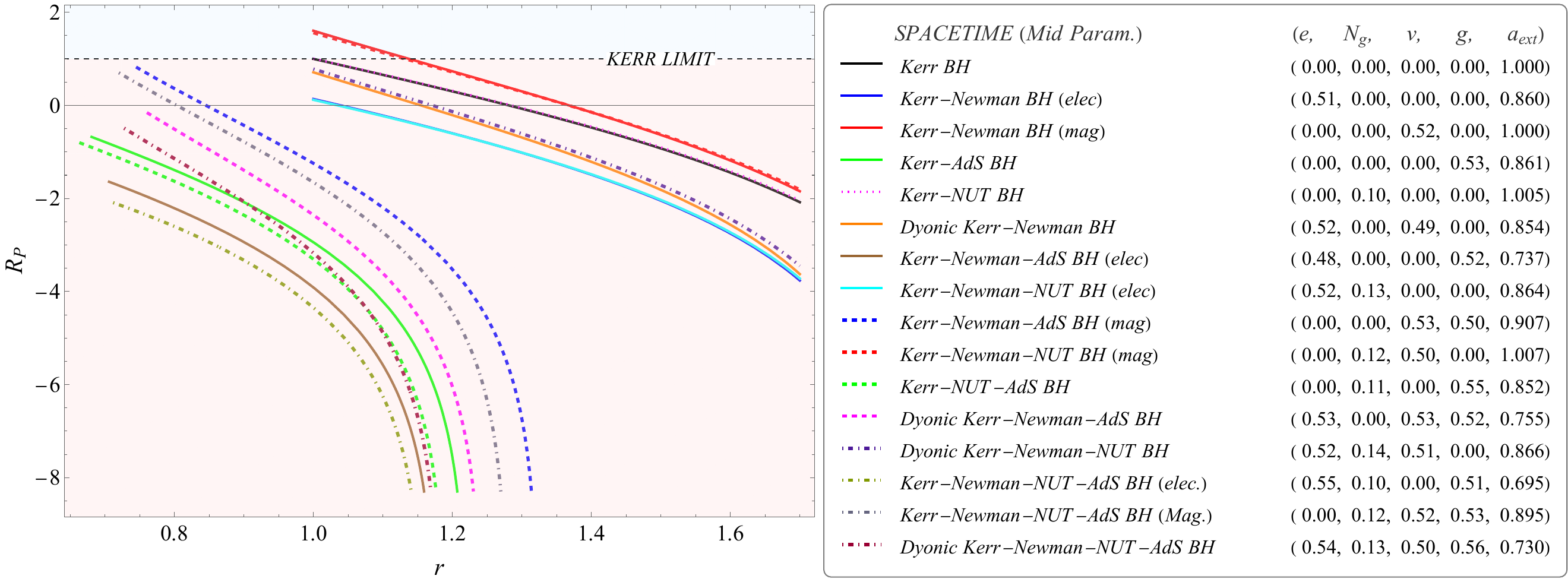}
    
    \vspace{0.0cm}

    \includegraphics[width=0.82\textwidth, height=5.25cm, keepaspectratio=false]{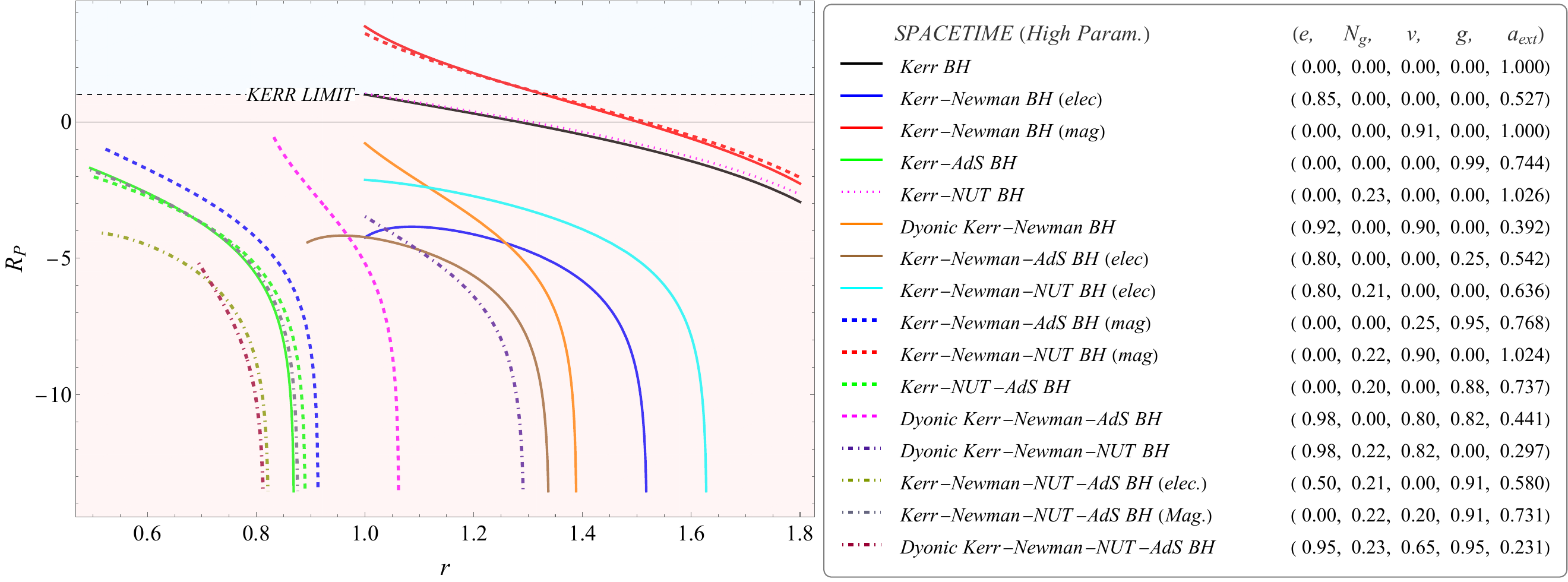}

     \vspace{0.0cm}

    \includegraphics[width=0.82\textwidth, height=5.25cm, keepaspectratio=false]{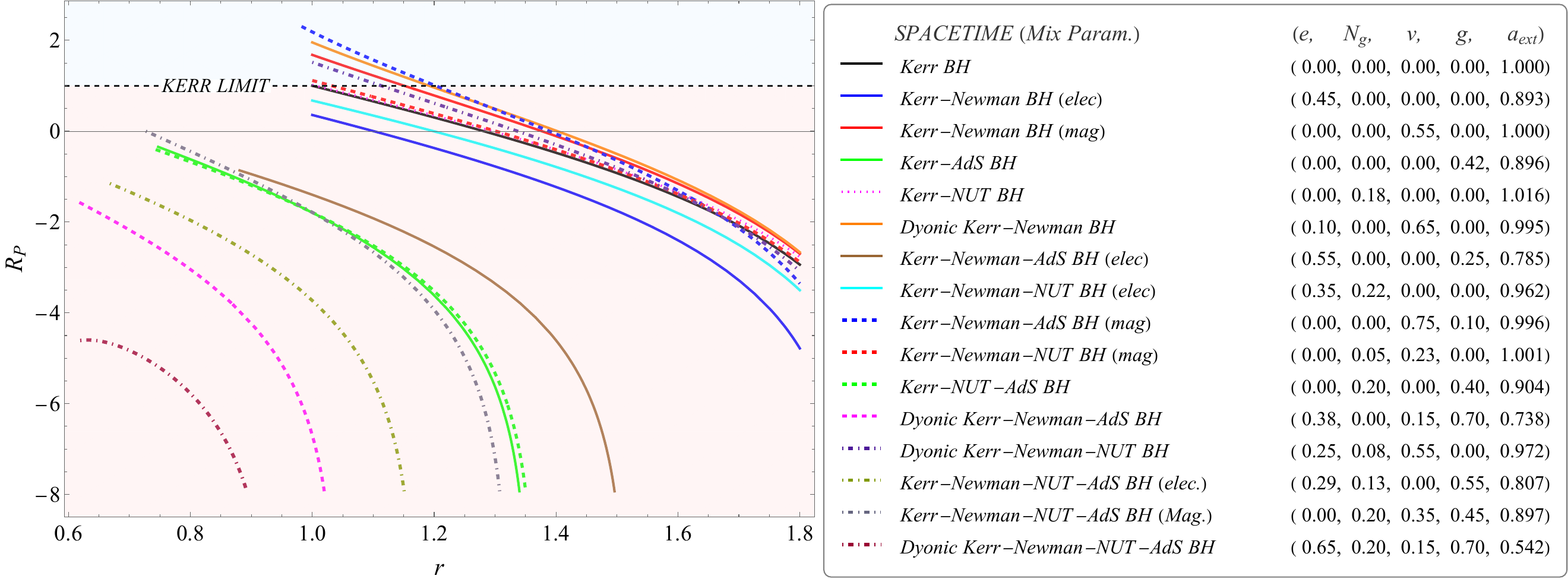}
    
    \caption{      \justifying{ Radial Dependence of Power ratio ${R}_{\mathcal{P}}$; various spacetimes plotted and compared with KBH for different parameter set ranges. The horizontal dashed line represents the Kerr power ratio limit, dividing the two regions (Blue and pink): the blue region is above the limit, and the pink region is below it. Whereas $M=1$, $(\sigma_0,\xi)$ $\equiv$ (100, 0) and the parameter range are given in Fig.~\ref{eff_vs_r_diff_g}}}
    \label{Plot: power_ratio_vs_r_diff_parameters}
\end{figure*}

where $\Omega$ is the angular velocity of the plasma. Similarly, the observable length of the current sheet becomes
\begin{equation}
L_{\rm obs} = \frac{L_c}{\sqrt{g_{\phi\phi}}}
= L_c \sqrt{\frac{(\mathcal{B}-aA)\Xi^2}{-R_g A^2 + \Theta_g \mathcal{B}^2 \sin^2\theta}}.
\end{equation}
These expressions represent simplified forms of the general observer-frame relations, adapted to the symmetries of the spacetime under consideration.
Combining these results, we obtain the observable Lundquist number
\begin{equation}
S_{\rm obs}(r,\theta)
=
\frac{L_c v_{out}}{\eta_r}
\cdot
\frac{\Xi }{\sqrt{R_g \Theta_g}\,\sin\theta}
\cdot
\frac{1}{1-\Omega\omega(r,\theta)}. 
\end{equation}
In terms of the intrinsic Lundquist number $S$, this becomes
\begin{equation}\label{S_obs_Dyonic1}
S_{\rm obs}(r,\theta)
=
S \cdot
\frac{\Xi }{\sqrt{R_g \Theta_g}\,\sin\theta}
\cdot
\frac{1}{1-\Omega\omega(r,\theta)}.
\end{equation}
In the Kerr limit ($N_g = 0$, $v = 0$, $g \to 0$), the above expression reduces to
\begin{equation}\label{S_obs_kerr1}
S_{\rm obs}^{\rm Kerr}
\sim
\frac{S}{\sqrt{\Delta}\,\sin\theta}
\cdot
\frac{1}{1-\Omega\omega_{\rm Kerr}},
\end{equation}
where, $\Delta=r^2-2Mr+a^2$. 
This form of $S_{\rm obs}^{\rm Kerr}$ contains the combined effect of gravitational redshift (through $\Delta$) and frame dragging (through $\omega_{\rm Kerr}$). However, the expression~\eqref{S_obs_kerr1} is not explicitly presented in earlier works; it is consistent with general relativistic magnetic reconnection models in Kerr BHs~\cite{Shen2024, ComissoAsenjo2018, Lyubarsky2005}. 
It is important to note that the factor $\sin\theta$ appearing in Eqs.~\eqref{S_obs_Dyonic1} and~\eqref{S_obs_kerr1} originate from the lapse function $\alpha$ (containing gravitational redshift and frame-dragging effects in the observer (ZAMO) frame). This dependence reflects the intrinsic structure of the rotating spacetime rather than the choice of current sheet orientation. The divergence of $S_{\rm obs}$ near the polar axis ($\theta \to 0$) is a coordinate effect associated with the degeneracy of the azimuthal direction and the breakdown of the ZAMO description and does not represent a physical singularity. Additionally, the Eqs.~\eqref{S_obs_Dyonic1} and~\eqref{S_obs_kerr1} demonstrate that these forms of $S_{\rm obs}$  are suitable for cases when the reconnection layers are embedded within disk-like plasma configurations, where the azimuthal direction is well-defined.

The observable reconnection rate is defined as
\begin{equation}
R_{\rm obs} = \frac{v_{\rm in,obs}}{v_{A,{\rm obs}}}.
\end{equation}
Using the observer-frame velocity transformation
\begin{equation}
v_{\rm obs} = \frac{v}{\alpha(1-\Omega\omega)},
\end{equation}
we obtain
\begin{equation}
R_{\rm obs}
=
\frac{v_{\rm in}}{\alpha(1-\Omega\omega)}
\Big/
\frac{v_{out}}{\alpha(1-\Omega\omega)}.
\end{equation}
The lapse function $\alpha$ and frame-dragging factor $(1-\Omega\omega)$ cancel identically, yielding
\begin{equation}
R_{\rm obs} = \frac{v_{\rm in}}{v_{out}} = R,
\end{equation}
consistent with the covariant formulation of magnetic reconnection in Ref.~\cite{Shen2024}. This demonstrates that spacetime curvature does not alter the intrinsic Sweet-Parker reconnection mechanism.

\begin{figure*}[htbp]
    \centering    
    \includegraphics[scale=0.37]{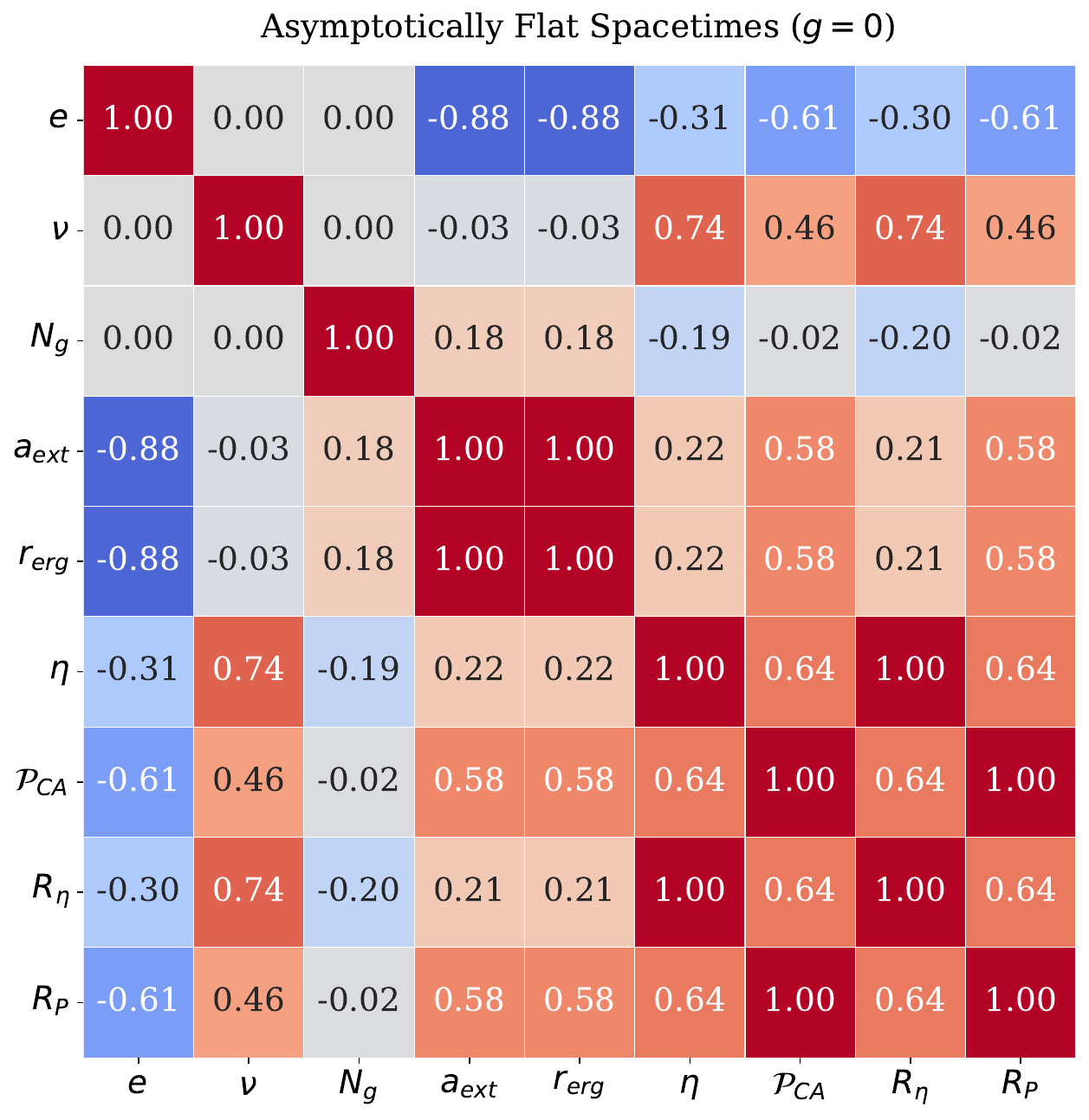}
    \hfill
     \includegraphics[scale=0.33]{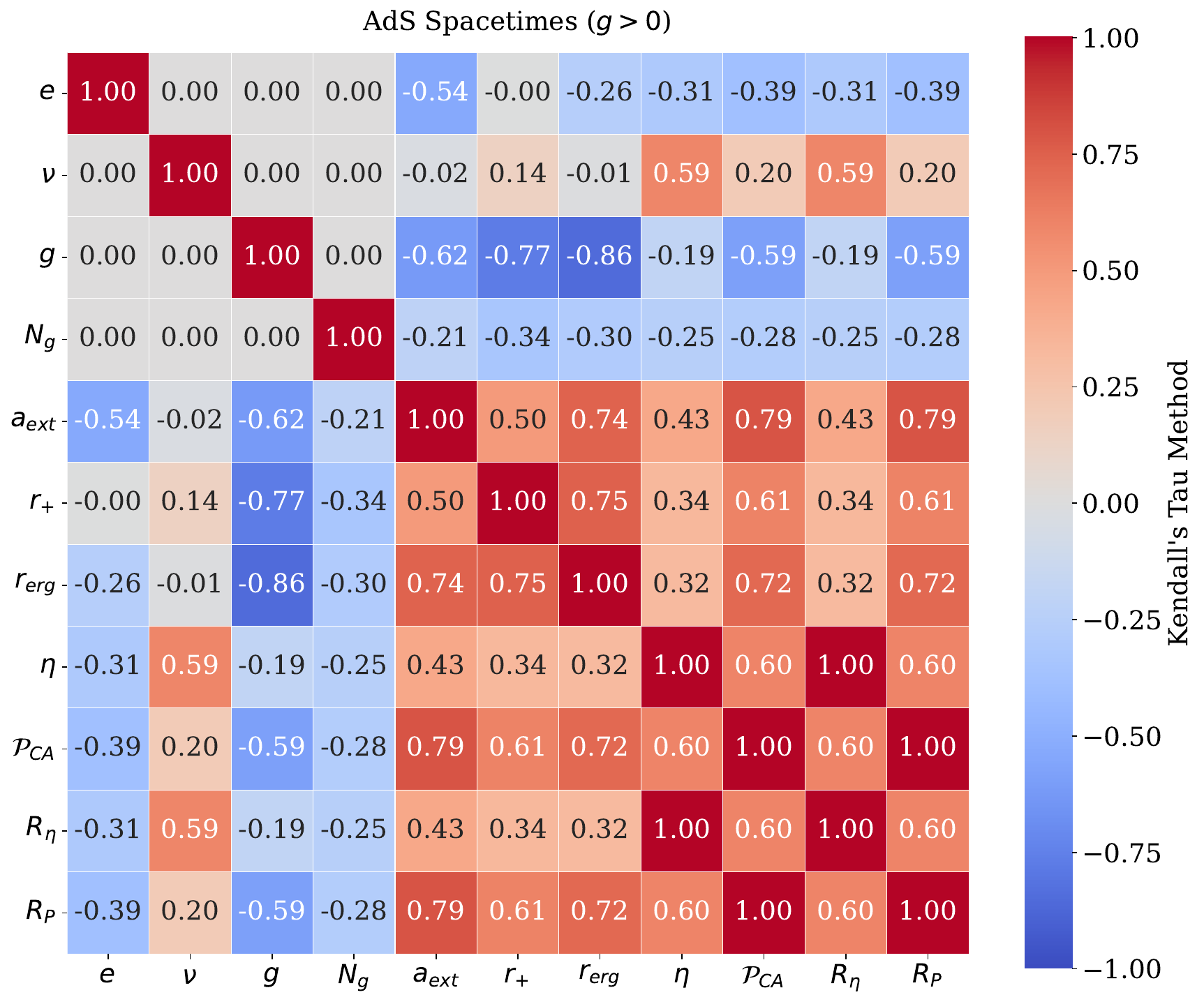}    
    \caption{      \justifying Correlation map relating the various parameters, data extracted from the Table~\ref{tab:extraction_bounds_combined}, where all the table data (low, mixed, and high) parameter data are combined to connect the relation with each other. Data sets from the table are separated into two regimes to prevent the masking of strong parameters by dormant parameters. }
    \label{Plot: Correlation map}
\end{figure*}

However, the observable Lundquist number $S_{\rm obs}$ acquires a non-trivial dependence on the spacetime geometry through the lapse function and frame-dragging effects. As a consequence, although the intrinsic relation $R \sim S^{-1/2}$ remains valid in the plasma rest frame, the corresponding observable quantities satisfy

\begin{equation}
R_{\rm obs} \neq S_{\rm obs}^{-1/2}.
\end{equation}
This deviation arises because $S_{\rm obs}$ incorporates geometric and kinematic effects that do not affect the intrinsic reconnection rate. Therefore, an observer inferring plasma properties from measured quantities would obtain an effective scaling that differs from the standard Sweet-Parker prediction.

\subsection{Observable scaling relation and geometric modification of Sweet-Parker reconnection}

We now derive an explicit relation between the observable reconnection rate $R_{\rm obs}$ and the observable Lundquist number $S_{\rm obs}$ in the Kerr-Newman-NUT-AdS BH spacetime. 
The intrinsic Sweet-Parker relation Eq.~(\ref{eq:local_scaling}) gives
\begin{equation}
R = S^{-1/2}
=
\left(\frac{S_{\rm obs}}{\mathcal{F}}\right)^{-1/2}
=
S_{\rm obs}^{-1/2}\,\mathcal{F}(r,\theta)^{1/2}.
\end{equation}
Using the fact that the reconnection rate is invariant under the observer transformation, $R_{\rm obs} = R$, we obtain the observable scaling relation
\begin{equation}
R_{\rm obs}(r,\theta)
=
S_{\rm obs}^{-1/2}
\,
\mathcal{F}(r,\theta)^{1/2}.
\end{equation}
This expression represents a modified Sweet-Parker scaling in curved spacetime. The factor $\mathcal{F}^{1/2}$ encodes the combined effects of gravitational redshift, frame dragging, and spacetime geometry. Specifically, the factor can be written as
\begin{equation}
\mathcal{F}^{1/2}
=
\left[
\frac{\Xi }{\sqrt{R_g \Theta_g}\,\sin\theta}
\right]^{1/2}
\cdot
\frac{1}{\sqrt{1-\Omega\omega(r,\theta)}}.
\end{equation}
The first term arises from the interplay of the lapse function and the spatial metric, while the second term captures the effect of frame dragging. Physically, $\mathcal{F}^{1/2}$ acts as an amplification factor that modifies the relation between observable quantities without altering the intrinsic plasma dynamics.

\begin{equation}
R_{\rm obs}(r,\theta)
=
S_{\rm obs}^{-1/2}
\left[
\frac{\Xi }{\sqrt{R_g \Theta_g}\,\sin\theta}
\cdot
\frac{1}{1-\Omega\,\omega(r,\theta)}
\right]^{1/2}.
\end{equation}

The above relation provides a modified observable scaling between the reconnection rate and the Lundquist number in curved spacetime. While the intrinsic Sweet-Parker scaling $R \sim S^{-1/2}$ remains valid locally, the observable quantities acquire additional geometric and kinematic corrections. In particular, the presence of the spacetime-dependent factor leads to a deviation from the standard scaling law, implying that $R_{\rm obs}$ can be significantly enhanced in regions of strong gravitational field and frame dragging.

To understand the physical implications of the observable reconnection rate scaling, we examine its limiting behaviour in three important regimes: the near-horizon region, the near-ergoregion, and the Kerr black hole limit for comparison, as follows:
\begin{enumerate}
\renewcommand{\labelenumi}{(\roman{enumi})}

\item \textbf{Near-horizon behavior:}
In the vicinity of the event horizon, the radial function satisfies $R_g \to 0$. Consequently,
\begin{equation}
\mathcal{F}^{1/2} \propto \frac{1}{(R_g)^{1/4}} \to \infty,
\end{equation}
which implies
\begin{equation}
R_{\rm obs} \to \infty.
\end{equation}
This indicates a strong enhancement of the observed reconnection rate near the horizon, even though the intrinsic reconnection physics remains unchanged.

\item \textbf{Near-ergoregion behavior:}
In regions where frame dragging becomes significant, the condition $1 - \Omega \omega \to 0$ can be approached. In this limit
\begin{equation}
\mathcal{F}^{1/2} \propto \frac{1}{\sqrt{1-\Omega\omega}} \to \infty,
\end{equation}
leading to
\begin{equation}
R_{\rm obs} \to \infty.
\end{equation}

This simply demonstrates that frame dragging can strongly amplify the observable reconnection rate, particularly within or near the ergoregion.
\item \textbf{Comparison with Kerr spacetime:}
In the Kerr limit ($v = 0$, $N_g = 0$, $g = 0$), the metric functions reduce to
\begin{equation}
R_g = r^2 - 2Mr + a^2=\Delta, \qquad \mathcal{B}-aA = r^2 + a^2 \cos^2\theta=\Sigma,
\end{equation}
and the frame-dragging angular velocity becomes
\begin{equation}
\omega = \frac{2aMr}{(r^2 + a^2)^2 - a^2 \Delta \sin^2\theta}=\omega_{\rm Kerr} .
\end{equation}
In this case, the observable scaling reduces to
\begin{equation}
R_{\rm obs}
=
S_{\rm obs}^{-1/2}
\left[
\frac{1}{\sqrt{\Delta}\sin\theta}
\cdot
\frac{1}{1-\Omega\, \omega_{\rm Kerr}}
\right]^{1/2}=R_{\rm obs}^{\rm Kerr}\, ,
\end{equation}
which is consistent with earlier studies of magnetic reconnection in Kerr spacetime \cite{Shen2024}.

\end{enumerate}


\begin{figure*}[htbp]
\includegraphics[width=0.48\textwidth]{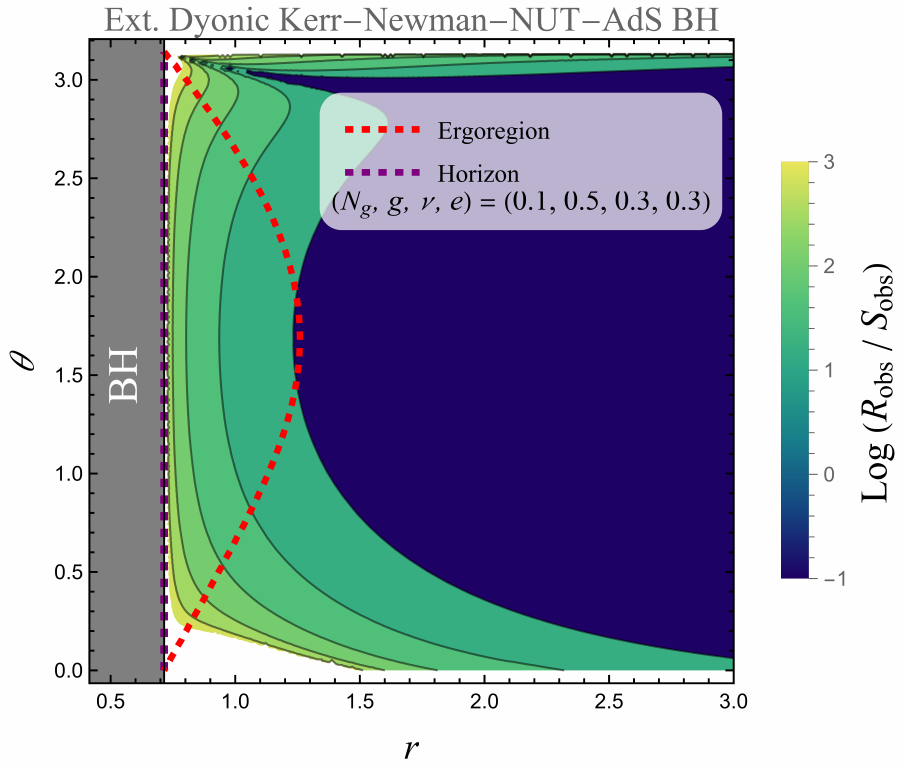}
\includegraphics[width=0.48\textwidth]{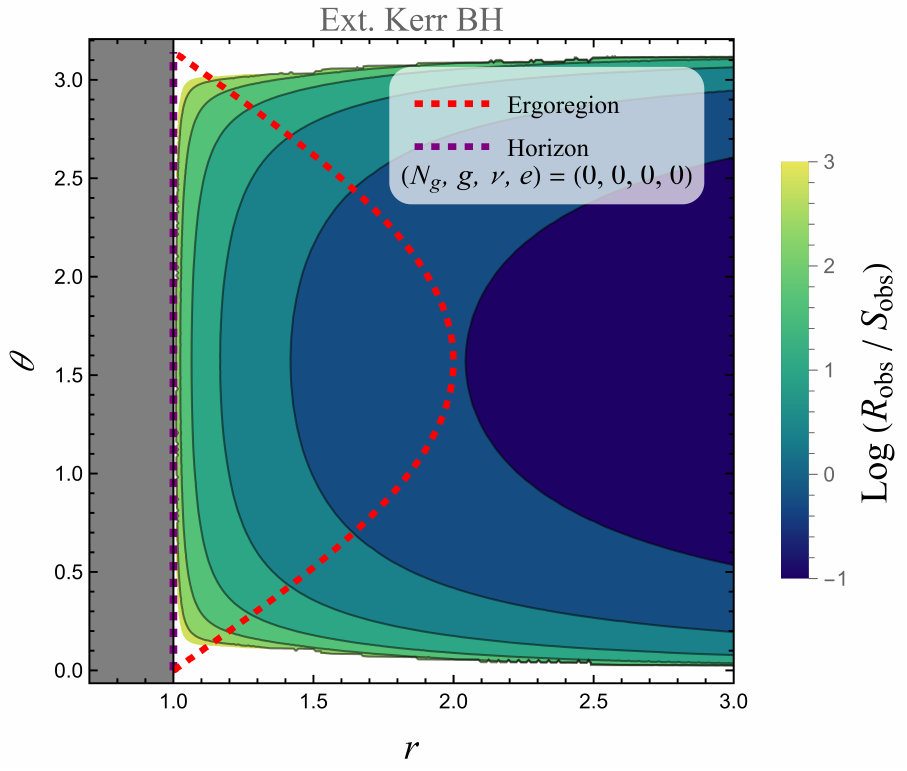}
\\[0.5em]

\caption{      \justifying
Comparison of the observable reconnection rate $R_{\rm obs}(r,\theta)$ for the extremal Dyonic Kerr-Newman-NUT-Ads BH and the extremal Kerr BH. 
Left panel: Extremal Dyonic Kerr-Newman-NUT-AdS BH with parameters $M=1.0$, $a=a_{\rm ext}$ (determined from the extremality condition), $e=0.3$, $v=0.3$, $N_g=0.1$, $g=0.5$, and $\Omega=0.2$.  
Right panel: Extremal Kerr BH with parameters $M=1.0$, $a=1.0$ (extremal limit), and $e=v=N_g=g=0$, while keeping $\Omega=0.2$ for consistency. The dashed vertical line (Purple) indicates the event horizon, while the dashed red curve marks the ergosphere boundary.}

\label{fig:ext_bh_comparison}
\end{figure*}

\begin{figure*}[htbp]
\centering
{\includegraphics[width=0.48\textwidth]{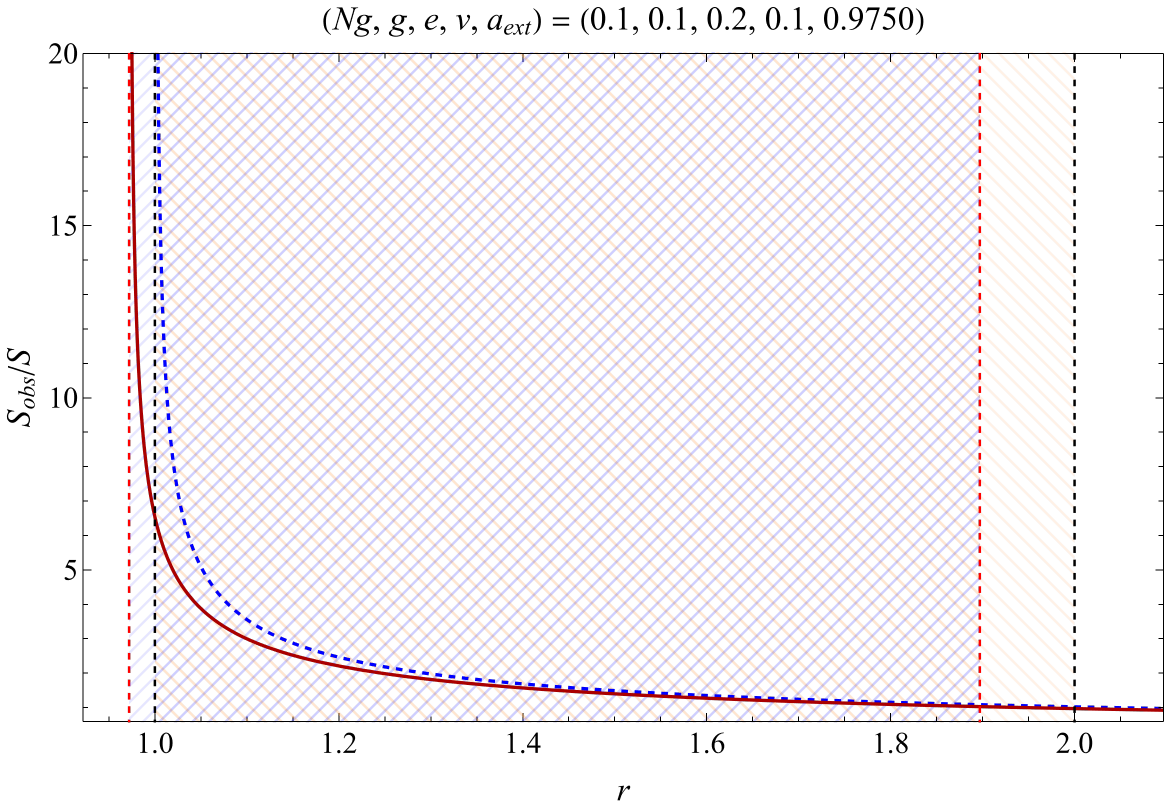}}
{\includegraphics[width=0.48\textwidth]{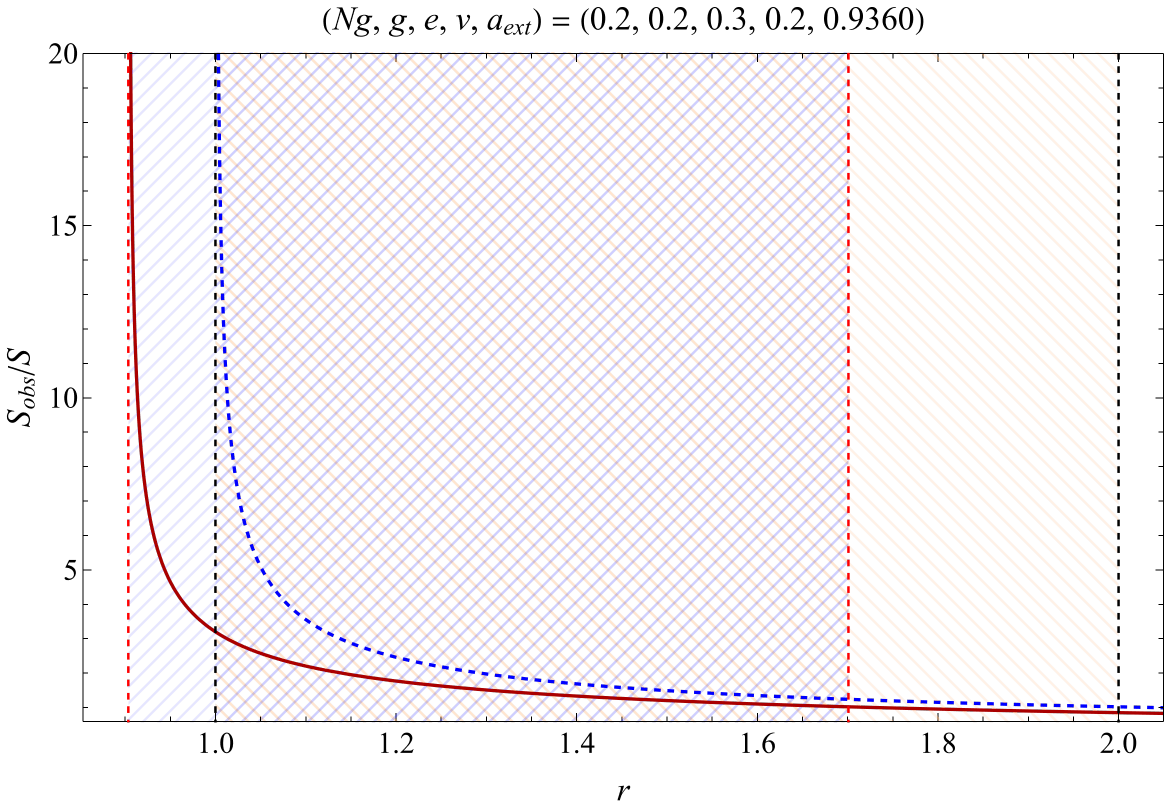}}
\vspace{0.3 cm}

{\includegraphics[width=0.48\textwidth]{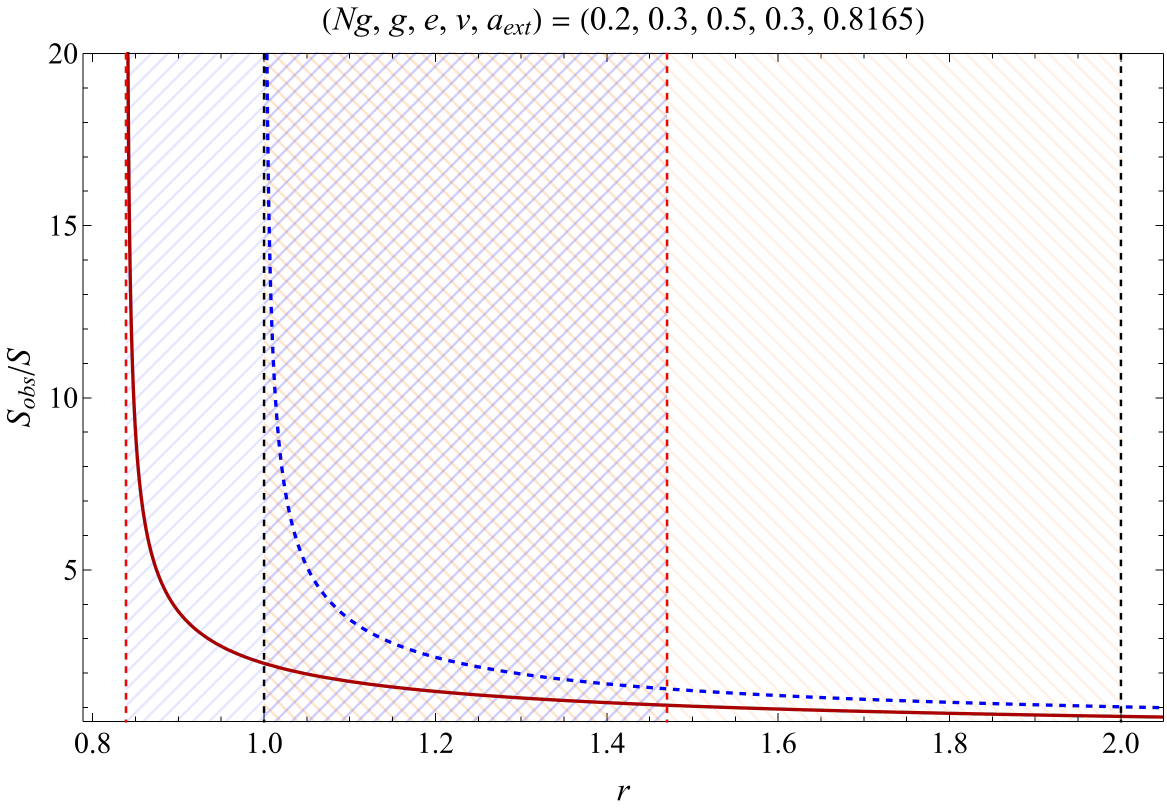}}
{\includegraphics[width=0.48\textwidth]{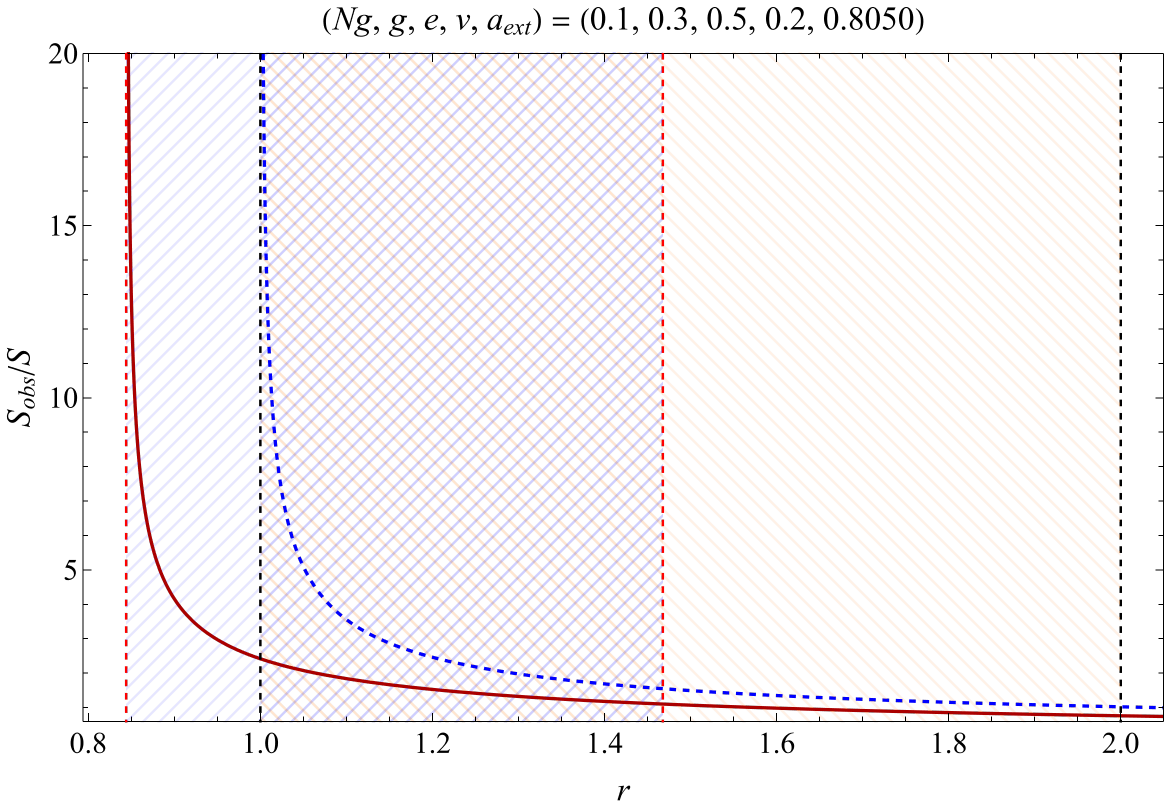}}

\caption{      \justifying{ Dependence of Normalized observable Lundquist Number ($S_{\rm obs}/S$) vs r on various spacetime parameters. The red solid line represents Kerr-Newman-NUT-AdS BH, and the dashed blue line denotes Kerr BH. The red and black vertical dashed lines on the left and right are the Event Horizon and the Static limit surface of Kerr-Newman-NUT-AdS BH AND Kerr BH, related to the solid line, respectively. The region between these two lines in red meshed region is the ergoregion where all phenomenon are taking place.}}
\label{fig: F_vs_r}
\end{figure*}

\section{Summary \& Conclusions}
\label{sec: summary}

Magnetic Reconnection has gained widespread acceptance as a mechanism for modeling the energy extraction process from plasmas near rapidly rotating BHs, where the magnetic reconnection leads to rapid changes in the magnetic field. In this paper, we considered the Dyonic Kerr-Newman-NUT-AdS BH as the host spacetime for the Comisso-Ansenjo magnetic reconnection process.  
\begin{itemize}
\item  We explored the spacetime metric by using parameter set ($N_g, g,e,v$), considering several distinct spacetimes, as shown in Table~\ref{Table: Various spacetimes}. The metric is rich in parameter space and highly nonlinear, thus restricting our focus to the extremal cases. The red curve in Fig.~\ref{Plot: extremal curve vs a} showed the variation of the extremal condition across different parameter spaces. This curve also separated the black hole region (i.e., the region where horizon conditions are satisfied) from the naked region (i.e., the no-horizon region) in the respective parameter space. Next, we calculated the other necessary quantities required to facilitate magnetic reconnection, i.e., the ergoregion and orbital conditions, which are essential in this model for magnetic reconnection and plasma acceleration, while keeping the singularity conditions satisfied for Dyonic Kerr-Newman-NUT-AdS BH. The angular deformation of the ergoregion resulted due to NUT charge and gauge constant as given in Eq. \ref{singularity}, leading to different singularity conditions at the two poles as seen in Fig~\ref{fig: dyonic Black Hole deformed}. 

     \item   {We derived the expression for energy per unit enthalpy for accelerated and decelerated plasma using the CA process in the framework of Dyonic Kerr-Newman-NUT-AdS BH. The addition of parameters, plasma magnetization and orientation angle ($\sigma_0,\xi$), respectively, made the framework richer. However, we mainly restricted our analysis to the equatorial plane (i.e., $\theta=\pi/2$) and strong-field regime for all the cases to get the maximum extracted energy, fixing $(\sigma_0,\xi)=(100,0)$ and comparing them with the extremal Kerr BH case.}
        
    \item  To satisfy the conditions for energy extraction, we plotted the region in the ($a,l$) plane in Fig.~\ref{Region Plot: spin vs r}, where our negative energy condition is satisfied. We observed various cases, the negative-energy region corresponds to different spin parameter values, as shown in Fig.~\ref{Region Plot: spin vs r}, $a$ varied between $a\in[0.6,1]$, which can be lower depending on the choice of parameters. We showed that the causality conditions are satisfied even in the slowly rotating case, indicating that Dyonic Kerr-Newman-NUT-AdS BHs can remain viable sources for energy extraction through the CA process. In addition, we found that the energy of decelerated plasma energy ($\epsilon _ {-} ^\infty$) becomes positive for some cases, which resulted in less extracted energy and power, as shown in Fig.~\ref{Plot: energy_pm vs r}.

\item The energy efficiency and the extracted power were examined in the extremal cases. To determine the parameter combinations yielding efficiencies $(\eta>1)$, we plotted Fig.~\ref{Plot: eff_vs_diff_parameters}, where $\eta \equiv \eta(N_g,g,v,e)$ is shown for different parameter choices. Each point in Fig.~\ref{Plot: eff_vs_diff_parameters} corresponds to an extremal configuration $(a,r)\equiv(a_{\mathrm{ext}},r_+)$. We found that the efficiency exhibits distinct behavior under different parameter settings. Based on these results, we selected representative parameter sets for further analysis and compared them with the Kerr BH, as summarized in Table~\ref{tab:extraction_bounds_combined}. We observed the variation of efficiency as the reconnection occurs further away from the BH horizon, as shown in Fig.~\ref{Plot: eff vs r different parameter values}. Consequently, when the parameter $v$ increases, $\eta$ increases. A monotonic decrease in efficiency was observed in all other parameter cases. This behavior is consistent with Fig.~\ref{Plot: energy_pm vs r}, where $\Delta\epsilon$ decreases for $r>r_+$, resulting in a decrease in $\eta$. Further, for the extracted power defined by Eq.~(\ref{eqn: Power extracted}), we found that higher efficiency does not necessarily correspond to higher extracted power. This happens because the extracted power mainly depends on the ergoregion's structure, which is influenced by parameter choices. We showed that the extracted power varied very differently for each parameter. The extracted power decreases as $r\rightarrow r_{\mathrm{erg}}$, but for sets where $g$ increases (as shown in Fig.~\ref{plot: power vs r for diff a vals}), the extracted power is lower.

\item   In Table~\ref{tab:extraction_bounds_combined}, we numerically compared various spacetimes with the Kerr BH in terms of efficiency, extracted power, efficiency ratio, and extracted power ratio, evaluated in the low, mixed, and high parameter regimes. To visually confirm how these ratios behave as a function of the radial distance $r$, we plotted Figs.~\ref{Plot: eff_ratio_vs_r_diff_ST} and~\ref{Plot: power_ratio_vs_r_diff_parameters}, where the blue regions indicate higher efficiency and extracted power ratios than the Kerr BH, while the red regions correspond to lower efficiency and extracted power ratios. It is worth noting that our analysis revealed that the Dyonic Kerr-Newman-NUT-AdS BH indeed had an advantage over the Kerr BH in many cases, i.e., for both slow and rapidly rotating regimes, especially when the reconnection occurred closer to the horizon. From Table~\ref{tab:extraction_bounds_combined}, we saw that the Kerr-Newman (mag.) BH, Kerr-Newman-NUT (mag.) BH, and Kerr-Newman-AdS (mag.) BH exhibited higher efficiency and extracted power. All the results, including the efficiency and power extracted from the Kerr BH, could be recovered by taking the appropriate limits of the Dyonic Kerr-Newman-NUT-AdS BH, i.e., $(N_g, g, v, e) \rightarrow 0$.

\item To get a clearer relationship between all parameters ($N_g,g,v,e$) and the extracted outputs ($a_{ext},r_E,\epsilon_{\pm}, \eta, \mathcal{P}_{CA}$), we utilized Kendall's rank correlation ($\tau$) (see Fig.\ref{Plot: Correlation map}). We distributed data points into two regimes: $g=0$ and $g\neq 0$. Two Kendall's correlation Maps were obtained and analyzed.
\begin{enumerate}[label=\Roman*.]
    \item ($g=0$): $v$ and $Ng$ act as boost parameters, enhancing the efficiency and extracted power, when $e$ is kept low.
    \item ($g\neq0$): Introduction of parameter $g$ breaks the degeneracy we saw in ($g=0$) case. $v$ solely acts a booster while other parameters damps the ($r_+,r_{erg},\eta,\mathcal{P}_{CA}$).
\end{enumerate}
We got a consolidated relationship for all parameters, and identified key trends containing boost and damped variables at a single point, without needing separate scattered plots for different spacetimes.

\item At last, we showed that the observable Lundquist number $S_{obs}$ in rotating BH spacetimes acquires an observer-dependent angular dependence through the lapse function as seen in Fig.~(\ref{fig:ext_bh_comparison}), leading to deviations from the standard Sweet-Parker scaling as seen in Fig.~\ref{fig: F_vs_r}, when expressed in terms of observable quantities. At last, to understand the physical implications, we examined its limiting behavior in different regimes, and it is perfectly consistent with earlier studies of Kerr BH spacetime.
\end{itemize}

}

\appendix
\section{Appendix: Roots and Extremality}
\subsection{Analytical Determination of the Horizon Roots}
\label{Analytic_roots}

Starting from the depressed quartic equation
\begin{equation}
r^4 + \zeta r^2 + \beta r + \Upsilon = 0,
\end{equation}
Ferrari’s method introduces an auxiliary parameter $y$ satisfying the resolvent cubic
\begin{equation}
y^3 - \zeta y^2 - 4\Upsilon y + (4\zeta\Upsilon - \beta^2)=0.
\end{equation}
whose real solution is $y_0$. Now, in order to calculate $y_0$, we start by
defining
\begin{equation}
y = z + \frac{\zeta}{3},
\end{equation}
the cubic reduces to the depressed form
\begin{equation}
z^3 + p z + q =0,
\end{equation}
with
\begin{align}
p &= -\frac{\zeta^2}{3} - 4\Upsilon, \\
q &= -\frac{2\zeta^3}{27} + \frac{4\zeta\Upsilon}{3} - \beta^2.
\end{align}

Using Cardano’s formula, the real solution is
\begin{equation}
z =
\sqrt[3]{-\frac{q}{2} + \sqrt{\Delta}}
+
\sqrt[3]{-\frac{q}{2} - \sqrt{\Delta}},
\end{equation}
where the discriminant is
\begin{equation}
\Delta =
\left(\frac{q}{2}\right)^2
+
\left(\frac{p}{3}\right)^3.
\end{equation}

Hence,
\begin{equation}
y_0 =
\frac{\zeta}{3}
+
\sqrt[3]{-\frac{q}{2} + \sqrt{\Delta}}
+
\sqrt[3]{-\frac{q}{2} - \sqrt{\Delta}}.
\end{equation}

Substituting $y_0$ into
\begin{equation}
R = \sqrt{\frac{\zeta^2}{4} - \Upsilon + y_0},
\end{equation}
and then into Eq.~(\ref{quarticroots}), yields the four analytical roots of $R_g(r)=0$.

\subsection{Extremality Condition}
\label{Extremality Condition}
The extremal configuration is determined by imposing
\begin{equation}
R_g(r_E)=0, 
\qquad 
R_g'(r_E)=0.
\end{equation}

Writing the radial function in compact form
\begin{equation}
R_g(r)= g^2 r^4 + \mathfrak{B} r^2 - 2Mr + C,
\end{equation}
with
\begin{align}
\mathfrak{B} &= 1 + g^2(a^2 + 6N_g^2 - 2\nu^2), \\
C &= a^2 + e^2 - N_g^2 + 3g^2N_g^2(a^2 - N_g^2),
\end{align}
its derivative is
\begin{equation}
R_g'(r)= 4g^2 r^3 + 2\mathfrak{B}r - 2M.
\end{equation}

From $R_g'(r_E)=0$, we obtain
\begin{equation}
M_E = 2g^2 r_E^3 + \mathfrak{B} r_E.
\label{mext}
\end{equation}

Substituting Eq.~(\ref{mext}) into $R_g(r_E)=0$ yields
\begin{equation}
3g^2 r_E^4 + \mathfrak{B} r_E^2 - C = 0.
\end{equation}

Defining $x = r_E^2$, the above reduces to a quadratic equation,
\begin{equation}
3g^2 x^2 + \mathfrak{B}x - C = 0,
\end{equation}
whose solution gives
\begin{equation}
r_E^2 =
\frac{-\mathfrak{B} \pm \sqrt{\mathfrak{B}^2 + 12g^2C}}{6g^2}.
\end{equation}

Choosing the positive branch gives the physical extremal radius
quoted in Eq.~(\ref{rEexplicit}). 

Substituting the physical branch into Eq.~(\ref{mext}), the extremal mass is obtained explicitly as

\begin{equation}
M_E =
\frac{1}{3}
\sqrt{
\frac{-\mathfrak{B} + \sqrt{\mathfrak{B}^2 + 12g^2C}}{6g^2}
}
\left(
2B + \sqrt{\mathfrak{B}^2 + 12g^2C}
\right).
\end{equation}

This expression provides the complete analytical extremality condition
\(
M = M_E(a,e,N_g,\nu,g)
\)
for which the outer and Cauchy horizons coincide, i.e.,
\(
r_+ = r_- = r_E
\).

\section*{Acknowledgements}
AJ acknowledges the financial support from the University Grants Commission (UGC) under the JRF scheme. Authors HN and PS would like to acknowledge financial support from the Anusandhan National Research Foundation (ANRF), New Delhi, under grant number CRG/2023/008980. The authors also acknowledge the research facilities at the IUCAA Center for Astronomy Research and Development (ICARD), HNB Garhwal University.

\bibliographystyle{JHEP}

\bibliography{biblio}
\end{document}